\RequirePackage{silence}
\documentclass[fleqn,usenatbib,useAMS]{mnras}

\usepackage{newtxtext,newtxmath}

\usepackage{graphicx}
\usepackage{amsmath}

\usepackage{amsfonts}
\usepackage{float}
\usepackage{bm}
\usepackage{ae,aecompl}
\usepackage{array}
\usepackage{soul}
\usepackage{mathtools}
\usepackage{multirow}
\usepackage[T1]{fontenc}
\usepackage[utf8]{inputenc}
\usepackage{booktabs}
\usepackage{graphicx}
\usepackage{subcaption}
\usepackage[normalem]{ulem}
\usepackage{anyfontsize}
\usepackage{xcolor}
\usepackage{orcidlink}
\usepackage{fontawesome5} 

\DeclareRobustCommand{\VAN}[3]{#2}
\let\VANthebibliography\thebibliography
\def\thebibliography{\DeclareRobustCommand{\VAN}[3]{##3}\VANthebibliography}

\DeclareRobustCommand{\appropto}{\mathrel{\vcenter{
		\offinterlineskip\halign{\hfil$##$\cr 
			\propto\cr\noalign{\kern2pt}\sim\cr\noalign{\kern-2pt}}}}}

\makeatletter
\DeclareRobustCommand*\bigcdot{\mathpalette\bigcdot@{2.4}}
\DeclareRobustCommand*\bigcdot@[2]{\mathbin{\vcenter{\hbox{\scalebox{#2}{$\m@th#1\bullet$}}}}}
\makeatother

\makeatletter
\DeclareRobustCommand*\diamond{\mathpalette\diamond@{1.6}}
\DeclareRobustCommand*\diamond@[2]{\mathbin{\vcenter{\hbox{\scalebox{#2}{$\m@th#1\blacklozenge$}}}}}
\makeatother

\DeclareRobustCommand*{\matr}[1]{\mathbfss{#1}}
\DeclareRobustCommand{\dd}{\mathop{}\!\mathrm{d}}

\defcitealias{Haslbauer_2020}{HBK20}
\defcitealias{Najera_2026}{N26}

\title[Constraints on the gravitational potential from the CMB and DESI DR2 BAO]{Constraints on the gravitational potential from DESI DR2 BAO and its implications for the local void scenario} 

\author[I. Banik et al.]{Indranil Banik$^{1}$\thanks{E-mail: \href{mailto:indranil.banik@port.ac.uk}{indranil.banik@port.ac.uk} (Indranil Banik)}\orcidlink{0000-0002-4123-7325}, Jos{\'e} Antonio N{\'a}jera$^{1}$\orcidlink{0000-0001-9738-7704} and Harry Desmond$^{1}$\orcidlink{0000-0003-0685-9791}\\
$^{1}$Institute of Cosmology and Gravitation, University of Portsmouth, Dennis Sciama Building, Burnaby Road, Portsmouth PO1 3FX, UK}

\date{Accepted XXX. Received YYY; in original form ZZZ}

\pubyear{\the\year{}}

\begin{document}
\label{firstpage}
\pagerange{\pageref{firstpage}--\pageref{lastpage}}
\maketitle

\begin{abstract} 
We constrain the difference in gravitational potential between our location and sources at $z \ga 0.3$ using datasets at those redshifts. Our motivation is that the Hubble tension might be caused by a local void, as suggested by galaxy number counts. This would increase the redshift through outflow and gravitational redshift (GR). Only the latter is important at high redshift, where a void contributes a fixed additional GR contribution of $z_0$ due to our location on a potential hill. This $z_0$ model has various subtle effects that were not previously considered, including a hotter CMB and reduced BAO scale $r_{\rm d}$. We test whether $z_0$ can have the previously expected value of 0.84\%, which was based on fitting void parameters to galaxy number counts and local $H_0$ measurements. Combining BBN, CMB, BAO, and CC datasets at $z > 0.5$, we find that $z_0 = -0.4^{+0.8}_{-0.9}\%$, which rises to $-0.1 \pm 0.7\%$ when extending our analysis down to $z > 0.29$. Although the results prefer the standard value of $z_0 = 0$, the best-fitting model with $z_0 = 0.84\%$ fits the data almost as well as $\Lambda$CDM, with $\Delta \chi^2 < 2$. We find that $\Lambda$CDM faces a $3.07\sigma$ BAO anomaly in the standard $(H_0 r_{\rm d}, \Omega_{\rm m})$ parameter space, where different regions are preferred by BAO and non-BAO datasets from $z > 0.29$. Fixing $z_0 = 0.84\%$ reduces this to $2.79\sigma$. This suggests that a local void large enough to solve the Hubble tension cannot be ruled out by higher-redshift datasets despite its novel impacts on them.

\end{abstract}



\section{Introduction}
\label{Introduction}

Although it has been clear for over a century that the Universe is expanding \citep{Wirtz_1922, Wirtz_1924}, the present rate of expansion remains a topic of major controversy known as the Hubble tension \citep{Valentino_2025, H0DN_2026}. This arises due to a mismatch between the observed rate at which redshift $z$ increases with distance $r$ in the local Universe, and the predicted rate of increase in the Lambda cold dark matter ($\Lambda$CDM) standard cosmological paradigm \citep*{Efstathiou_1990, Ostriker_Steinhardt_1995} calibrated using observations of the early Universe. If the universe is taken to be homogeneous and isotropic, we must have that
\begin{eqnarray}
    cz' ~=~ \dot{a} ~=~ H_0 \, ,
    \label{eq:Hubble_law}
\end{eqnarray}
where $c$ is the speed of light, $z' \equiv \dd z/ \dd r$ in the local Universe, $a$ is the cosmic scale factor normalised to unity today, overdots denote time derivatives, 0 subscripts denote quantities at the present epoch, and $H \equiv \dot{a}/a$ is the Hubble parameter \citep*{Mazurenko_2025}.

We can predict $H_0$ from unrelated observations in several ways, but the technique most commonly used nowadays involves fitting the precisely observed power spectrum of anisotropies in the cosmic microwave background (CMB) using the $\Lambda$CDM model. The power spectrum has characteristic oscillatory features which provide a wealth of information about the baryon--photon plasma in the primordial Universe \citep{Hu_1996, Eisenstein_1998}. Modern observations cover up to ten peaks in the CMB power spectrum \citep{Planck_2020, Louis_2025, SPT_2026}. This is enough to fix the $\Lambda$CDM parameters to high precision, leading to a clear prediction that $H_0 = H_0^{\rm CMB} = 67.19 \pm 0.38$~km/s/Mpc \citep{SPT_2026}.

This precise prediction is in $>7\sigma$ tension with the ``community consensus'' estimate of $cz' = 73.50 \pm 0.81$~km/s/Mpc reported recently by the $H_0$ Distance Network Collaboration \citep{H0DN_2026}. This estimate combines several popular methods and takes into account their covariances, though it does not properly account for selection effects and homogeneous Malmquist bias, the larger volume per unit distance at larger distances due to sources being distributed in 3D \citep{Desmond_2026, Stiskalek_2026_Cepheids, Stiskalek_2026_MW}. A wide variety of different techniques to construct the local distance ladder give numerically similar estimates of $cz'$ \citep{Scolnic_2023}. The most precise results rely on using the period-luminosity relation or Leavitt Law of Cepheid variable stars \citep{Leavitt_1912} to calibrate Type~Ia supernovae (SNe~Ia), which are then assumed to be physically similar out to $z = 0.15$ \citep{Riess_2022_comprehensive, Breuval_2024}. Single-step techniques such as masers \citep*{Pesce_2020, Barua_2025} and Type~II SNe \citep{Vogl_2025} have also been claimed to produce $cz'$ values above $H_0^{\rm CMB}$. This is also the case with a two-rung geometry-to-Cepheids ladder \citep{Stiskalek_2026_Cepheids}. For a comprehensive review of the Hubble tension and possible solutions, we refer the reader to \citet{Valentino_2025}.

Given the evidence that $cz' > H_0^{\rm CMB}$, many workers have tried to find a way to explain the CMB anisotropies in a cosmology with higher $H_0$, for instance with early dark energy \citep{Poulin_2019, Poulin_2023} or primordial magnetic fields \citep*{Mirpoorian_2025}. Whatever the precise details, the lack of new physics in the late universe implies that $\dot{a}(z)$ should be 9\% above that predicted by the CMB-calibrated $\Lambda$CDM cosmology for $z \la 1000$. This generic prediction runs into several difficulties. For instance, a faster expansion rate throughout nearly the entirety of cosmic history would reduce the age of the universe from the usual 13.8~Gyr down to $\approx 13.8/1.09 = 12.7$~Gyr \citep{Bernal_2021}. This is inconsistent with the ages of the oldest Galactic stars and globular clusters \citep{Valcin_2020, Bernal_2021, Montalban_2021, Valcin_2021, Limberg_2022, Xiang_2022, Cimatti_2023, Nepal_2024, Souza_2024, Lundkvist_2025, Valcin_2025, Xiang_2025, Banik_2026_age, Shariat_2026, Tomasetti_2026, Valcin_2026}. While these results are based on absolute stellar ages, similar conclusions can be drawn from differential ages between galaxies at different redshifts \citep{Cogato_2024, Guo_2025}. In general, probes of the expansion rate at intermediate epochs line up well with the \emph{Planck} cosmology \citep[for a review, see][]{Banik_2025_cosmology}. Moreover, one might expect that substantial changes to pre-recombination physics would distort the CMB power spectrum in some way. However, there are no anomalies in the CMB power spectrum that might be associated with the hypothesised new physics prior to recombination \citep{Planck_2020, Tristram_2024, Calabrese_2025, SPT_2026}. There are several additional challenges to this approach \citep{Vagnozzi_2023}, which may also introduce some tension with Big Bang nucleosynthesis (BBN) given that CMB fits with higher $H_0$ are usually associated with a higher baryon density \citep*{Giovanetti_2026, Launders_2026}.

Since the CMB only probes the late Universe through the comoving distance to recombination, it is possible to distort the expansion history such that this integral constraint is preserved but $H_0$ is higher. A crucial constraint on such models is baryon acoustic oscillation (BAO) measurements from DESI data release~2 \citep[DESI~DR2;][]{DESI_2025}. The BAO provides a statistical standard ruler whose comoving size $r_{\rm d}$ was fixed by the sound horizon at early times \citep{Hu_1996, Eisenstein_1998}. By observing the angular scale and redshift depth of the BAO ruler, we can place tight constraints on the expansion history at intermediate epochs \citep{Chen_2024_BAO, Banik_2025_BAO}. A recent exploration of several background solutions to the Hubble tension found that all the considered models struggle when confronted with the BAO (\citealt{Najera_2026}, hereafter \citetalias{Najera_2026}). The more physically motivated models generally struggled more, while the phenomenological (``Phen'') models performed better. A major result was that the Phen models can succeed to a large extent, but the deviation from $\Lambda$CDM behaviour should be restricted to rather low redshifts, far below the scale where dark energy causes $\ddot{a}$ to switch sign. It is unclear why $H(z)$ would start deviating from $\Lambda$CDM only at such low redshifts after working well for billions of years. This would require very unusual behaviour for the dark energy equation of state in the context of General Relativity, though the empirically reconstructed behaviour might be natural in some modified gravity theory.

Given these difficulties with solving the Hubble tension through new physics prior to recombination or a modified background expansion history at late times, we must seriously question the validity of Equation~\ref{eq:Hubble_law}, the only common thread in all the proposals mentioned above. The only way to violate Equation~\ref{eq:Hubble_law} is if some of the observed redshift arises from sources other than homogeneous cosmic expansion. $cz'$ can be inflated if there are outward peculiar velocities or we are located on a potential hill, creating gravitational redshift (GR) contributions as photons from the distant universe struggle against gravity to reach our detectors \citep[equation~1 of][]{Zhao_2013_GR}. Both arguments require that we live in a local underdensity or void. While these non-cosmological contributions to $z$ would generally decay as we get to larger $z$, an increasing trend is possible over a limited range of radii, depending on the void properties and our vantage point. A void of the required size and depth is not possible in $\Lambda$CDM, which predicts too little cosmic variance in $cz'$ to explain the Hubble tension this way \citep{Wu_2017, Camarena_2018}. Therefore, the local void scenario still requires new physics -- but not at the background level. The main requirement is enhanced growth of structure on scales $\ga 100$~Mpc, which we suggest is related to a modified behaviour of gravity on these scales.

Unlike most solutions to the Hubble tension which were proposed only after it became apparent, a local void was proposed many decades prior based on optical galaxy number counts \citep{Maddox_1990, Shanks_1990}. The void was later claimed in near-infrared observations covering 90\% of the sky, which suggest that we live inside the Keenan-Barger-Cowie (KBC) void \citep*{Keenan_2013}. There is evidence for the KBC void across the electromagnetic spectrum, from X-rays \citep{Bohringer_2015, Bohringer_2020} to radio wavelengths \citep*{Rubart_2013, Rubart_2014} and in the infrared \citep{Huang_1997, Busswell_2004, Frith_2003, Frith_2005, Frith_2006, Keenan_2013, Whitbourn_2014, Whitbourn_2016, Wong_2022}. Since the underlying observable is the galaxy luminosity density or source number density in redshift space, these results are not dependent on what value is assumed for the local $cz'$. The increasing comoving luminosity density towards larger $z$ is difficult to understand through selection effects, which would instead be expected to favour the detection of nearby galaxies. The observations can be understood in a homogeneous universe only if we assume extreme evolution of the galaxy population at $z \la 0.1$ followed by much milder evolution out to $z = 1$ \citep{Wong_2022}. Another argument for a local underdensity is that the mass fraction of galaxies residing in their stellar haloes increases with redshift \citep{Tao_2026}. Because this halo fraction traces the merger history of galaxies, it can serve as a probe of the local galaxy number density, as noted by those authors.

The local void scenario was explored in some detail by \citealt*{Haslbauer_2020} (hereafter \citetalias{Haslbauer_2020}). Those authors evolved a semi-analytic model of a small initial underdensity from $z = 9$ to the present, enhancing the gravitational field using an approach inspired by Milgromian dynamics \citep[MOND;][]{Milgrom_1983}. It proved possible to obtain a joint fit to the local $cz'$ and the redshift space comoving density profile of the KBC void, whose apparent density contrast in redshift space is inconsistent with $\Lambda$CDM expectations at $>6\sigma$ confidence \citepalias{Haslbauer_2020}. More recently, their best-fitting models were claimed to provide a good match to BAO observations over the last 20 years \citep{Banik_2025_BAO}. One can conversely infer the existence of a local void similar to that claimed by \citet{Keenan_2013} without using galaxy number counts if a background \emph{Planck} cosmology is assumed and local $cz'$ constraints are applied \citep*{Futamase_2026}. These and other studies on the local void model were recently reviewed in \citet{Banik_2026_void}.

Just like background solutions to the Hubble tension, a local void would necessarily introduce `spillover' effects at high $z$, by which we mean that some impact on cosmology is inevitable besides inflating the local $cz'$. Our main goal in this contribution is to check if the predicted spillover effects of a local void are consistent with the latest observations, which by now are quite precise yet remain close to $\Lambda$CDM expectations. While some of the effects we will consider have been previously discussed \citepalias{Haslbauer_2020}, they have not been calculated in detail or compared to the latest observations, while other effects were not recognised. This is crucial to addressing high precision BAO observations, which were claimed to provide strong support for the \citetalias{Haslbauer_2020} void model compared to the homogeneous \emph{Planck} cosmology \citep{Banik_2025_BAO}. However, we identify several major deficiencies with this study:
\begin{enumerate}
    \item The cosmological parameters were fixed to values that best fit the CMB, but slightly varying them while remaining consistent allows a much improved fit to the BAO;
    \item The CMB perceived by an average observer at the present epoch must be slightly hotter than measured at Earth if we live on a potential hill \citepalias[section~5.3.3 of][]{Haslbauer_2020}. Those authors argued that slight changes to the cosmological parameters should allow a good CMB fit to be preserved, but these changes were not taken into account when predicting the BAO;
    \item Related to the above, $r_{\rm d}$ would be slightly lower with a hotter CMB because we generally expect the physical size of the BAO ruler to remain about the same -- but a hotter CMB implies less expansion since decoupling. Combined with the reduced $H_0$ that a hotter CMB requires \citep*{Ivanov_2020_TCMB}, there would be a reduction in the $H_0 r_{\rm d}$ product, which is crucial in BAO studies.
\end{enumerate}


To test the local void model at high $z$, we discuss how it affects high-$z$ observables in Section~\ref{Methods}, where we also describe the observational constraints. Our results are presented in Section~\ref{Analysis} and discussed in Section~\ref{Discussion}. We conclude in Section~\ref{Conclusions}. Appendix~\ref{T_without_z0} provides useful fitting formulae for the recombination and decoupling temperatures in $\Lambda$CDM as a function of the baryon and CDM densities. Appendix~\ref{Neutrino_density_evolution} discusses the evolution of the neutrino density, while Appendix~\ref{CMB_compressed_likelihood} discusses the accuracy of the compressed CMB likelihood approach for handling CMB constraints on models with standard physics prior to recombination.

\section{Methods and observational constraints}
\label{Methods}

In this contribution, we focus on testing the local void model at high redshift because this greatly simplifies the setup. In general, a local void distorts the distance--redshift relation by creating additional non-cosmological contributions to the redshift through peculiar velocity and GR. Equation~52 of \citetalias{Haslbauer_2020} shows that the observed redshift $z$ satisfies
\begin{eqnarray}
    1 + z ~=~ \frac{1}{a \left( t \right)} \overbrace{\sqrt{\frac{c + v_{\mathrm{int}}}{c - v_{\mathrm{int}}}}}^{\text{Doppler}} \overbrace{\exp \left( \frac{1}{c^2} \int g_{\mathrm{void}} \dd r \right)}^{\text{GR}} \, ,
    \label{eq:z_contributions}
\end{eqnarray}
where $a$ was the cosmic scale factor at the emission time $t$ of an observed photon, $g_{\mathrm{void}}$ is the outward gravitational field induced by the void due to it having less density than the cosmic mean, and $v_{\mathrm{int}}$ is the outflow velocity in the reference frame of the void, which may be moving as a whole.

A local void would have little physical effect on the high-redshift universe. This allows us to greatly simplify the effect of a local void by neglecting peculiar velocity. Moreover, since GR would mostly arise at rather small distances, we can assume that the GR contribution remains fixed as we observe sources at ever larger redshift. We discuss the validity of these approximations in Section~\ref{z0_validity}. With these simplifications, the relation between the observed redshift $z$ and the cosmological redshift $z_{\rm c} \equiv a^{-1} - 1$ becomes
\begin{eqnarray}
    1 + z ~=~ \left( 1 + z_{\rm c} \right) \left( 1 + z_0 \right),
    \label{eq:z_0}
\end{eqnarray}
where $z_0$ is a model parameter capturing the dimensionless height of the local gravitational potential hill created by the void. For a void that can solve the Hubble tension jointly with galaxy number counts, we expect that $z_0 = 0.84\%$ \citepalias[section~5.3.3 of][]{Haslbauer_2020}. Our main goal is to freely vary $z_0$ and then check if its inferred value is consistent with this estimate, which was made prior to DESI without considering BAO data. We also fix $z_0$ to the prediction of \citetalias{Haslbauer_2020} and assess the impact on the considered datasets. Since it is quite possible that somewhat different void profiles can also fit the available observations, we then consider a reduced value of $z_0 = 0.5\%$. Finally, we perform control $\Lambda$CDM analyses by setting $z_0 = 0$.

In the following sections, we discuss the validity of the $z_0$ approximation for the impact of a local void (Section~\ref{z0_validity}). We then describe how $z_0$ affects cosmological observables at various epochs assuming a standard background expansion history, which we briefly recap in Section~\ref{FRW_with_neutrinos}. The main impact at low $z$ is that the cosmological redshift is slightly smaller than the observed redshift (Sections~\ref{CC_impact} and \ref{BAO_distance_impact}). In the early universe, the $z_0$ term has important implications for our interpretation of the CMB anisotropies (Section~\ref{CMB}) and the sound horizon at recombination and decoupling, changing the calibration of CMB and BAO observations, respectively (Section~\ref{Sound_horizon_impact}).

\subsection{Validity of the \texorpdfstring{$z_0$}{z₀} approximation}
\label{z0_validity}

\begin{figure*}
    \centering
    \includegraphics[width=\linewidth]{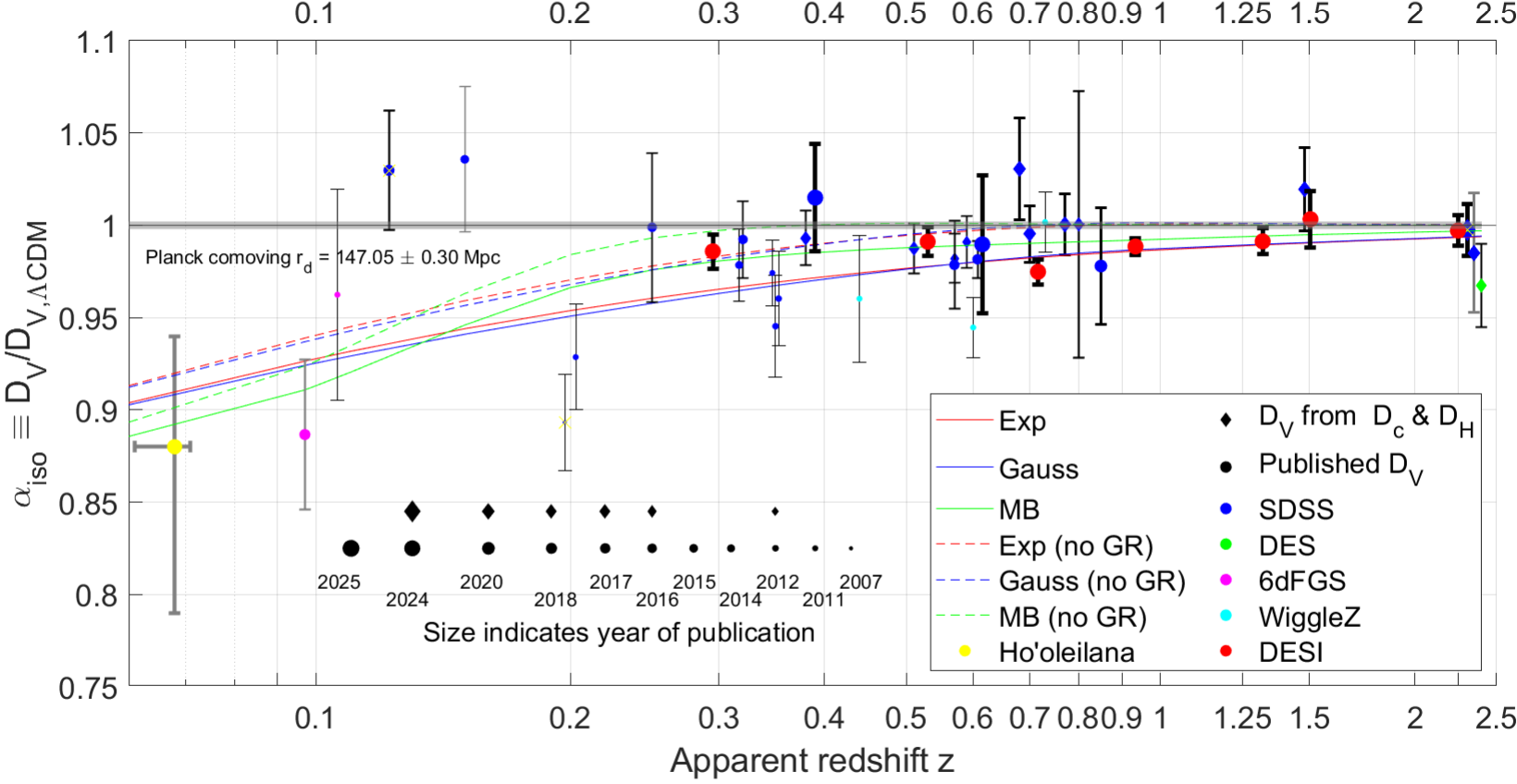}
    \caption{Ratio between the isotropically averaged BAO distance scale $D_{\rm V}$ and the $\Lambda$CDM prediction based on \emph{Planck} observations of the CMB \citep{Planck_2020}. The shaded grey horizontal band shows a 0.2\% uncertainty on the $\Lambda$CDM prediction due to the \emph{Planck} uncertainty on $r_{\rm d}$ \citep{Planck_2020}, though this does not properly account for the uncertainties (see the text). The coloured lines show predictions of the best-fitting void models from \citetalias{Haslbauer_2020} with an initial underdensity profile that is exponential (red), Gaussian (blue), or Maxwell-Boltzmann (green). Solid lines show predictions including both non-cosmological redshift contributions in Equation~\ref{eq:z_contributions}, while the dashed lines in the same colour show the corresponding results without the GR contribution. This highlights that the dominant non-cosmological contribution is the Doppler effect at low $z$, while GR dominates at high $z$. Points with uncertainties show BAO measurements over the last 20 years, with different colours indicating different surveys and larger markers indicating a more recent result, as indicated towards the bottom left. Reproduced from figure~4 of \citet{Banik_2025_BAO}.}
    \label{D_V_graph}
\end{figure*}

Our approximate treatment of the impact of a local void (Equation~\ref{eq:z_0}) is only expected to be valid at sufficiently high $z$. To estimate where our $z_0$ approximation first becomes accurate, we use Figure~\ref{D_V_graph} to reproduce figure~4 of \citet{Banik_2025_BAO}. This shows the ratio $\alpha_{\rm iso}$ between $D_{\rm V}/r_{\rm d}$ and its predicted value in the $\Lambda$CDM model calibrated using \emph{Planck} observations of the CMB \citep{Planck_2020}, where $D_{\rm V}$ is the isotropically averaged BAO distance (Equation~\ref{eq:D_V}). Unlike in the present study, uncertainties in cosmological parameters due to uncertainties in CMB observations were included only by means of a 0.2\% error margin in $r_{\rm d}$. This does not account for uncertainty in the predicted expansion history at low $z$, which inflates uncertainties from the CMB due to the need to extrapolate forward over many Gyr and $>3$ orders of magnitude in $a$.

The important aspect of Figure~\ref{D_V_graph} for our purposes is the comparison between the thin solid and dashed lines in the same colour, with different colours indicating different initial underdensity profiles. The solid lines show model predictions including both non-cosmological sources of redshift in Equation~\ref{eq:z_contributions}, while the dashed lines show predictions considering only the special relativistic Doppler effect -- excluding the impact of GR. The similarity of the predictions with and without GR shows that at low $z$, the dominant source of non-cosmological redshift is predicted to be the Doppler effect. The reason is that sources here are at a similar potential height to the void centre, but the gravitational fields are still able to create some outflow, which is necessary to solve the Hubble tension. The opposite occurs at high $z$, where outflows alone only negligibly distort the $\Lambda$CDM prediction. This is apparent from the dashed lines without GR lying almost on top of the $\Lambda$CDM expectation ($\alpha_{\rm iso} = 1$). However, the solid lines including GR are still appreciably different from $\Lambda$CDM. It is therefore clear that GR is the dominant source of non-cosmological redshift at high $z$. This is because sources very far out are not directly affected by outflow from a local void. Its only impact is an observational effect due to the need for photons from such sources to climb the local gravitational potential hill in order to reach our detectors.

Figure~\ref{D_V_graph} shows that GR is more important than the Doppler effect at $z \ga 0.3$ for all three density profiles considered by \citetalias{Haslbauer_2020}. We therefore consider data from a `nominal redshift range' of $z > 0.5$, where GR completely dominates over the Doppler effect of outflow velocities. This allows us to address 12 of the 13 BAO measurements from DESI~DR2 \citep{DESI_2025}. We complement this by considering an `extended redshift range' of $z > 0.29$ for most analyses. This simplifies the comparison with other studies that typically include all 13 BAO measurements. We argue that the $z_0$ approximation should still work fairly well at this lower redshift because GR is still dominant over the Doppler effect at $z \ga 0.3$. Sources at $z = 0.29 - 0.5$ would be partway up the local potential hill, reducing the GR contribution compared to that for a very distant source. On the other hand, outflows are obviously important at $z \approx 0.3$ in the local void model. These would increase the non-cosmological redshift via the Doppler effect (Equation~\ref{eq:z_contributions}). We can suppose that there would be an approximate cancellation between the impact of a reduced GR contribution and an unmodelled outflow contribution to $z$. The results for our extended redshift range should clearly be taken with some caution given the approximate nature of this argument, but we note that the $z_0$ approximation would become valid at lower $z$ with a smaller void than proposed in \citetalias{Haslbauer_2020}, as suggested based on peculiar velocities \citep*{Stiskalek_2025_void}. This would make the $z_0$ approximation more accurate at the redshifts considered here. Bearing these caveats in mind, we show variants to most of our analyses in which we consider data at $z > 0.29$. The extra data from $z = 0.29-0.5$ tightens the constraints somewhat compared to our nominal analysis.

\subsection{Friedmann equation with neutrinos}
\label{FRW_with_neutrinos}

By construction, the local void model does not alter the standard $a(t)$. Considering dark energy, matter, radiation, and neutrinos, the Friedmann equation for $E \equiv H/H_0$ is as follows \citepalias{Najera_2026}:
\begin{eqnarray}
    E^2 \left( a \right) = \Omega_\Lambda + \frac{\Omega_{\rm bc}}{a^3} + \frac{\Omega_\gamma}{a^4} + \frac{\Omega_{\nu, {\rm ur}, i}}{a^4} \left[ 1 + \sum_{i=A,B} \sqrt{1 + \left( \frac{a}{a_{{\rm nr},i}} \right)^2} \right],
    \label{eq:Friedmann}
\end{eqnarray}
where $\Omega_x$ refers to the present fraction of the cosmic critical density $3H_0^2/(8\mathrm{\pi} G)$ in component $x$ (which is $\Lambda$, $\mathrm{bc}$, $\gamma$, or $\nu$ for dark energy, baryons + CDM, photons, or neutrinos, respectively), $\Omega_{\nu, {\rm ur}, i}$ is the present contribution of each neutrino species $i$ if these neutrinos were massless, while $a_{{\rm nr}, i}$ is the scale factor when they transition from ultra-relativistic to non-relativistic. We label the neutrino species $A$, $B$, and $C$ in descending order of rest mass. The final term only includes species $A$ and $B$ due to our assumption that neutrinos of species $C$ are massless, therefore behaving like radiation. For the massive species ($A$ and $B$), we find $a_{\rm nr, i}$ using the approach discussed in Appendix~\ref{Neutrino_density_evolution}, which also recaps the ratio between photon and neutrino energy densities in the ultra-relativistic limit. We follow the same outline as appendix~B of \citetalias{Najera_2026}, but unlike their study, we assume different masses for the different neutrino species to better conform to terrestrial neutrino experiments. We follow the usual convention of keeping the neutrino masses as low as possible, so the neutrino species are assumed to have $m_{\nu, i} = \left( 50, 10, 0 \right)$~meV/$c^2$. This places the transition from relativistic to non-relativistic behaviour at $z \approx 100$ (20) for species $A$ ($B$), while species $C$ is massless. The matter density parameter $\Omega_{\rm m}$ applicable to lower redshift datasets includes the massive neutrino contribution (Equation~\ref{Omega_M}). The photon energy density follows from standard blackbody physics assuming a photon temperature of $T_{\rm CMB}$ (Equation~\ref{eq:T_FIRAS_adjustment}). Motivated by inflationary theory, we only consider flat models, neglecting the curvature contribution.

\subsection{Cosmic chronometers (CCs)}
\label{CC_impact}

The stellar populations in passively evolving galaxies serve as excellent CCs \citep{Jimenez_2002, Moresco_2018, Moresco_2020, Cogato_2024, Moresco_2024, Guo_2025}. The relative age between galaxies at different redshifts constrains the expansion history even more directly than the distance--redshift relation, since there is no need to convert distances into lookback times through division by $c$. Observers report CC results in the form
\begin{eqnarray}
    H_{\rm CC} \left( z \right) ~=~ -\frac{\dot{z}}{1 + z} \, ,
    \label{eq:H_CC_def}
\end{eqnarray}
which is generally not the same as $H(z)$. Assuming a fixed GR contribution to the redshift, applying the chain rule to Equation~\ref{eq:z_0} tells us that
\begin{eqnarray}
    \dd z ~=~ \left( 1 + z_0 \right) \dd z_{\rm c} \, .
    \label{eq:dz_Jacobian}
\end{eqnarray}
This inflates both $\dot{z}$ and $(1 + z)$ by the same factor of $(1 + z_0)$, which therefore cancels out. Taking into account that observations at redshift $z$ are actually probing the apparent expansion rate at the cosmological redshift $z_{\rm c} < z$, Equation~\ref{eq:H_CC_def} reduces to
\begin{eqnarray}
    H_{\rm CC} \left( z \right) ~=~ H \left( z_{\rm c} \right).
    \label{eq:H_CC}
\end{eqnarray}

We compare predictions for $H_{\rm CC}(z)$ with a widely used compilation of 32 observations \citep{Moresco_2020} supplemented by a very recent measurement at $z = 0.12$ \citep*{Wang_2026}. However, not all 33 of these CC measurements are used in our analyses because the $z_0$ model is not a valid approximation to a local void at low redshift (Section~\ref{z0_validity}). This requires us to consider only those CC measurements beyond the redshift floor imposed for each analysis (0.29 or 0.5).

\subsection{Comoving and Hubble distances from BAO}
\label{BAO_distance_impact}

The BAO signal provides both a redshift depth and an angular scale \citep{Banik_2025_BAO}. Equation~\ref{eq:dz_Jacobian} shows that the redshift depth is inflated by a factor of $\left( 1 + z_0 \right)$, reducing the inferred Hubble distance $D_H \equiv c/H$.
\begin{eqnarray}
    D_H \left( z \right) ~=~ \frac{D_H \left( z_{\rm c} \right)}{1 + z_0} \, .
    \label{eq:DH}
\end{eqnarray}

The angular BAO scale is used to constrain the comoving distance $D_{\rm c}$ to the redshift of the tracer used, where
\begin{eqnarray}
    D_{\rm c} ~\equiv~ c \int_0^{z_{\rm c}} \frac{d\tilde{z}_c}{H \left( \tilde{z}_c \right)} \, .
    \label{eq:Dc}
\end{eqnarray}
This remains true in the $z_0$ model, provided we calculate the $D_{\rm c}$ integral in terms of $z_{\rm c}$ rather than $z$. This is because the BAO constraint on $D_{\rm c}$ tells us the comoving distance to $z_{\rm c}$ in the void-free cosmology, implying that $H(z_{\rm c})$ must be used in the integral rather than $H(z)$.

We use the 13 BAO measurements from DESI~DR2 \citep{DESI_2025}. In our nominal analysis considering only data at $z > 0.5$, we neglect the lowest redshift measurement of $D_{\rm V}/r_{\rm d}$, where $D_{\rm V}$ is the isotropically averaged BAO distance.
\begin{eqnarray}
    D_{\rm V} \left( z \right) ~\equiv~ \left[ z D^2_{\rm c} \left( z \right) D_H \left( z \right) \right]^{1/3}.
    \label{eq:D_V}
\end{eqnarray}
The other measurements tell us $\left( D_{\rm c}/r_{\rm d}, D_H/r_{\rm d} \right)$ and their covariance at six redshifts, leading to a total of $1 + 6 \times 2 = 13$ measurements and a block-diagonal covariance matrix. In our extended analyses using data at $z > 0.29$, we consider all 13 measurements. Since BAO measurements must be calibrated using some assumed $r_{\rm d}$, we recalculate it at each step of our analysis using the approach described in Section~\ref{Sound_horizon_impact}.

\subsection{CMB temperature and anisotropies}
\label{CMB}

The GR contribution in Equation~\ref{eq:z_contributions} would also affect CMB photons. This would make the CMB appear to be slightly cooler than it actually is. In other words, an idealised observer at the present epoch (not on a potential hill or in a potential well) would measure a CMB temperature of
\begin{eqnarray}
    T_{\rm CMB} ~=~ T_{\rm FIRAS} \left( 1 + z_0 \right),
    \label{eq:T_FIRAS_adjustment}
\end{eqnarray}
where $T_{\rm FIRAS} = 2.72548 \pm 0.00057$~K is the temperature observed by the Far-Infrared Absolute Spectrophotometer (FIRAS) instrument onboard the \emph{Cosmic Background Explorer} (\emph{COBE}) satellite \citep{Fixsen_2009}. We neglect the negligible uncertainty on $T_{\rm FIRAS}$.

The impact of a hotter CMB has been explored in a few previous studies \citep{Yoo_2019, Ivanov_2020_TCMB}. The main impact is that since the recombination temperature $T_\star$ is almost fixed, there must have been less expansion since recombination. To preserve the physics in the primordial universe, the present baryon and matter densities must be correspondingly increased. Therefore, the standard CMB constraint on the physical baryon density $w_{\rm b}$ must be reinterpreted as a constraint on
\begin{eqnarray}
    \widetilde{w}_{\rm b} ~\equiv \frac{w_{\rm b}}{\left( 1 + z_0 \right)^3} \, ,
    \label{eq:wbct}
\end{eqnarray}
where $h \equiv H_0$ in units of 100~km/s/Mpc, $w_x \equiv \Omega_x h^2$ for any component $x$, and the tilde denotes division by $(1 + z_0)^3$. This argument also applies to $w_{\rm bc} \equiv w_{\rm b} + w_{\rm c}$, the total matter contribution at early times. To match the constraint on the angular scale of the acoustic peaks, a slightly lower $H_0$ is required for a hotter CMB \citep{Ivanov_2020_TCMB}. Those authors showed that with these changes, the primary CMB anisotropies are unaltered because the physics in the early universe and the comoving distance to recombination remain the same. This means the CMB power spectrum is barely affected. There is a subtle effect on large angular scales, which we return to in Section~\ref{Void_implications}.

Since a core objective of the local void model is to avoid introducing new physics at early times, we can follow standard techniques when comparing the void model to the CMB. We do this using the compressed CMB likelihood \citep{Lemos_2023}, a widely used approximation that preserves most of the information contained in the full CMB likelihood \citep{DESI_2025}. We demonstrate the accuracy of this framework in Appendix~\ref{CMB_compressed_likelihood}. Within this framework, the CMB constraints are expressed in terms of the parameter vector $(100 \, \theta_\star, w_{\rm b}, w_{\rm bc})$, where
\begin{eqnarray}
    \theta_\star ~=~ \frac{r_\star}{D_{\rm c} \left( a_\star \right)} \, ,
    \label{eqn:Dc_star}
\end{eqnarray}
where $r$ denotes the comoving sound horizon and $\star$ subscripts denote recombination. The comoving distance to it is found using Equation~\ref{eq:Friedmann} for $H(z)$, while the sound horizon then is found using Equation~\ref{eq:r_star}.

We obtain our adopted CMB constraints on $(100 \, \theta_\star, \widetilde{w}_{\rm b}, \widetilde{w}_{\rm bc})$ from the chains in \cite{SPT_2026}, which combines \emph{Planck}~2020 \citep{Planck_2020}, SPT-3G \citep{SPT_2026}, and ACT~DR6 \citep{Louis_2025}. The observed values of these parameters and their associated covariance matrix are as follows:
\begin{eqnarray}
    &&\mu_\text{obs}(100 \, \theta_\star, \widetilde{w}_{\rm b}, \widetilde{w}_{\rm bc})^{\rm T} ~=~ (1.04161, \, 0.02240, \, 0.14265), \nonumber \\
    &&\matr{C}_{\rm CMB} ~=~ 10^{-9} \times {\setlength{\arraycolsep}{3pt}
    \begin{bmatrix}
    53.8482 & 1.10723 & -28.7005 \\
    1.10723 & 9.20447 & -17.6016 \\
    -28.7005 & -17.6016 & 843.132
    \end{bmatrix}.}
    \label{CMB_covariance_matrix}
\end{eqnarray}
These are slightly updated from the results stated in appendix~B of \citetalias{Najera_2026} because that referenced an older version of \citet{SPT_2026}, but here we use its final published version in which the authors correct a previous coding error.

A higher matter density necessarily implies higher $H(a)$ in the matter-dominated era. To compensate and preserve $D_{\rm c} \left( a_\star \right)$, we must reduce $H(a)$ at late times. This is possible through a reduced dark energy density. We therefore expect a hotter CMB to be associated with reduced $H_0$ \citep{Ivanov_2020_TCMB}. However, we will find that the $H_0$ shift is quite small compared to uncertainties in the local $cz'$. A more serious issue is that the reduced $H_0$ reinforces the reduced $r_{\rm d}$ (Section~\ref{Sound_horizon_impact}) to yield a smaller $H_0 r_{\rm d}$ product, which is tightly constrained by BAO.


\subsection{CMB and BAO sound horizons}
\label{Sound_horizon_impact}

One of the most precisely constrained quantities in cosmology is $\theta_\star$, the ratio between the comoving distance to recombination and the comoving sound horizon at that epoch. We compute $r_\star$ using the exact analytic result from \cite{Hu_1996}:
\begin{eqnarray}
    r_\star ~=~ \frac{2c}{3} \sqrt{\frac{3 \, a_{\rm eq}}{\Omega_{\rm bc} H_0^2 \mathcal{R}_{\rm eq}}} \ln \left( \frac{\sqrt{1 + \mathcal{R}_\star} + \sqrt{\mathcal{R}_\star + \mathcal{R}_{\rm eq}}}{1 + \sqrt{\mathcal{R}_{\rm eq}}} \right),
    \label{eq:r_star}
\end{eqnarray}
where $\star$ subscripts denote recombination, eq subscripts denote matter--radiation equality (treating neutrinos as radiation; Equation~\ref{eq:R_nu_gamma}), bc refers to the combination of baryons and CDM, and $\mathcal{R} \equiv 3\rho_{\rm b}/(4\rho_\gamma)$ is related to the sound speed, which depends on the ratio between the baryon density $\rho_{\rm b}$ and the photon density $\rho_\gamma$ \citepalias[see appendix~B of][]{Najera_2026}.

To solve Equation~\ref{eq:r_star}, we need to know $\mathcal{R}_\star$ and thus the recombination scale factor $a_\star$. We do not obtain this following the approach of \citetalias{Najera_2026} as it relies on an ancient fitting formula, albeit rescaled to better line up with modern results. Instead, we run the Cosmic Linear Anisotropy Solving System \citep*[\textsc{class};][]{CLASS} over a small grid of parameters centred approximately where we expect the posterior inference to peak based on preliminary calculations. For reasons described in Section~\ref{CMB}, we set up our grid in terms of the parameters $(\widetilde{w}_{\rm b}, \widetilde{w}_{\rm bc}, z_0)$. For each step in our grid, we record the recombination temperature $T_\star$, which should reduce the sensitivity to $z_0$. We also record the decoupling temperature $T_{\rm d}$, which corresponds to the end of the `drag epoch' when photons have significant momentum exchange with baryons. We use d subscripts to denote quantities evaluated at this epoch, when the BAO $r_{\rm d}$ becomes fixed.

We obtain $a_\star$ from $T_\star$ using the relation
\begin{eqnarray}
    a_\star ~=~ \frac{T_{\rm CMB}}{T_\star} \, ,
\end{eqnarray}
where $T_{\rm CMB}$ is found using Equation~\ref{eq:T_FIRAS_adjustment}. There is an analogous relation between $a_{\rm d}$ and $T_{\rm d}$, which is important for the BAO. We find $r_{\rm d}$ by applying Equation~\ref{eq:r_star} using $\mathcal{R}_{\rm d}$ instead of $\mathcal{R}_\star$. We note that this equation cannot be applied at any later stage in cosmic history because it assumes tight coupling between photons and baryons.

Since $T_\star$ and $T_{\rm d}$ do not change much across our parameter grid, we approximate these temperatures as depending quadratically on the model parameters. Thus, we assume that
\begin{eqnarray}
    T_\star ~=~ T_\star^{\rm ref} + \bm{d}^{\rm T} \bm{g}_\star + \frac{\bm{d}^{\rm T} \matr{H}_\star \bm{d}}{2} \, ,
    \label{eq:T_star}
\end{eqnarray}
where $T_\star^{\rm ref} = 2970.24$~K, $\bm{d} \equiv \Theta - \Theta_{\rm ref}$ is the difference between the model parameters and the reference values at the grid centre, $\bm{g}_\star$ is the gradient of $T_\star$ with respect to the considered parameters, and $\matr{H}_\star$ is the Hessian. $T_{\rm d}$ is found using an analogous procedure. The grid we use is centred on
\begin{eqnarray}
    \Theta_{\rm ref} ~\equiv~
    \begin{pmatrix}
    \widetilde{w}_{\rm b} \\ 
    \widetilde{w}_{\rm bc} \\ 
    z_0
    \end{pmatrix}_{\rm ref}
    ~=~
    \begin{pmatrix}
    0.0225 \\
    0.1450 \\
    0.0100
    \end{pmatrix}.
    \label{eq:Theta_ref}
\end{eqnarray}

Using our \textsc{class} grid, we find that
\begin{eqnarray}
    \bm{g}_\star = \begin{pmatrix}
    -3187.35 \\ 
    194.00 \\ 
    -3.90
    \end{pmatrix}, ~~
    \matr{H}_\star = \begin{bmatrix}
    265184.96 & -7854.34 & 144.61 \\
    -7854.34 & -491.81 & -9.32 \\
    144.61 & -9.32 & 23.46
    \end{bmatrix}.
    \label{eq:g_H_recombination}
\end{eqnarray}
Across our parameter grid, the maximum absolute error in $T_\star$ is only 1.1~mK, while the root mean square (rms) error is only 0.24~mK. Given that $T_\star \approx 2970$~K, our quadratic approximation to $T_\star$ is more than sufficient for our purposes.

We obtain $T_{\rm d}$ similarly to Equation~\ref{eq:T_star}. Since we extract both $T_\star$ and $T_{\rm d}$ from CLASS when running it, $\Theta_{\rm ref}$ remains the same as in Equation~\ref{eq:Theta_ref}. The reference temperature is now only $T_{\rm d}^{\rm ref} = 2892.72$~K, while the gradient and Hessian become
\begin{eqnarray}
    \bm{g}_{\rm d} = \begin{pmatrix}
    5926.80 \\
    203.29 \\
    -2.37
    \end{pmatrix}, ~~
    \matr{H}_{\rm d} = \begin{bmatrix}
    -189596.68 & -4765.72 & 108.90 \\
    -4765.72 & -624.69 & -5.34 \\
    108.90 & -5.34 & 17.11
    \end{bmatrix}.
    \label{eq:g_H_decoupling}
\end{eqnarray}
The maximum absolute error in $T_{\rm d}$ across our parameter grid is only 1.0~mK, while the rms error is only 0.34~mK. This is again a negligibly small fraction of $T_{\rm d} \approx 2890$~K.

We expect that $T_{\rm d} < T_\star$ because the recombination redshift is defined as the maximum of the visibility function, which roughly speaking requires photons to be unaffected by baryons. Since there are far fewer baryons than photons, this is relatively easy to obtain, so it occurs earlier in cosmic history. However, decoupling requires baryons to no longer be `dragged' by photons. Since there are far more photons than baryons, the ionisation fraction has to drop far lower to reach this stage. Therefore, baryon--photon decoupling occurs later in cosmic history than recombination.

\subsection{BBN}
\label{BBN}

In the first few minutes after the Big Bang, the whole Universe was hot enough to fuse hydrogen nuclei into deuterium and helium, along with trace amounts of metals \citep{Alpher_1948, Tytler_2000, Cyburt_2016, Pettini_2026}. The precise details of BBN have a large impact on the final yields of non-trivial nuclei, making their measured primordial abundances a sensitive probe of the baryon:photon ratio $\eta$.

Since BBN provides a tight constraint on $\eta$, studies which report a BBN constraint on $w_{\rm b}$ must assume some value for the present photon temperature. This is presumably $T_{\rm FIRAS}$, which is nowadays typically assumed without question. We therefore assume that any reported BBN constraints on $w_{\rm b}$ are actually constraints on $\widetilde{w}_{\rm b}$. Since the CMB also constrains $\widetilde{w}_{\rm b}$ (Section~\ref{CMB}), a slightly hotter CMB would not affect the agreement in standard theory between $w_{\rm b}$ inferred from BBN and that from the CMB anisotropies \citep{Cyburt_2016}.

Applying the \textsc{primat} nuclear reaction code \citep{Pitrou_2018} to BBN assuming the primordial deuterium abundance $(D/H)_{\rm p} = \left( 2.527 \pm 0.030 \right) \times 10^{-5}$ \citep*{Cooke_2018} and the primordial helium abundance $Y_{\rm p} = 0.2449 \pm 0.004$ \citep*{Aver_2015}, the baryon density $\widetilde{w}_{\rm b} = 0.021915 \pm 0.000215$ \citep{Giovanetti_2025_BBN}. This result is based on the purely BBN section of their table~B1 with fixed $N_\mathrm{eff}$, the effective number of neutrino species. For simplicity, we have symmetrised their very slightly asymmetric error bars while fixing their reported $1\sigma$ lower and upper limits on $\widetilde{w}_{\rm b}$. We adopt this constraint on $\widetilde{w}_{\rm b}$ by adding the corresponding $\chi^2$ contribution.



\section{Analysis and results}
\label{Analysis}

We conduct several analyses to understand the impact of $z_0$. The most obvious is to let $z_0$ be a free parameter, which we infer. Since we have a clear \emph{a priori} expectation that $z_0 \approx 0.84\%$ \citepalias[section~5.3.3 of][]{Haslbauer_2020}, we also run analyses where we fix $z_0$ to 0.84\%. Given the results of \citet{Stiskalek_2025_void} suggesting a smaller void, we also consider the case $z_0 = 0.5\%$. $\Lambda$CDM behaviour is recovered by fixing $z_0 = 0$. We also extend our main analysis by including SNe~Ia from the recent Dark Energy Survey (DES) Dovekie catalogue \citep{DES_2024_SNe, Popovic_2026}. The prior ranges of the model parameters are summarised in Table~\ref{tab:priors}.

\begin{table}
    \centering
    \begin{tabular}{lc}
        \toprule
        Parameter & Prior range \\ \midrule
        $w_{\rm b}$ & (0.02, 0.025) \\
        $w_{\rm bc}$ & (0, 0.5) \\
        $H_0$ (km/s/Mpc) & (50, 90) \\
        $z_0$ & $(-0.1, 0.1)$ \\ \bottomrule
    \end{tabular}
    \caption{The range for the uniform prior we adopt on each model parameter.}
    \label{tab:priors}
\end{table}

\subsection{Nested sampling of model parameters}
\label{Nested_sampling}

We use the nested sampling algorithm \texttt{dynesty}\footnote{\url{https://dynesty.readthedocs.io/en/v2.1.5/}} \citep{Skilling_2004, Skilling_2006, Feroz_2009, Speagle_2020, Koposov_2024} to compute the posterior probability and the Bayesian evidence $Z$, given by
\begin{eqnarray}
    Z \left( \mathcal{M} \right) ~=~ \int P(\theta) \, P(\mathbf{D}|\theta, \mathcal{M}) \dd\theta,
    \label{eq:Bayesian_evidence}
\end{eqnarray}
where $\theta$ is the parameter vector, $P(\theta)$ is the prior, and $P(\mathbf{D}|\theta, \mathcal{M})$ is the likelihood of data $\mathbf{D}$ given model $\mathcal{M}$ with parameters $\theta$. We take 500 live points and use a stopping criterion of $\Delta \ln \hat{Z} = 0.01$, where the hat denotes that this is the approximate remaining $Z$, not the true $Z$. We will see later that this tolerance is sufficient because the different models considered have much larger differences in $Z$. We plot the $1\sigma$ and 2$\sigma$ contours using the \texttt{GetDist} package\footnote{\url{https://getdist.readthedocs.io/en/latest/}} \citep{Lewis_2025_GetDist}. Posterior convergence was confirmed by checking that the number of effective samples was $N_{\mathrm{eff}} > 2000$ for all analyses.

Our full nested sampling analyses consider the parameters $(w_{\rm b}, w_{\rm bc}, H_0, z_0)$. We do not infer $\Omega_{\rm m}$ but rather compute it by adding the baryon + CDM contribution to that of the neutrinos which are non-relativistic at low $z$.
\begin{eqnarray}
    w_{\rm m} ~=~ w_{\rm bc} + \sum_{i = A, B} w_{\nu, i} \, .
    \label{Omega_M}
\end{eqnarray}
Late-time observables like BAO and CC are sensitive to $\Omega_{\rm m}$, while early Universe observables like the CMB are sensitive to $\Omega_{\rm bc}$ as the neutrinos are still relativistic then.

\begin{figure}
    \centering
    \includegraphics[width=\linewidth]{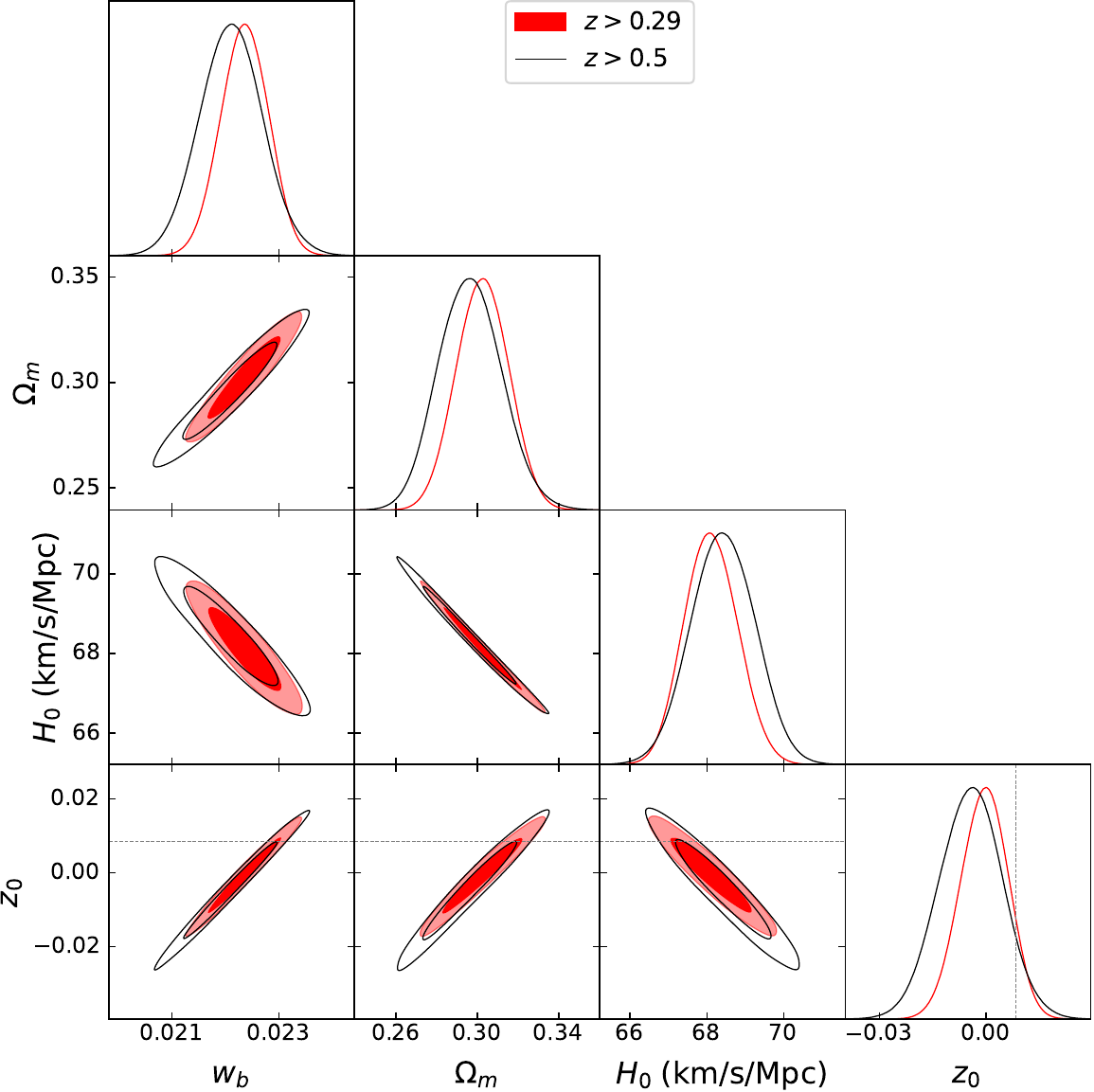}
    \caption{Triangle plot showing the inferred parameters with data from $z > 0.5$ (open black contours) or $z > 0.29$ (filled red contours). We show the $1\sigma$ and $2\sigma$ error ellipses in panels involving two parameters. The dotted line shows the \emph{a priori} prediction that $z_0 = 0.84\%$ \citepalias[section~5.3.3 of][]{Haslbauer_2020}.}
    \label{fig:Triangle}
\end{figure}

Figure~\ref{fig:Triangle} shows the triangle plot resulting from our full analysis, with open black contours showing our nominal redshift range and filled red contours showing our extended redshift range. The previously published expectation that $z_0 = 0.84\%$ is shown as a dotted line on panels involving $z_0$. It is clear that the posterior inference on $z_0$ is consistent with 0.84\%, though it peaks at a very slightly negative $z_0$. Using the extended redshift range reduces the uncertainties, but interestingly the inferred $z_0$ rises slightly, so the mild tension with the predicted 0.84\% remains about the same. This is because the DESI~DR2 measurement of $D_{\rm V}/r_{\rm d}$ at $z = 0.295$ lines up better with the void model, as we discuss later (Section~\ref{BAO_reconstruction}).

The degeneracies apparent in Figure~\ref{fig:Triangle} are to be expected from previous analyses. For instance, we expect a negative correlation between $\Omega_{\rm m}$ and $h$ due to the shape of the CMB power spectrum providing a tight constraint on $w_{\mathrm m}$ and the opposite impacts of $\Omega_{\rm m}$ and $h$ on $\theta_\star$ \citep*{Percival_2002, Kable_2019}. As discussed in Section~\ref{CMB}, we also expect a negative correlation between $z_0$ and $h$ because a hotter CMB implies a higher matter density and thus higher expansion rate at high $z$, which must be compensated for by a lower expansion rate at low $z$ to maintain the same comoving distance to recombination \citep[see also][]{Ivanov_2020_TCMB}. These degeneracies are indeed apparent from our results, explaining the induced positive correlation between $z_0$ and $\Omega_{\rm m}$.

\renewcommand{\arraystretch}{1.2}
\begin{table*}
    \centering
    \begin{tabular}{lcccc}
        \toprule
        & \multicolumn{4}{c}{Assumed value of $z_0$} \\
        Parameter & Free $z_0$ & 0 ($\Lambda$CDM) & 0.005 & 0.0084 \\ \midrule
        \multicolumn{5}{c}{$z > 0.5$ dataset} \\ \midrule
        $w_{\rm b}$ & $0.02211^{+0.00057}_{-0.00059}$ & $0.02239 \pm 0.00008$ & $0.02273 \pm 0.00009$ & $0.02295 \pm 0.00008$ \\
        $\Omega_{\rm m}$ & $0.296 \pm 0.015$ & $0.303 \pm 0.003$ & $0.313 \pm 0.004$ & $0.319 \pm 0.004$ \\
        $H_0$ (km/s/Mpc) & $68.42^{+0.83}_{-0.82}$ & $68.04 \pm 0.26$ & $67.59 \pm 0.25$ & $67.29^{+0.26}_{-0.24}$ \\
        $z_0$ & $-0.004^{+0.008}_{-0.009}$ & -- & -- & -- \\ \midrule \midrule
        \multicolumn{5}{c}{$z > 0.29$ dataset} \\ \midrule
        $w_{\rm b}$ & $0.02236^{+0.00046}_{-0.00044}$ & $0.02239 \pm 0.00008$ & $0.02272 \pm 0.00009$ & $0.02295 \pm 0.00009$ \\
        $\Omega_{\rm m}$ & $0.303 \pm 0.013$ & $0.303 \pm 0.003$ & $0.313 \pm 0.003$ & $0.319 \pm 0.004$ \\
        $H_0$ (km/s/Mpc) & $68.11 \pm 0.69$ & $68.07 \pm 0.23$ & $67.58^{+0.24}_{-0.24}$ & $67.25^{+0.26}_{-0.23}$ \\
        $z_0$ & $-0.001 \pm 0.007$ & -- & -- & -- \\ \bottomrule
    \end{tabular}
    \caption{Inferred parameter values under different assumed $z_0$ scenarios for the nominal redshift range (top) and extended redshift range (bottom).}
    \label{tab:parameters}
\end{table*}
\renewcommand{\arraystretch}{1.0}

The inferred model parameters are summarised in Table~\ref{tab:parameters}. The reported central value is the mean across the \texttt{dynesty} chain, while the $1\sigma$ confidence interval is taken to be the range between percentiles 15.87 and 84.13. The first column shows results with free $z_0$, while other columns show results with $z_0$ fixed to 0 ($\Lambda$CDM), 0.5\%, or 0.84\%. Fixing $z_0$ reduces the uncertainties considerably, but it also slightly shifts the inferred values due to the parameter degeneracies discussed above. The inferred $z_0$ is slightly negative with our nominal redshift range, but it is almost exactly 0 with our extended redshift range. In both cases, the prediction of 0.84\% \citepalias{Haslbauer_2020} is just outside the $1\sigma$ range.

\subsection{Reconstructed BAO distance scale}
\label{BAO_reconstruction}

To gain further insight into the impact of $z_0$ on BAO predictions, we use our \texttt{dynesty} chain to obtain the posterior predictive distribution (PPD) of $D_{\rm V}/r_{\rm d}$ at each redshift in the range (0.1, 2.5). This allows us to obtain the mean value of $D_{\rm V}/r_{\rm d}$ and its $1\sigma$ confidence interval, which we again approximate using percentiles 15.87--84.13. For the $2\sigma$ confidence interval, we use percentiles 2.275--97.725.

Since $D_{\rm V}/r_{\rm d}$ spans a large range in values over such a wide redshift range, it is common to divide $D_{\rm V}/r_{\rm d}$ at each redshift by the value in some fiducial cosmology. For this, we use the mean value of the PPD of $D_{\rm V}/r_{\rm d}$ in the $\Lambda$CDM model ($z_0 = 0$). This approach is similar to using the best-fitting $\Lambda$CDM model as our fiducial cosmology when calculating $\alpha_{\rm iso}$, which is defined as $D_{\rm V}/r_{\rm d}$ divided by the fiducial value. However, our approach involves obtaining $D_{\rm V}/r_{\rm d}$ from a fiducial framework ($\Lambda$CDM) instead of a fiducial model ($\Lambda$CDM with best-fitting values of its parameters). We note that the fiducial $D_{\rm V}/r_{\rm d}$ has no bearing on our statistical results as it is used only for plotting purposes. We will see that using a fiducial $\Lambda$CDM cosmology with parameters that best fit all considered high-redshift datasets leads to important differences compared to considering just the CMB (Figure~\ref{D_V_graph}).

\begin{figure}
    \centering
    \includegraphics[width=\linewidth]{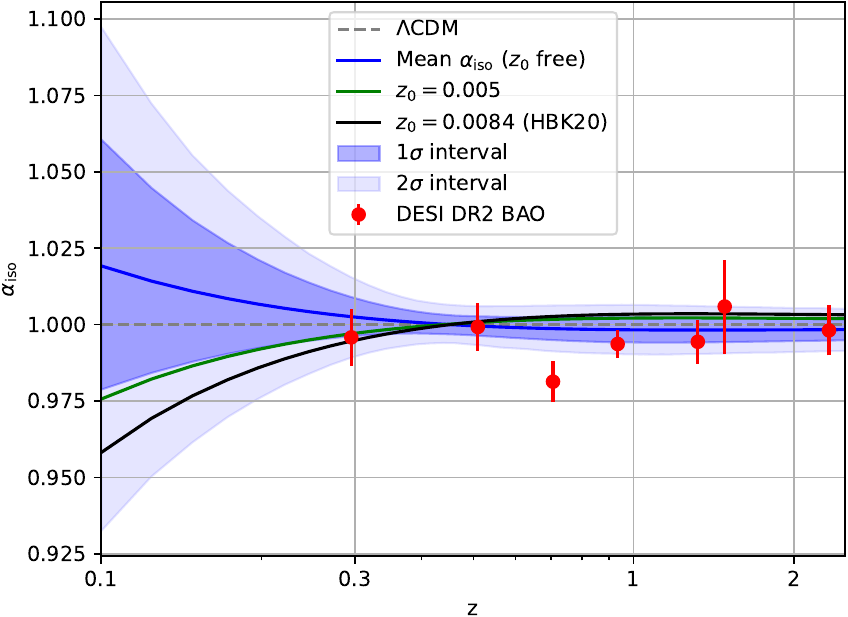}
    \caption{Reconstructed $\alpha_{\rm iso}(z) \equiv D_{\rm V}/r_{\rm d}$ in units of its fiducial value at each redshift, which here is the mean value from our $\Lambda$CDM nested sampling analysis using data at $z > 0.5$ (see the text). The solid blue line shows the mean result using our free $z_0$ analysis with the same data, while the shaded blue bands show the $1\sigma$ and $2\sigma$ uncertainties using Gaussian equivalent percentiles. The red data points with uncertainties show DESI~DR2 measurements \citep{DESI_2025}. The solid green line shows the best-fitting model with $z_0 = 0.5\%$, while the solid black line assumes instead the previously expected value of $z_0 = 0.84\%$ \citepalias{Haslbauer_2020}.}
    \label{fig:alpha_iso}
\end{figure}

Figure~\ref{fig:alpha_iso} shows our reconstructed $\alpha_{\rm iso}(z)$ using our analysis with free $z_0$. The solid blue line shows the mean of the PPD on $D_{\rm V}/r_{\rm d}$ at each $z$, while the dark and light blue bands show the $1\sigma$ and $2\sigma$ confidence intervals, respectively. The DESI data points are shown in red. Since the $\Lambda$CDM model is used as the fiducial cosmology, the $\Lambda$CDM prediction is by definition $\alpha_{\rm iso} = 1$, which we highlight as a dotted grey line. This fits most of the data points quite well, underscoring the lack of a clear BAO anomaly in $\Lambda$CDM when considering only high-redshift datasets. Although a BAO anomaly is widely discussed in the literature \citep{DESI_2025} and often attributed to evolution of the dark energy density in a theoretically problematic manner \citep{Lewis_2025}, the anomaly is not apparent in DESI data space \citepalias{Najera_2026}. It is apparent in the commonly used $(H_0 r_{\rm d}, \Omega_{\rm m})$ parameter space, an issue we return to later (Section~\ref{BAO_anomaly}).

A crucial aspect of Figure~\ref{fig:alpha_iso} is the prediction of the best-fitting model where $z_0$ is fixed to various values. The predicted run of $\alpha_{\rm iso}$ for the 0.5\% case is shown using the solid green curve, while the solid black curve shows results with $z_0$ fixed to the previously expected 0.84\% \citepalias{Haslbauer_2020}. In both cases, the results are in good agreement with the reconstructed $\alpha_{\rm iso}(z)$ when assuming a free $z_0$. There is a slight mismatch with DESI~DR2 at $z = 0.7$ and 0.9, but this is also the case for $\Lambda$CDM. Raising $z_0$ to 0.5\% or even 0.84\% only slightly worsens the tension here.

\begin{table}
    \centering
    \begin{tabular}{lcccc}
        \toprule
        Observable & \multicolumn{4}{c}{Assumed value of $z_0$} \\
        (d.o.f.) & 0 ($\Lambda$CDM) & $-0.0044$ (best) & 0.005 & 0.0084 \\ \midrule
        \multicolumn{5}{c}{$z > 0.5$ dataset} \\ \midrule
        BBN (1) & 4.855 & 4.919 & 4.782 & 4.733 \\
        CMB (3) & 6.070 & 6.501 & 5.603 & 5.300 \\
        BAO (12) & 15.307 & 14.612 & 16.669 & 17.950 \\
        CC (14) & 11.561 & 11.514 & 11.625 & 11.675 \\ \midrule
        Total (30) & 37.793 & 37.547 & 38.680 & 39.659 \\ \midrule \midrule
        \multicolumn{5}{c}{$z > 0.29$ dataset} \\ \midrule
        & 0 ($\Lambda$CDM) & $-5.4 \times 10^{-4}$ (best) & 0.005 & 0.0084 \\ \midrule
        BBN (1) & 4.905 & 4.921 & 4.764 & 4.668 \\
        CMB (3) & 6.494 & 6.606 & 5.507 & 4.885 \\
        BAO (13) & 14.989 & 14.854 & 16.791 & 18.611 \\
        CC (23) & 13.417 & 13.418 & 13.429 & 13.458 \\ \midrule
        Total (40) & 39.805 & 39.799 & 40.492 & 41.623 \\ \bottomrule
    \end{tabular}
    \caption{The $\chi^2$ contributions from each considered observable and the total for each assumed $z_0$ scenario, shown here considering data in our nominal redshift range (top) and extended redshift range (bottom). The numbers in brackets in the first column show how many d.o.f. each observable contributes.}
    \label{tab:chi_sq}
\end{table}

The minimum $\chi^2$ of each considered model is shown in Table~\ref{tab:chi_sq}. The parameters are usually optimised to minimise $\chi^2$ with fixed $z_0$, but in the central column, $z_0$ is also varied and its best-fitting value is indicated. Each row shows the contribution to the total $\chi^2$ from each considered observable, for which we give the number of degrees of freedom (d.o.f.). All $\chi^2$ contributions and the total are clearly acceptable for $\Lambda$CDM, as might be expected from Figure~\ref{fig:alpha_iso} and the low data space BAO--CMB tension \citepalias{Najera_2026}. Unlike the Hubble tension in which the CMB and local $cz'$ cannot be fit simultaneously, we can adequately fit the CMB and BAO datasets in $\Lambda$CDM. However, the parameters required for this are clearly not exactly the same as those that best fit the CMB. Indeed, fixing the parameters to those in the \emph{Planck} cosmology \citep{Planck_2020} leads to almost $3\sigma$ tension with DESI~DR2 \citep{Banik_2025_BAO}. This shows the continued importance of improving CMB constraints to better assess the BAO anomaly \citep{SPT_2026}.

As might be expected from the slightly negative inferred values of $z_0$, fixing it to a positive value of 0.5\% or 0.84\% slightly increases the total $\chi^2$. Interestingly, this improves the agreement with the CMB, especially for our extended redshift range. However, this is more than offset by increased $\chi^2$ from BAO. Even so, the differences in $\chi^2$ are rather small, especially with BBN and CC. Changes to the total $\chi^2$ do not clearly favour any of the considered $z_0$ values over any other. This highlights that $z_0 = 0.84\%$ lies within present uncertainties. We can envisage that a slightly different void model with $z_0 = 0.5\%$ might also fit low-$z$ constraints like the local $cz'$. This would be in even better agreement with the high-redshift datasets considered here.

\subsection{Quantifying the BAO anomaly}
\label{BAO_anomaly}

To quantify whether the BAO data can be predicted by a model with parameters constrained using the non-BAO datasets considered here, we follow the approach of \citetalias{Najera_2026}. Those authors defined two metrics known as the data space and parameter space tensions, the former working directly in the DESI data space (12 or 13~d.o.f.) and the latter in the $(H_0 r_{\rm d}, \Omega_{\rm m})$ parameter space (2~d.o.f.), which is commonly used in BAO analyses because these are the two main quantities which can easily be extracted. We avoid using non-standard parameter spaces to quantify the tension faced by each model, since this can create look-elsewhere effects associated with considering multiple possible metrics.

The data space tension captures the extent to which the individual DESI measurements can be predicted by a model that has been fitted to other data. Usually only the CMB is considered, but since there is little tension between it and CC \citep{Guo_2025} or BBN \citep{Pettini_2026}, we use all non-BAO datasets to constrain the parameters of each model when predicting the BAO angular scale and redshift depth. Uncertainties on both the DESI observations and model predictions must be considered when calculating the total $\chi^2$, which is then converted into a Gaussian equivalent tension for 12 or 13~d.o.f. depending on whether we are using our nominal or extended redshift range, respectively. The parameter space tension is calculated similarly, but this has only 2~d.o.f. Following \citetalias{Najera_2026}, we recommend that the maximum of the data and parameter space (MDPS) tension be used to efficiently quantify the BAO anomaly in any model, since both contain useful information and neither should be large in the correct model. In the community, the focus is usually on the parameter space tension, which is the MDPS tension for $\Lambda$CDM.

\begin{table}
    \centering
    \setlength{\tabcolsep}{4pt}
    \begin{tabular}{lcccc}
        \toprule
        Assumed $z_0$ & 0 & $-0.0044$ & 0.005 & 0.0084 \\ \midrule
        \multicolumn{5}{c}{Nominal redshift range} \\ \midrule
        Data space $\chi^2$ (12 d.o.f.) \rule{0pt}{9pt} & 22.14 & 22.35 & 22.91 & 24.03 \\
        Data space tension & $2.10\sigma$ & $2.12\sigma$ & $2.19\sigma$ & $2.32\sigma$ \\ [5pt]
        Parameter space $\chi^2$ (2 d.o.f.) & 11.88 & 13.40 & 11.44 & 11.38 \\
        Parameter space tension & $\boldsymbol{3.01\sigma}$ & $\boldsymbol{3.23\sigma}$ & $\boldsymbol{2.94\sigma}$ & $\boldsymbol{2.93\sigma}$ \\ \midrule \midrule
        \multicolumn{5}{c}{Extended redshift range} \\ \midrule
        Assumed $z_0$ & 0 & $-5.4 \times 10^{-4}$ & 0.005 & 0.0084 \\ \midrule
        Data space $\chi^2$ (13 d.o.f.) \rule{0pt}{9pt} & 22.31 & 22.22 & 23.18 & 24.56 \\
        Data space tension & $1.95\sigma$ & $1.94\sigma$ & $2.06\sigma$ & $2.22\sigma$ \\ [5pt]
        Parameter space $\chi^2$ (2 d.o.f.) & 12.29 & 12.25 & 11.07 & 10.51 \\
        Parameter space tension & $\boldsymbol{3.07\sigma}$ & $\boldsymbol{3.06\sigma}$ & $\boldsymbol{2.88\sigma}$ & $\boldsymbol{2.79\sigma}$ \\ \bottomrule
    \end{tabular}
    \caption{BAO tensions for our nominal redshift range (top) and extended redshift range (bottom), showing the data space and parameter space $\chi^2$ and corresponding Gaussian equivalent tension for each assumed $z_0$ scenario. The MDPS tension is highlighted in bold in each column.}
    \label{tab:MDPS_tension}
    \setlength{\tabcolsep}{6pt}
\end{table}

The top part of Table~\ref{tab:MDPS_tension} shows the BAO tension in data and parameter space for our nominal redshift range. Given that models with $z_0 = 0.5\%$ or even 0.84\% are barely distinguishable from $\Lambda$CDM at the level of current observations, it is unsurprising that just as with $\Lambda$CDM, these non-standard $z_0 > 0$ scenarios have more tension in parameter space than in data space. The MDPS tension for $\Lambda$CDM is $3.01\sigma$. Fixing $z_0$ to the previously expected level of 0.84\% \citepalias{Haslbauer_2020} reduces the MDPS tension to $2.93\sigma$. This slightly worsens the data space tension from $2.10\sigma$ to $2.32\sigma$, which is quite small in both cases. There appears to be an interesting tradeoff between the tension in data and parameter space. The local void model appears able to navigate this subtlety in such a way that the MDPS tension is reduced, thereby slightly alleviating the conventional measure of the BAO anomaly.

The reduced MDPS tension in the local void model is even more apparent in the bottom part of Table~\ref{tab:MDPS_tension}, which shows the BAO tension for our extended redshift range. In this case, $\Lambda$CDM faces a tension of $3.07\sigma$, but this is reduced to $2.79\sigma$ with $z_0 = 0.84\%$. Note that since $z_0$ is fixed in both $z_0$ scenarios, the parameter uncertainties are almost identical (Table~\ref{tab:parameters}). The reduced MDPS tension must then arise from assuming $z_0 > 0$ genuinely shifting the regions of $(H_0 r_{\rm d}, \Omega_{\rm m})$ parameter space preferred by the BAO and non-BAO datasets considered here.


Our estimated MDPS tension of $3.07\sigma$ for $\Lambda$CDM is higher than the $2.65\sigma$ reported by \citetalias{Najera_2026}. This is partly due to the removal of low-redshift data required by the $z_0$ approximation to the local void model. However, when assessing $\Lambda$CDM, we can include low-redshift CC measurements. We have excluded them simply to ensure that the same data is used when comparing different $z_0$ scenarios. To gain a better assessment of the BAO anomaly faced by $\Lambda$CDM, we extend the non-BAO dataset to include all 33 considered CC measurements \citep{Moresco_2020, Wang_2026}. In this case, the data space tension becomes $1.94\sigma$, while the parameter space tension becomes $3.05\sigma$. Both tensions are very slightly lower than for our extended redshift range. The MDPS tension of $3.05\sigma$ is higher than the previously reported $2.65\sigma$ \citepalias{Najera_2026}. This could be due to a much more modern calculation of the recombination temperature (Appendix~\ref{T_without_z0}) and a more accurate treatment of neutrinos (Appendix~\ref{Neutrino_density_evolution}).

\subsection{Bayesian evidence}
\label{Bayesian_evidence}

\begin{table}
    \centering
    \begin{tabular}{lcc}
        \hline
        Assumed & \multicolumn{2}{c}{Dataset considered for analysis} \\
        $z_0$ value & $z > 0.5$ (nominal) & $z > 0.29$ \\ \hline
        0 ($\Lambda$CDM) & -- & -- \\
        Free & $-1.99 \pm 0.17$ & $-2.32 \pm 0.17$ \\
        0.005 & $-0.37 \pm 0.16$ & $-0.20 \pm 0.16$ \\
        0.0084 & $-0.82 \pm 0.16$ & $-0.96 \pm 0.16$ \\
        \hline
    \end{tabular}
    \caption{The natural logarithmic Bayesian evidence for different datasets and assumed $z_0$ scenarios (Equation~\ref{eq:Bayesian_evidence}). Results are shown relative to that for the corresponding $\Lambda$CDM model, whose uncertainty of 0.16 is not included here.}
    \label{tab:Bayesian_evidence}
\end{table}

Since our analysis infers a very slightly negative $z_0$ while a local void requires a positive value, the data considered here do not by themselves prefer the addition of $z_0$ as an extra model parameter. To quantify this, we compute the Bayesian evidence of each model (Equation~\ref{eq:Bayesian_evidence}). The results in Table~\ref{tab:Bayesian_evidence} show that adding $z_0$ as a free parameter reduces $Z$ by around 2 natural units (nats). If $z_0$ is fixed to 0.5\%, there is a loss of only 0.37~nats, which rises to 0.82~nats when fixing $z_0$ to the previously expected 0.84\%. These very small differences in $Z$ indicate that the considered data are presently not discriminating enough to tell whether $z_0 = 0.84\%$, but this remains allowed.

In both our analysis and that of \citetalias{Najera_2026}, local $cz'$ measurements are not included because $\Lambda$CDM cannot fit these simultaneously with the other datasets, making it unphysical to calibrate the $\Lambda$CDM parameters using, e.g., both the CMB and the local $cz'$. However, this approach should be followed in models which can solve the Hubble tension, since obviously data should generally be included if at all possible. The local $cz'$ constraint is indirectly included in our results through the choice of $z_0 = 0.84\%$ \citepalias{Haslbauer_2020}. In the future, we hope to directly address local $cz'$ constraints using a more advanced void model with outflows, which must be included to address the distance ladder.

\subsection{Results including DES-Dovekie SNe}
\label{Results_including_SNe}

We extend our main analysis by including likely SNe~Ia from the DES-Dovekie catalogue \citep{DES_2024_SNe, Popovic_2026}. Even with our extended redshift range, the redshift floor of 0.29 is high enough to exclude all external or legacy SNe in the catalogue, helping to ensure its homogeneity by involving only the DES photometric filter system. Since the SNe are not always spectroscopically confirmed, we restrict to only those SNe with a $\geq 90\%$ probability of being Type~Ia.

The predicted luminosity distance $D_L$ is found as follows:
\begin{eqnarray}
    D_L \left( z_{\rm HD}, z_{\rm hel} \right) ~=~ \left( 1 + z_{\rm hel} \right) D_c \left( z_{\rm c} \right),
\end{eqnarray}
where the heliocentric redshift $z_{\rm hel}$ is corrected for small-scale peculiar velocities to obtain $z_{\rm HD}$, which is then used to estimate the cosmic redshift $z_{\rm c}$ by applying Equation~\ref{eq:z_0}. As with other datasets, we only consider SNe with sufficiently large $z_{\rm HD}$ for the $z_0$ approximation to be accurate. We use the predicted $D_L$ to obtain a predicted distance modulus, which we compare to the observed value using the published precision or inverse covariance matrix. Since the DES catalogue lacks any absolute calibration, we add a magnitude offset with wide uniform prior to the published distance moduli, which assume a particular arbitrary calibration. Since this offset is arbitrary, we marginalise over it. We note that although the SNe~Ia are uncalibrated, combining with other datasets assuming the validity of $\Lambda$CDM at early times effectively does impose a calibration via the inverse distance ladder.

\begin{figure}
    \centering
    \includegraphics[width=\linewidth]{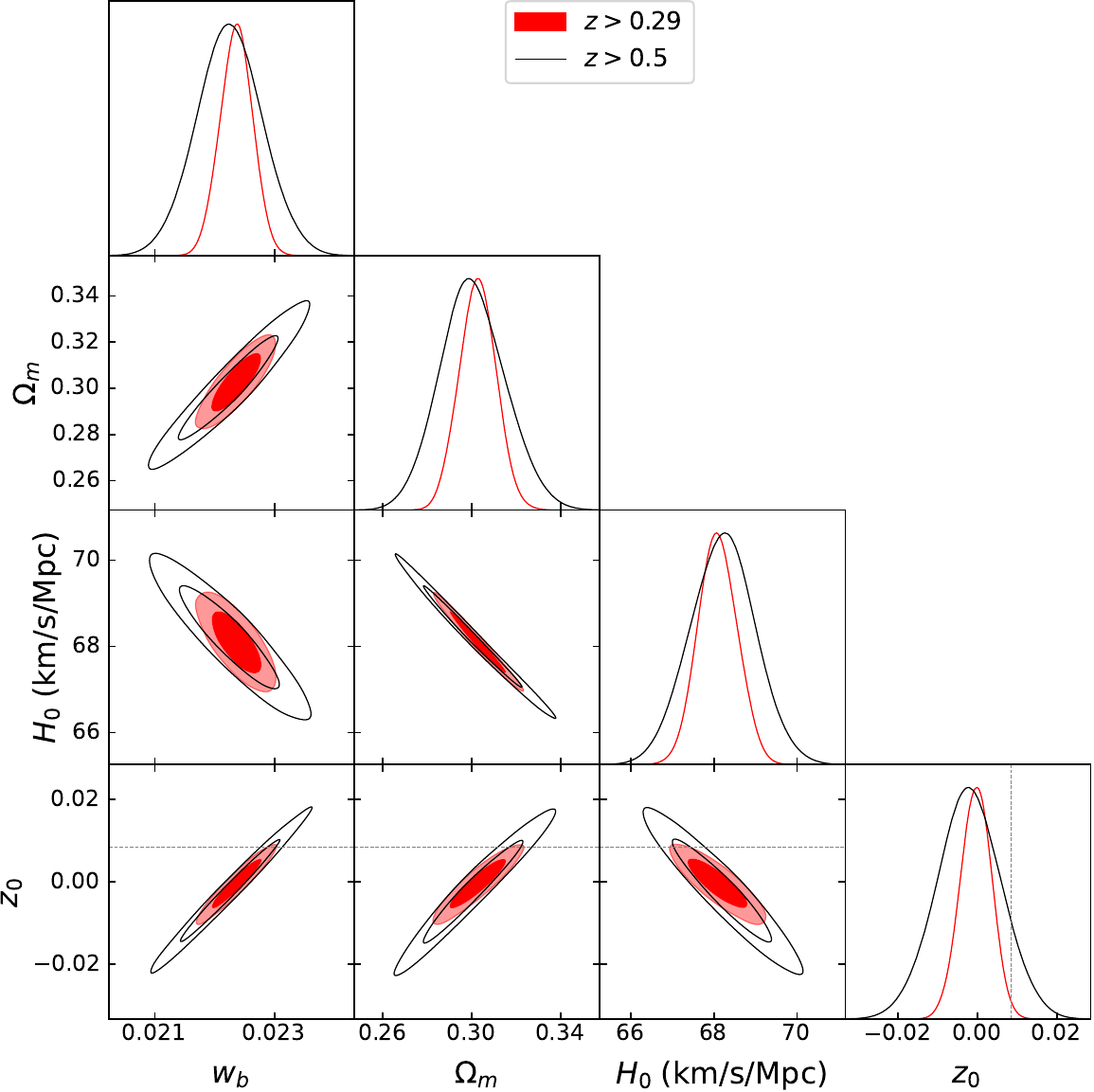}
    \caption{Triangle plot shown similarly to Figure~\ref{fig:Triangle} including also DES-Dovekie SNe~Ia. The posteriors are now tighter than before, but the previously predicted $z_0 = 0.84\%$ is still allowed at just over $2\sigma$.}
    \label{fig:Triangle_SN}
\end{figure}

Figure~\ref{fig:Triangle_SN} shows the triangle plot from our nested sampling run with free $z_0$. The parameters are summarised in Table~\ref{tab:parameters_SN}. Including DES-Dovekie SNe~Ia leads to tighter constraints, especially with our extended redshift range. However, the central parameter values do not change much, with the $z_0$ posterior still peaking close to 0. This leads to just over $2\sigma$ tension with the previously predicted value of 0.84\% \citepalias{Haslbauer_2020}. The posterior likelihood of an even higher $z_0$ is 1.20\%, indicating that $z_0 = 0.84\%$ cannot be excluded at the 99\% confidence level, even if we use our extended redshift range and include DES-Dovekie SNe~Ia. If we do not include SNe, the likelihood that $z_0 > 0.84\%$ rises to $P(z_0 > 0.0084) = 9.00\%$ for our extended redshift range, indicating that SNe disfavour the previously predicted $z_0$. Interestingly, this is not the case for our nominal redshift range, where including DES-Dovekie SNe~Ia raises $P(z_0 > 0.84\%)$ from 7.27\% to 9.74\%. This is because although SNe do tighten the error bars, they also increase the inferred value of $z_0$ for the nominal redshift range, while the opposite occurs for the extended redshift range. Given that the $z_0$ approximation might be inaccurate for the extended redshift range (Section~\ref{z0_validity}), a more detailed analysis is required to assess whether SNe favour the local void scenario.

\renewcommand{\arraystretch}{1.2}
\begin{table*}
    \centering
    \begin{tabular}{lcccc}
        \toprule
        & \multicolumn{4}{c}{Assumed value of $z_0$} \\
        Parameter & Free $z_0$ & 0 ($\Lambda$CDM) & 0.005 & 0.0084 \\
        \midrule
        \multicolumn{5}{c}{$z > 0.5$ dataset} \\
        \midrule
        $w_{\rm b}$ & $0.02225 \pm 0.00054$ & $0.02239 \pm 0.00009$ & $0.02272 \pm 0.00008$ & $0.02295 \pm 0.00009$ \\
        $\Omega_{\rm m}$ & $0.300^{+0.015}_{-0.014}$ & $0.304 \pm 0.004$ & $0.313 \pm 0.004$ & $0.319 \pm 0.004$ \\
        $H_0$ (km/s/Mpc) & $68.21^{+0.76}_{-0.80}$ & $68.02^{+0.26}_{-0.27}$ & $67.57 \pm 0.25$ & $67.28 \pm 0.25$ \\
        $z_0$ & $-0.0022 \pm 0.0081$ & -- & -- & -- \\ \midrule \midrule
        \multicolumn{5}{c}{$z > 0.29$ dataset} \\
        \midrule
        $w_{\rm b}$ & $0.02236 \pm 0.00027$ & $0.02239 \pm 0.00009$ & $0.02272 \pm 0.00008$ & $0.02294 \pm 0.00009$ \\
        $\Omega_{\rm m}$ & $0.303 \pm 0.008$ & $0.303 \pm 0.003$ & $0.313 \pm 0.003$ & $0.320 \pm 0.003$ \\
        $H_0$ (km/s/Mpc) & $68.09^{+0.48}_{-0.47}$ & $68.05 \pm 0.24$ & $67.53^{+0.24}_{-0.24}$ & $67.18^{+0.24}_{-0.23}$ \\
        $z_0$ & $-0.0004 \pm 0.0039$ & -- & -- & -- \\
        \bottomrule
    \end{tabular}
    \caption{Inferred parameter values under different assumed $z_0$ scenarios for the nominal redshift range (top) and extended redshift range (bottom), including DES-Dovekie SNe~Ia in the analysis. This leads to tighter constraints (c.f. Table~\ref{tab:parameters}).}
    \label{tab:parameters_SN}
\end{table*}
\renewcommand{\arraystretch}{1.0}

\subsection{Results excluding the LRG2 BAO measurement}
\label{Results_excluding_LRG2}

The low measured $D_{\rm V}/r_{\rm d}$ at $z = 0.7$ slightly reduces the inferred $z_0$ (Figure~\ref{fig:alpha_iso}). This may explain why the best-fit $z_0 < 0$ with our nominal redshift range. The inferred $z_0$ is not as negative with our extended redshift range, presumably due to the slightly low $D_{\rm V}/r_{\rm d}$ measurement at $z = 0.295$ pushing up $z_0$. We note that other $D_{\rm V}/r_{\rm d}$ measurements at $z \approx 0.7$ are all higher than reported by DESI~DR2, typically by around 2--3\% (Figure~\ref{D_V_graph}). If the DESI~DR2 measurement of $D_{\rm V}/r_{\rm d}$ at $z = 0.7$ is indeed revised upwards in the future and ends up more in line with pre-DESI results, this would reduce the tension for both $\Lambda$CDM and the $z_0 = 0.84\%$ model. It would also increase the inferred $z_0$, which might then be in better agreement with the predicted 0.84\%.

\renewcommand{\arraystretch}{1.2}
\begin{table}
    \centering
    \begin{tabular}{lcc}
        \hline
        & \multicolumn{2}{c}{Dataset considered for analysis} \\
        & $z > 0.5$ (nominal) & $z > 0.29$ \\ \hline
        Without SNe & $-0.003 \pm 0.011$ & $0.004 \pm 0.007$ \\
        With SNe & $-0.0004^{+0.0095}_{-0.0093}$ & $0.0010^{+0.0038}_{-0.0040}$ \\
        \hline
    \end{tabular}
    \caption{The inferred value of $z_0$ in analysis variants excluding the LRG2~BAO measurement from DESI~DR2 at $z = 0.7$, motivated by previous measurements at a similar redshift being systematically higher (Figure~\ref{D_V_graph}). Different rows show results without and with DES-Dovekie SNe~Ia (Section~\ref{Results_including_SNe}). Compared to results including the LRG2~BAO measurement (Tables~\ref{tab:parameters} and \ref{tab:parameters_SN}), the inferred $z_0$ is slightly higher, especially for our extended redshift range without SNe. However, $z_0$ remains consistent with 0 in all cases, especially once SNe are included.}
    \label{tab:z0_noLRG2}
\end{table}
\renewcommand{\arraystretch}{1.0}

Given the somewhat anomalous nature of the luminous red galaxy (LRG) 2 BAO measurement from DESI~DR2 at $z = 0.7$, we redo our free $z_0$ analysis without the BAO angular scale and redshift depth measurements at this redshift. The $z_0$ inference is summarised in Table~\ref{tab:z0_noLRG2} for our nominal and extended redshift without and with DES-Dovekie SNe~Ia. Although the results are consistent with $z_0 = 0$ in all cases, the inferred $z_0$ is slightly higher as a result of excluding the LRG2~BAO measurement (c.f. Table~\ref{tab:parameters}) -- as one might expect from Figure~\ref{fig:alpha_iso}. The preference for higher $z_0$ is especially apparent if we neglect SNe and use our extended redshift range. However, once SNe are included, the $z_0$ posterior peak remains close to 0 even if we exclude LRG2. Removing this point has only a modest impact on the results with SNe. Even so, it will still be important to revisit our analysis once updated BAO measurements become available.

\section{Discussion}
\label{Discussion}

\subsection{Implications for the local void scenario}
\label{Void_implications}

Our results demonstrate that the latest $z > 0.5$ measurements provide a sub-percent constraint on any fixed non-cosmological contribution to the redshift of distant sources. In the local void scenario, it was expected that this contribution would be $z_0 = 0.84\%$ \citepalias[section~5.3.3 of][]{Haslbauer_2020}. This remains plausible for both our nominal and extended redshift ranges (bottom right panel of Figure~\ref{fig:Triangle}). Interestingly, the most likely $z_0$ inferred by our analysis increases slightly in the latter case, lining up better with the predicted 0.84\%. This shows that the spillover effects of a local void at high $z$ remain consistent with the latest observations. However, our results prefer the $\Lambda$CDM value of $z_0 = 0$, causing the Bayesian evidence to disfavour the addition of $z_0$ as an extra model parameter (Table~\ref{tab:Bayesian_evidence}).

Our reconstructed $\alpha_{\rm iso}(z)$ shows that $\Lambda$CDM can adequately match the available BAO observations, as shown by the horizontal line at unity passing through most of the DESI~DR2 measurements on Figure~\ref{fig:alpha_iso}. The fiducial cosmology used to normalise the predicted $D_{\rm V}/r_{\rm d}$ differs slightly from the parameters that best fit the CMB, but the total $\chi^2$ is still quite acceptable (Table~\ref{tab:chi_sq}). Fixing $z_0$ to 0.5\% or 0.84\% slightly increases the total $\chi^2$, but this remains well within acceptable bounds. Interestingly, the higher $\chi^2$ arises despite a better fit to the CMB. This is because of a worse fit to DESI~DR2, as shown in Figure~\ref{fig:alpha_iso}.

The mild tension faced by both $\Lambda$CDM and the considered $z_0$ models is clearly largely driven by the anomalously low $D_{\rm V}/r_{\rm d}$ measurement at $z = 0.7$, though the low measurement at $z = 0.9$ also contributes somewhat. While there are few pre-DESI measurements at $z = 0.9$ with which to compare, there are several measurements at $z = 0.7$ (Figure~\ref{D_V_graph}). These measurements are generally about 2--3\% higher than reported by DESI~DR2. A recent reanalysis of DESI~DR1 using a covariance matrix estimated from mock catalogues also gives a result about 2\% higher than if DESI~DR2 results are linearly interpolated to the same redshift \citep{Farshad_2026}. This makes it likely that the DESI~DR2 measurement at $z = 0.7$ will be revised upwards by about this amount in future data releases, reducing the mild tension faced by both $\Lambda$CDM and the void models. We note that the inferred $z_0$ is clearly dragged down by the low DESI measurement at $z = 0.7$, since reducing $z_0$ to negative values provides a better fit to this point (Figure~\ref{fig:alpha_iso}). We therefore expect the posterior inference on $z_0$ to shift upwards in the future, which might then line up even better with the prediction of the local void scenario. We explore this possibility by inferring $z_0$ without the LRG2~BAO measurement (Section~\ref{Results_excluding_LRG2}). This does indeed increase the inferred $z_0$, but including DES-Dovekie SNe~Ia reduces it to almost exactly 0.

Our results in Figure~\ref{fig:alpha_iso} show little visible sign of the controversial BAO anomaly, a claimed internal inconsistency between high-redshift datasets in $\Lambda$CDM. However, this has to be assessed following the usual scientific method of making predictions in advance of the observations used to test a model -- or at least by setting up the model without using those observations. This procedure was discussed in detail in section~5 of \citetalias{Najera_2026}. The corresponding results in the $z_0$ model are presented in Section~\ref{BAO_anomaly}. With our nominal redshift range, $\Lambda$CDM faces an MDPS tension of $3.01\sigma$, which rises to $3.07\sigma$ with our extended redshift range. Focusing on the latter because it includes all 13 DESI~DR2 BAO measurements, we find that the local void model can reduce this tension down to $2.79\sigma$ if $z_0 = 0.84\%$. This is despite a slight increase in data space tension from $1.95\sigma$ to $2.22\sigma$, but this is not the MDPS tension (Table~\ref{tab:MDPS_tension}). Our results imply that although the local void model slightly worsens the fit to the data in terms of $\chi^2$ and ability to predict the DESI observations, it reduces the gap between BAO and non-BAO datasets in the standard $(H_0 r_{\rm d}, \Omega_{\rm m})$ parameter space commonly used for BAO problems. The reduced parameter space tension cannot be explained by inflated uncertainties because these are almost the same regardless of whether we fix $z_0 = 0$ or 0.84\% (Table~\ref{tab:parameters}). The latter gives a lower MDPS tension, but BAO and non-BAO datasets are inconsistent at the 99\% confidence level in both scenarios. However, if we double the probability of consistency to account for the look-elsewhere effect associated with considering the maximum of two tension metrics, then the reduction in parameter space tension that arises from fixing $z_0 = 0.84\%$ instead of 0 is sufficient to reconcile BAO and non-BAO datasets from $z > 0.29$ at the 99\% confidence level. In short, despite the slightly worse fit to the data if $z_0 = 0.84\%$, the residuals with respect to observations look less systematic in standard ways of viewing BAO problems.

The local void model with $z_0 = 0.84\%$ performs quite well compared to the background solutions considered by \citetalias{Najera_2026}. Their table~5 shows that each model with best-fitting values of its parameters usually predicts a BAO contribution to $\chi^2$ exceeding 20, often by a substantial margin. This is not ideal for 13~d.o.f. The only exceptions are the Phen models, which assume a simple algebraic addition to the $\Lambda$CDM $H(z)$ (see their equation~31). The Phen, exp model achieves a BAO $\chi^2$ of 19.24, while the Phen, sech model achieves 18.31. Our Table~\ref{tab:chi_sq} shows that the best-fitting model with $z_0 = 0.84\%$ achieves a similar BAO $\chi^2$ contribution of 18.61. This could be reduced further if the void parameters are optimised including BAO datasets, since it is quite possible that a different model with reduced $z_0$ might still adequately fit the local $cz'$ and galaxy number counts. For instance, if $z_0 = 0.5\%$, the BAO contribution to $\chi^2$ drops to 16.79. Even without this, it is clear that the void model provides an adequate fit for 13~d.o.f.

To perform a comparison in terms of the MDPS BAO tension, we need to mimic our procedure of fixing $z_0$ to the best-fitting value based on non-BAO datasets. For this, we rerun the BAO tension calculation in \citetalias{Najera_2026} for two of the Phen models, fixing their two non-standard parameters to best-fitting values on the basis of all the non-BAO datasets they considered. The third Phen model in their study is not shown here because it cannot solve the Hubble tension (the CMB contribution to $\chi^2$ is 26.7 from 3~d.o.f.). The results for the other two models are shown in Table~\ref{tab:MDPS_tension_Phen}. As with $\Lambda$CDM, the MDPS tension comes from parameter space and is just under $2\sigma$ for both Phen models. By this measure, the local void model with our extended redshift range and $z_0 = 0.84\%$ performs slightly worse, since it faces an MDPS tension of $2.79\sigma$, also from parameter space (bottom part of Table~\ref{tab:MDPS_tension}). Moreover, the data space tension is slightly higher in the void model compared to the Phen models from \citetalias{Najera_2026}. These small differences make it difficult to tell which is better from an observational perspective.


\begin{table}
    \centering
    \setlength{\tabcolsep}{4pt} 
    \begin{tabular}{lcc}
        \hline
        & \multicolumn{2}{c}{Phen model \citepalias{Najera_2026}} \\
        BAO tension metric & Exp & Sech \\ \hline
        Data space $\chi^2$ (13 d.o.f.) \rule{0pt}{9pt} & 18.98 & 21.55 \\
        Data space tension & $1.54\sigma$ & $1.86\sigma$ \\ [5pt]
        Parameter space $\chi^2$ (2 d.o.f.) & 5.00 & 6.03 \\
        Parameter space tension & $\boldsymbol{1.74\sigma}$ & $\boldsymbol{1.97\sigma}$ \\ 
        \hline
    \end{tabular}
    \caption{BAO tensions for the Phen models considered by \citetalias{Najera_2026} which can solve the Hubble tension. The MDPS tension is highlighted in bold in each column.}
    \label{tab:MDPS_tension_Phen}
    \setlength{\tabcolsep}{6pt} 
\end{table}

We suggest that the Phen models in \citetalias{Najera_2026} empirically capture how a local void distorts the distance--redshift relation. While their models can of course be interpreted as background solutions, this raises fine-tuning issues because the non-standard addition to $H(z)$ must decay rather rapidly with redshift, as captured by their $z_1$ parameter. Those authors infer that $z_1 < 0.20$ at high confidence (see their table~3). This is far below the redshift where dark energy changes the sign of $\ddot{a}$, raising the question of why departures from $\Lambda$CDM should arise only at such recent epochs. If this is the case, it would imply rather unusual behaviour for the effective dark energy equation of state (see their figure~4). As argued by those authors, the low inferred $z_1$ is more naturally interpreted as capturing the impact of a local void rather than a genuine adjustment to the background $a(t)$.

While our goal here is to address the consistency of the local void model at high $z$, it is clear that its main impact would be at lower $z$, where a more detailed model would be required \citepalias{Haslbauer_2020}. This would be particularly important if the DESI BAO measurements can be extended to lower $z$, where a larger deviation from $\Lambda$CDM is expected. Indeed, a generic consequence of a low-redshift solution to the Hubble tension is that $\alpha_{\rm iso} \approx 0.92$ at low $z$ because inflating $cz'$ by 9\% would correspondingly reduce the distance to any given $z$ \citep{Banik_2025_BAO}. Other tests of a local void are also possible at low $z$ \citep{Banik_2026_void}. For instance, it is possible to reconstruct $\dot{a}(z)$ assuming $\Lambda$CDM at high $z$ by allowing a flexible evolution for the dark energy equation of state \citep{Jia_2023, Jia_2025a, Jia_2025b}. These results can be cast in terms of how much deviation there is from the CMB-calibrated $\Lambda$CDM expectation using the concept of $H_0(z)$, the value of $H_0$ that would be inferred from extrapolating data in a narrow redshift range centred on $z$ down to $z = 0$ assuming $\Lambda$CDM  \citep{Krishnan_2020, Krishnan_2021}. The predicted $H_0(z)$ curve in the void model \citep{Mazurenko_2025} is in good agreement with the observational reconstruction \citep{Jia_2025a}. More recent results indicate even better agreement \citep{Jia_2025b, Jia_2026}. Importantly, the tendency for the reconstructed $H_0(z)$ to reach the community consensus on the local $cz'$ as $z \to 0$ occurs without applying local $cz'$ constraints \citep{Jia_2025b}. Those authors only use uncalibrated SNe~Ia combined with BAO assuming a standard $r_{\rm d}$, which effectively calibrates SNe~Ia via the inverse distance ladder. Similar results can be obtained by taking the opposite approach, using the local $cz'$ without assuming $\Lambda$CDM at high $z$ \citep{Lopez_2025}. These results make it likely that the solution to the Hubble tension lies at low $z$, though they do not directly address whether this is through a local void or a late-Universe background solution.

One unusual aspect of the CMB that cannot easily be understood through modified $a(t)$ is the preference for a much lower $H_0$ and hotter CMB when the FIRAS prior is dropped \citep{Ivanov_2020_TCMB}. Their Figure~3 shows that if we fix $(\widetilde{w}_{\rm b}, \widetilde{w}_{\rm bc}, \theta_\star)$, we can alter $T_{\rm CMB}$ without impacting the primary CMB anisotropies, but there is a small secondary effect on the power spectrum around a multipole moment of $\ell \approx 10$. This is due to the integrated Sachs-Wolfe (ISW) effect \citep{Sachs_1967}, which depends on the balance between matter and dark energy densities at late times. Since this is the only discernible impact on the CMB power spectrum if we move along the usual degeneracy between $H_0$ and $T_{\rm CMB}$ corresponding to fixed $(\widetilde{w}_{\rm b}, \widetilde{w}_{\rm bc}, \theta_\star)$, it seems likely that the inference of a hotter CMB is related to some difficulty faced by $\Lambda$CDM with the growth of structure on large scales. We note that the CMB temperature anomaly would be even larger than reported by \citet{Ivanov_2020_TCMB} if they had allowed $\Omega_{\rm m} > 1$, since the posteriors on $H_0$ and $T_{\rm CMB}$ were clearly truncated by this prior, which lacks theoretical motivation (the cosmological constant could be negative). The large deviation from conventional estimates of $H_0$ and $\Omega_{\rm m}$ is not a sign that these parameters could really be drastically different from $\Lambda$CDM, but instead indicates that it faces a problem with the CMB power spectrum in the temperature-sensitive part at $\ell \approx 10$. This may be related to the enhanced growth of structure on large scales required for a local void to solve the Hubble tension. Since the details of the CMB power spectrum at these low multipoles depend on as yet unknown details of structure formation on large scales, our main analyses neglect the fact that altering $T_{\rm CMB}$ at fixed $(\widetilde{w}_{\rm b}, \widetilde{w}_{\rm bc}, \theta_\star)$ slightly alters the predicted CMB power spectrum at low $\ell$. The information content here is in any case far lower than in the primary CMB anisotropies, which are unaffected at fixed $(\widetilde{w}_{\rm b}, \widetilde{w}_{\rm bc}, \theta_\star)$ \citep{Ivanov_2020_TCMB}.

To obtain a more detailed prediction for the CMB power spectrum at low $\ell$, we would require a deeper theory where KBC-like voids are common, yet the background expansion history and early universe physics remain largely unchanged. Models that increase the frequency of extreme voids would inevitably also increase the frequency of extreme overdensities, perhaps explaining the properties of the El Gordo massive galaxy cluster collision better than $\Lambda$CDM, which struggles to explain its high redshift, mass, and collision velocity \citep*{Asencio_2021, Asencio_2023}. It has even been proposed that relaxing the assumption of independence between modes on different wavelengths could jointly solve the Hubble tension and the cosmic radio dipole anomaly, an unexpectedly large dipole amplitude in radio source number counts at cosmological distances \citep{Secrest_2021, Secrest_2022, Dam_2023, Wagenveld_2024, Oayda_2026}. Both tensions could be assigned to $\la 2\sigma$ cosmic variance if it is enhanced in this way \citep{Gandhi_2026}.

\subsection{Implications for theories other than a local void}
\label{Non_void_implications}

The limits that we place on $z_0$ limit the scope for theories that enhance the growth of structure too much, as that would make it unlikely for $z_0$ to be as small as our results imply \citep[c.f.][]{Russell_2026}. Indeed, the main reason the \citetalias{Haslbauer_2020} model predicts $z_0$ at an almost observable level is their use of MOND-inspired equations to enhance the gravitational field of a given (under)density distribution on scales $\ga 100$~Mpc. The \citetalias{Haslbauer_2020} model thus has both a larger and deeper underdensity than allowed in $\Lambda$CDM, and stronger outward gravity even for the same underdensity.

To illustrate this, we can consider the predicted $z_0$ if there is a standard relation between the gravitational field and the density distribution sourcing it. Though we assume that density variations exist only in the matter component, our derivation works similarly if the underdensity is instead in the dark energy distribution \citep{Nunes_2006}. The gravitational field $g$ at distance $r$ from a spherically symmetric fractional underdensity $\delta$ is
\begin{eqnarray}
    g ~=~ \frac{4 \mathrm{\pi} G r \overline{\rho} \delta}{3} \, ,
    \label{g_void}
\end{eqnarray}
where $\overline{\rho}$ is the cosmic average matter density. Assuming the void has size $r_v$ and integrating Equation~\ref{g_void} once to obtain the central potential $\Phi_0$, we can estimate that
\begin{eqnarray}
    z_0 ~=~ \frac{\Phi_0}{c^2} ~=~ \frac{1}{4} \left( \frac{r_v}{r_H} \right)^2 \Omega_\mathrm{m} \, \delta \, ,
    \label{z0_void}
\end{eqnarray}
where $r_H \equiv c/H_0$ and we used the fact that $\overline{\rho} \equiv 3H_0^2\Omega_{\rm m}/(8 \mathrm{\pi}G)$. Assuming that $H_0 = 70$~km/s/Mpc and $\Omega_{\rm m} = 0.3$, our constraint that $\left| z_0 \right| \la 0.01$ (Table~\ref{tab:parameters}) implies that
\begin{eqnarray}
    \left| \delta \right| ~\la~ 2.5 \left( \frac{1~\mathrm{Gpc}}{r_v} \right)^2 \, .
\end{eqnarray}
This constraint is very weak for any structure that can be described as local or non-cosmological, indicating that our limit on $\left| z_0 \right|$ is not particularly helpful in standard gravity. This is the case even if we assume the density variation actually lies in the dark energy, since that would just replace $\Omega_{\rm m} \to \Omega_\Lambda$, leaving our result qualitatively unchanged.

Our constraints on $z_0$ are far more relevant to modified theories of gravity in which it is stronger than in standard theory on very large scales. The ultimate detection of $z_0$ at the $\mathcal{O} \left( 10^{-3} \right)$ level would suggest that the effective $G$ relevant to the growth of structure on scales $\ga 100$~Mpc is different from the $G$ entering the Friedmann equation. In the context of the local void model, our $z_0$ constraint places a limit on the void size because $z_0 \appropto r_v^2$ (Equation~\ref{z0_void}). Incorporating the techniques developed here into the \citetalias{Haslbauer_2020} analysis would naturally rectify its tendency to drift towards very large $r_v$ (see their appendix~C).

We have assumed that any local structure would be contained within $z \la 0.3$, enabling us to use the $z_0$ approximation (Equation~\ref{eq:z_0}). This would not be true in scenarios where there is a very long wavelength perturbation imprinted by inflationary physics. In this case, we could make the opposite approximation that any local potential hill would extend out much further than the most distant BAO data point at $z = 2.3$. In this case, the $z_0$ term would still affect the CMB and the $r_{\rm d}$ calculation, but it should not be considered when calculating CC and BAO predictions. This is because there would be little potential difference between us and any sources at $z < 2.3$. Our analysis would need to be redone with this revised structure to test such models, though it is possible that stronger constraints can be obtained from the kinematic Sunyaev-Zel'dovich (kSZ) effect \citep{Sunyaev_1980}. kSZ constraints still allow a void similar to the KBC void, but a much larger and deeper void is excluded \citep*{Ding_2020, Cai_2025}. A similar argument would apply to a local overdensity, which would create a kSZ signal of the opposite sign. Constraints from the kSZ effect probe peculiar velocities generated by a potential gradient, while $z_0$ constraints directly probe potential differences through impacts like a hotter predicted CMB and reduced $r_{\rm d}$, issues we have considered in some detail.

Although our focus has been on GR, there is also a kinematic contribution to the redshift. This is generally expected to be small at the distances considered here, even in the local void scenario \citep{Banik_2025_BAO}. However, it has been proposed that the cosmic radio dipole anomaly is caused by a substantial velocity between the quasar rest frame and the CMB, which would create angular variation in number counts due to special relativistic beaming effects \citep*{Singal_2023, Bashir_2026}. This kinematic interpretation is by no means necessary even if the cosmic radio dipole anomaly is confirmed, since the dipole in number counts could be intrinsic \citep{Wagenveld_2025, Gandhi_2026}. In the kinematic interpretation, there would be an interesting interplay with BAO observations because these generally cover only a small portion of the sky. A coherent bulk flow on such large scales could then create an extra contribution to the redshift, distorting the relation between redshift and the BAO observables. However, even a bulk flow of 1000~km/s is still only $0.003 \, c$, so the effect ought to be well below our uncertainty on $z_0$. This is especially true given that much of the impact of $z_0$ on our analysis stems from the hotter predicted CMB, which would not arise in this scenario given CMB observations cover the whole sky. It is also unclear how a coherent bulk flow could arise on the very large scales relevant to DESI~DR2 BAO observations, though this should be tested by checking for anisotropy in the BAO angular scale and redshift depth once deep BAO surveys cover more of the sky.

\section{Conclusions}
\label{Conclusions}

There is now a considerable body of evidence that the local $cz' > H_0^{\rm CMB}$ (Section~\ref{Introduction}). This Hubble tension might arise due to outflow from a local void \citep[as reviewed in][]{Banik_2026_void}. As with all models to inflate the local $cz'$, there are inevitably spillover effects at high $z$. We test the consistency of the local void model using the wealth of high-$z$ data that has recently become available, including BBN, CMB, CC, and especially BAO measurements from DESI~DR2 \citep{DESI_2025}. It was previously claimed that the local void model from \citetalias{Haslbauer_2020} provides a better fit than $\Lambda$CDM to BAO measurements over the last 20 years \citep{Banik_2025_BAO}. We assess this claim in more detail by including several effects that those authors did not account for.

To simplify our analysis, we approximate that a local void only affects the redshift through a fixed additional GR contribution of $z_0$, keeping only the GR term in Equation~\ref{eq:z_contributions}. Figure~\ref{D_V_graph} shows that this is a good approximation at $z > 0.5$, where a local void would have little physical effect. Its main impact on observations of such distant sources would be through the GR contribution arising from photons being repelled by the outward gravity of a local void. We therefore approximate a local void using the $z_0$ model (Equation~\ref{eq:z_0}). We argue in Section~\ref{z0_validity} that this may well still be reasonably accurate down to $z = 0.29$, allowing us to compare with all 13 DESI~DR2 BAO measurements and benefit from several additional CC measurements (Section~\ref{Analysis}).

Our main result is that $z_0 = -0.004^{+0.008}_{-0.009}$ with data from our nominal redshift range, rising to $z_0 = -0.001 \pm 0.007$ for our extended redshift range. In both cases, the results are in good agreement with the prior expectation that $z_0 = 0.0084$ based on fitting the local $cz'$ and galaxy number counts without considering BAO datasets \citepalias[section~5.3.3 of][]{Haslbauer_2020}. A slightly better fit can be obtained in a different void model with $z_0 = 0.005$ if it still matches the local constraints. This might arise due to a smaller void \citep[as suggested by][]{Stiskalek_2025_void}. In either case, the best-fitting models are in good agreement with DESI~DR2 (Figure~\ref{fig:alpha_iso}) and provide an acceptable $\chi^2$ (Table~\ref{tab:chi_sq}).

It has been claimed that $\Lambda$CDM suffers from an internal tension between BAO and other high-redshift datasets. This BAO anomaly arises when considering the ability of $\Lambda$CDM to predict BAO data on the basis of non-BAO datasets \citepalias{Najera_2026}. Applying their approach to data from our extended redshift range so that all DESI~DR2 measurements can be addressed, we find that $\Lambda$CDM faces an MDPS tension of $3.07\sigma$ from the standard $(H_0 r_{\rm d}, \Omega_{\rm m})$ parameter space. In the local void model with $z_0 = 0.84\%$, this decreases to $2.79\sigma$, which now lies within the 99\% credible region if the tail probability is doubled to account for the look-elsewhere effect associated with considering the maximum of two tension metrics. This is not the case for $\Lambda$CDM. The BAO anomaly is not apparent in data space, where $\Lambda$CDM faces a tension of only $1.95\sigma$ for our extended redshift range. Interestingly, the void model actually increases this to $2.22\sigma$, indicating that it navigates the subtle tradeoffs between data and parameter space tensions to reduce the MDPS tension, which is conventionally used to quantify the BAO anomaly.

Fixing $z_0$ to the 0.84\% level predicted by a model previously shown to fit local constraints quite well \citepalias[section~5.3.3 of][]{Haslbauer_2020}, we find an MDPS BAO tension of only $2.79\sigma$ and acceptable $\chi^2$ contributions from BBN, CMB, BAO, and CC in the best-fitting model (Table~\ref{tab:chi_sq}). Despite the non-standard prediction of a slightly hotter CMB and reduced $r_{\rm d}$, fixing $z_0 = 0.84\%$ does not cause conflict with the high-redshift datasets considered here. It is also possible that a different void model could acceptably fit local constraints with reduced $z_0$ of perhaps 0.5\%, which would fit the data better but mildly increase the MDPS tension to $2.88\sigma$. We therefore conclude that the local void model remains consistent with high-redshift datasets despite their high precision and at best modest deviations from the best-fit $\Lambda$CDM model.

These results motivate a more thorough analysis of the void scenario beyond the high-redshift approximation adopted here, since its primary impact occurs at low redshift, where a more detailed model including outflows is required to assess its viability in light of the Hubble tension. A valuable dataset in this context is the CF4 galaxy catalogue of redshifts and redshift-independent distances \citep{Tully_2023_CF4}. This was used to show that any local void in a background \emph{Planck} cosmology should be smaller and perhaps deeper than claimed by \citetalias{Haslbauer_2020}, but the local $cz'$ over the 100--600~Mpc range typically used to measure it can still be as high as 73~km/s/Mpc within the $1\sigma$ error bar for one of the three considered void density profiles \citep{Stiskalek_2025_void}. We are also exploring if the rapid return to the \emph{Planck} cosmology predicted by the local void scenario is consistent with DESI~CC measurements at $z < 0.2$ \citep{Wang_2026}.

\section*{Acknowledgements}

The authors are grateful to the anonymous referees for comments which helped to substantially improve this manuscript.

\section*{Funding}

All authors are supported by Royal Society University Research Fellowship grant 211046.

\section*{Author contributions}

Conceptualization, I. B. and J. A. N.; methodology, I. B. and J. A. N.; software, I. B. and J. A. N.; validation, I. B. and J. A. N.; formal analysis, J. A. N.; investigation, J. A. N.; resources, H. D.; writing---original draft preparation, I. B.; writing---review and editing, I. B., J. A. N., and H. D.; visualization, J. A. N.; supervision, I. B. and H. D.; project administration, H. D.; funding acquisition, H. D. All authors have read and agreed to the published version of the manuscript.


\section*{Conflicts of interest}

The authors declare no conflict of interest.

\section*{Data Availability}

Our code is publicly available \href{https://github.com/antonionajeraq/z0_local_void_BAO.git}{\faGithub}. This includes the main cosmological analysis and scripts to extract \textsc{class} recombination and decoupling temperatures over a parameter grid, fit the results using a quadratic form, and then print out the rms and maximum absolute residual. The datasets used in this work are publicly available and come from the cited publications.

\begin{appendix}

\section{The recombination and decoupling temperatures in \texorpdfstring{$\Lambda$}{L}CDM}
\label{T_without_z0}

Most solutions to the Hubble tension are at the background level, in which case $z_0 = 0$ and $T_{\rm CMB} = T_{\rm FIRAS}$. For the benefit of readers, here we provide fitting formulae for the recombination and decoupling temperatures as a function of $(w_{\rm{b}}, w_{\rm bc})$, with our grid still centred on (0.0225, 0.1450) as before. Though $z_0$ is no longer a parameter, $T_\star$ should still be found using Equation~\ref{eq:T_star}, with $T_{\rm d}$ found analogously.

The reference recombination temperature is now $T_{\star}^{\rm ref} = 2970.29$~K, while its gradient and Hessian become
\begin{eqnarray}
    \bm{g}_{\star} = \begin{pmatrix}
    -3188.92 \\
    194.09
    \end{pmatrix}, ~~
    \matr{H}_{\star} = \begin{bmatrix}
    265960.27 & -7860.93 \\
    -7860.93 & -488.36
    \end{bmatrix}.
    \label{eq:g_H_recombination_noz0}
\end{eqnarray}

The reference decoupling temperature is now $T_{\rm d}^{\rm ref} = 2892.74$~K, while its gradient and Hessian become
\begin{eqnarray}
    \bm{g}_{\rm d} = \begin{pmatrix}
    5925.64 \\
    203.34
    \end{pmatrix}, ~~
    \matr{H}_{\rm d} = \begin{bmatrix}
	-189076.80 & -4767.87 \\
    -4767.87 & -626.49
    \end{bmatrix}.
    \label{eq:g_H_decoupling_noz0}
\end{eqnarray}

These results are very similar to those used in our main analysis if we simply set $z_0 = 0$, but fixing this when calculating our \textsc{class} grid improves the accuracy slightly in studies which assume homogeneity. This is evident in that the maximum (rms) error in $T_\star$ across our grid is now only 0.68 (0.23)~mK, while the maximum (rms) error in $T_{\rm d}$ is only 0.80 (0.32)~mK.

\section{Evolution of the neutrino density}
\label{Neutrino_density_evolution}

We assume a standard background cosmology similarly to \citetalias{Najera_2026}, but we make a small refinement to the handling of neutrinos in their appendix~B. Since the neutrino energy density is defined relative to that in photons, we briefly recap the photon energy density. Expressed as an equivalent mass density, this is
\begin{eqnarray}
    \rho_\gamma ~=~ \frac{\mathrm{\pi}^2 k^4}{15 \hbar^3 c^5} T_\text{CMB}^4 \, ,
\end{eqnarray}
where $k$ is Boltzmann's constant, $\hbar$ the reduced Planck constant, and $T_\text{CMB}$ the temperature of the CMB (Equation~\ref{eq:T_FIRAS_adjustment}). The photon number density $n_\gamma$ follows from standard blackbody physics.
\begin{eqnarray}
    n_\gamma ~=~ \frac{2 \, \zeta(3)}{\mathrm{\pi}^2} \left(\frac{k T_\text{CMB}}{\hbar c} \right)^3,
\end{eqnarray}
where $\zeta(3) = 1.202$ is the Riemann zeta function with argument 3.

Neutrinos behave like photons at high redshift. The ultra-relativistic neutrino density in species $i$ is
\begin{eqnarray}
    \rho_{\nu, \mathrm{ur}, i} ~\equiv~ \rho_\gamma R_{\nu_i/ \gamma} \, .
\end{eqnarray}
We designate the neutrino species as $A$, $B$, and $C$ in descending order of rest mass. The ur subscript indicates that the ratio $R_{\nu_i/ \gamma}$ between the photon density and that in neutrino species $i$ is applicable only in the early universe, when neutrinos are ultra-relativistic. Using quantum statistics, we get that
\begin{eqnarray}
    R_{\nu_i/\gamma} ~=~ \frac{7}{8} \left( \frac{4}{11} \right)^{4/3} \frac{N_\mathrm{eff}}{3} \, ,
    \label{eq:R_nu_gamma}
\end{eqnarray}
where $N_\mathrm{eff} = 3.044$ is effectively the number of neutrino species around the time of recombination \citep{Akita_2020, Froustey_2020, Bennett_2021}.

In the non-relativistic regime, the neutrino density in species $i$ is given by
\begin{eqnarray}
    \rho_{\nu, i} ~=~ n_{\nu, i} m_{\nu, i} \, ,
\end{eqnarray}
where $m_{\nu, i}$ is the neutrino mass of flavour $i$ and $n_{\nu, i}$ is the number density of such neutrinos, which is given by
\begin{eqnarray}
    n_{\nu, i} ~=~ \frac{3}{11} n_\gamma \left( \frac{N_\mathrm{eff}}{3} \right)^{0.68}.
\end{eqnarray}
If the neutrinos were thermalised, the exponent here would be $3/4$. We use a smaller exponent of 0.68 to ensure that in the non-relativistic limit, $w_\nu = \sum_i m_{\nu, i} c^2/(93.14 \, \mathrm{eV})$ \citep{Lesgourgues_2012}. The lower exponent arises because the energy from electron--positron annihilation mostly goes into the high-energy tail of the neutrino energy distribution, which does not have enough time to thermalise before neutrino decoupling.

Unlike photons which are always relativistic, massive neutrinos of species $i$ become non-relativistic when the cosmic scale factor is
\begin{eqnarray}
    a_{nr, i} ~=~ \frac{1}{\sqrt{\left( \left. \dfrac{\rho_{\nu, i}}{\rho_{\nu, \mathrm{ur}, i}} \right|_{a = 1} \right)^2 - 1}}.
    \label{eqn:a_nr}
\end{eqnarray}

We use the following form for the transition between relativistic and non-relativistic behaviour:
\begin{eqnarray}
    \rho_{\nu, i} = \frac{\left. \rho_{\nu, \mathrm{ur}, i} \right|_{a = 1}}{a^4} \sqrt{1 + \left( \frac{a}{a_{\mathrm{nr}, i}} \right)^2} \, .
    \label{Neutrino_transition_function}
\end{eqnarray}
This only applies to massive neutrinos, which consequently have a present neutrino density $\rho_{\nu, i} \gg \rho_{\nu, \mathrm{ur}, i}$, what the neutrino density would have been if the neutrinos were massless. For any massless species, the behaviour is identical to radiation:
\begin{eqnarray}
    \rho_{\nu, i} = \frac{\left. \rho_{\nu, \mathrm{ur}, i} \right|_{a = 1}}{a^4} \, .
\end{eqnarray}

\citetalias{Najera_2026} assumed for simplicity that all the neutrino species have the same mass, so $m_{\nu, i} = \left( 20, 20, 20 \right)$~meV/$c^2$. We replace this with the much more realistic assumption that $m_{\nu, i} = \left( 50, 10, 0 \right)$~meV/$c^2$, as indicated by terrestrial neutrino experiments if neutrino rest masses are kept as low as possible, implying we need to use the normal hierarchy or mass ordering. The neutrino mass sum therefore remains $\sum m_\nu = 60$~meV/$c^2$, matching \citetalias{Najera_2026} and most other studies.

Equation~\ref{Neutrino_transition_function} can be generalised to
\begin{eqnarray}
    \rho_{\nu, i} ~=~ \frac{\left. \rho_{\nu, \mathrm{ur}, i} \right|_{a = 1}}{a^4} \left[ 1 + \left( \frac{a}{a_{\mathrm{nr}, i}} \right)^{k_\nu} \right]^{1/k_\nu} \, .
    \label{Neutrino_transition_function_knu}
\end{eqnarray}
Following \citetalias{Najera_2026}, we have assumed that the transition of neutrinos from relativistic to non-relativistic has a sharpness of $k_\nu = 2$. We can obtain a better fit to detailed calculations using Fermi-Dirac statistics if we use $k_\nu = 1.83$ \citep[equation~26 of][]{Komatsu_2011}. However, we will see below that this makes very little difference to the results. We therefore assume that $k_\nu = 2$, which is much cheaper computationally because it avoids the use of fractional powers.

\section{Accuracy of the compressed CMB likelihood}
\label{CMB_compressed_likelihood}

An important improvement over \citetalias{Najera_2026} is that the recombination temperature is found using a quadratic fit to \textsc{class} results rather than based on an ancient fitting formula \citep{Hu_1996}. To better line up with modern results and include three decades of progress on the atomic physics constants and other factors relevant to recombination, \citetalias{Najera_2026} adapted the formula slightly by changing a fixed factor of $1048 \to 1045$ in their equation~A10. In our approach, such a `fix' ought not to be necessary, but it is still important to check the accuracy of our approach in some way.

\renewcommand{\arraystretch}{1.2}
\begin{table}
    \centering
    \begin{tabular}{ccc}
        \hline
        Max. $m_\nu$ & \multicolumn{2}{c}{Transition sharpness} \\
        (meV/$c^2$) & $k_\nu = 1.83$ & $k_\nu = 2$ \\
        \hline
        60 & $67.20^{+0.37}_{-0.41}$ & $67.18 \pm 0.39$ \\
        50 & $67.18^{+0.41}_{-0.36}$ & $67.19^{+0.41}_{-0.36}$ \\
        20 & $67.24 \pm 0.39$ & $67.27^{+0.38}_{-0.41}$ \\
        \hline
    \end{tabular}
    \caption{The inferred value of $H_0$ with fixed $\sum m_\nu = 60$~meV/$c^2$ but different masses for the most massive neutrino, as indicated in the first column. In the 50~meV/$c^2$ case, another species has a mass of 10~meV/$c^2$, while the third species is massless. All neutrinos are equally massive in the 20~meV/$c^2$ case. The sharpness of the relativistic to non-relativistic transition is governed by the parameter $k_\nu$ (Equation~\ref{Neutrino_transition_function_knu}).}
    \label{tab:H0_neutrinos}
\end{table}
\renewcommand{\arraystretch}{1.0}

We infer $H_0$, $w_{\rm b}$, and $w_{\rm bc}$ in $\Lambda$CDM using the compressed CMB likelihood (Equation~\ref{CMB_covariance_matrix}). Since the preferred values and covariance matrix are obtained from the chains in \citet{SPT_2026} without using their reported $H_0$, we can compare this to our $H_0$ inference. The results are shown in Table~\ref{tab:H0_neutrinos} for different assumed values of $k_\nu$ and the neutrino masses, with $\sum m_\nu = 60$~meV/$c^2$ in all cases. The first column shows $m_{\nu, A}$, the mass of the most massive neutrino species. We assume this is the only massive neutrino, species $A$ and $B$ are the only massive neutrinos, or that all three species are equally massive \citepalias[as assumed in][]{Najera_2026}. Our nominal assumption is that $m_{\nu, A} = 50$~meV/$c^2$ and $m_{\nu, B} = 10$~meV/$c^2$, with species $C$ being massless. Our results are barely affected by whether we choose $k_\nu = 1.83$ or 2, so we choose the latter to reduce complexity.

The \citet{SPT_2026} implicitly assumed that species $B$ and $C$ are massless, recovering $H_0 = 67.19 \pm 0.38$~km/s/Mpc in this case. Our compressed CMB likelihood approach is able to recover this value and uncertainty to an accuracy of 0.01~km/s/Mpc, a negligibly small difference. This demonstrates the reliability of our much simpler system of equations, which bypasses calls to \textsc{class} in the likelihood function and does not involve comparison to the full CMB power spectrum. Although we assume that $m_{\nu, A} = 50$~meV/$c^2$ and $m_{\nu, B} = 10$~meV/$c^2$, the results are barely affected by assuming instead that only species $A$ is massive while keeping $\sum m_\nu$ fixed. However, there is a noticeable difference if we assume that all three species are equally massive -- despite fixing $\sum m_\nu$. This means that although details of the neutrino relativistic to non-relativistic transition are far less important for cosmology than $\sum m_\nu$, it is still important to assign the vast majority of $\sum m_\nu$ to just one species, as indicated by terrestrial neutrino experiments if we make the usual assumption that neutrino masses should be kept as low as possible. This assumption is called the normal mass ordering, which can be distinguished from the inverted mass ordering with a few years of data from the recently constructed Jiangmen Underground Neutrino Observatory \citep[JUNO;][]{JUNO_2025}.

\end{appendix}

\bibliographystyle{mnras_max10}
\bibliography{z0m_bbl}

@ARTICLE{Nunes_2006,
       author = {{Nunes}, N.~J. and {Mota}, D.~F.},
        title = "{Structure formation in inhomogeneous dark energy models}",
      journal = {MNRAS},
         year = 2006,
        month = may,
       volume = {368},
       number = {2},
        pages = {751-758},
          doi = {10.1111/j.1365-2966.2006.10166.x},
archivePrefix = {arXiv},
       eprint = {astro-ph/0409481},
 primaryClass = {astro-ph},
       adsurl = {https://ui.adsabs.harvard.edu/abs/2006MNRAS.368..751N}}

@ARTICLE{Wirtz_1922,
       author = {{Wirtz}, C.},
        title = "{Einiges zur Statistik der Radialbewegungen von Spiralnebeln und Kugelsternhaufen}",
      journal = {Astronomische Nachrichten},
         year = 1922,
        month = apr,
       volume = {215},
        pages = {349},
          doi = {10.1002/asna.19212151703},
       adsurl = {https://ui.adsabs.harvard.edu/abs/1922AN....215..349W}}

@ARTICLE{Wirtz_1924,
       author = {{Wirtz}, C.},
        title = "{De Sitters Kosmologie und die Radialbewegungen der Spiralnebel}",
      journal = {Astronomische Nachrichten},
         year = 1924,
        month = jul,
       volume = {222},
       number = {2},
        pages = {21},
          doi = {10.1002/asna.19242220203},
       adsurl = {https://ui.adsabs.harvard.edu/abs/1924AN....222...21W}}

@ARTICLE{Zhao_2013_GR,
       author = {{Zhao}, HongSheng and {Peacock}, John A. and {Li}, Baojiu},
        title = "{Testing gravity theories via transverse Doppler and gravitational redshifts in galaxy clusters}",
      journal = {Physical Review D},
         year = 2013,
        month = aug,
       volume = {88},
       number = {4},
          eid = {043013},
        pages = {043013},
          doi = {10.1103/PhysRevD.88.043013},
archivePrefix = {arXiv},
       eprint = {1206.5032},
 primaryClass = {astro-ph.CO},
       adsurl = {https://ui.adsabs.harvard.edu/abs/2013PhRvD..88d3013Z}}

@ARTICLE{Komatsu_2011,
       author = {{Komatsu}, E. and {Smith}, K.~M. and {Dunkley}, J. and {Bennett}, C.~L. and {Gold}, B. and {Hinshaw}, G. and {Jarosik}, N. and {Larson}, D. and {Nolta}, M.~R. and {Page}, L. and {Spergel}, D.~N. and {Halpern}, M. and {Hill}, R.~S. and {Kogut}, A. and {Limon}, M. and {Meyer}, S.~S. and {Odegard}, N. and {Tucker}, G.~S. and {Weiland}, J.~L. and {Wollack}, E. and {Wright}, E.~L.},
        title = "{Seven-year Wilkinson Microwave Anisotropy Probe (WMAP) Observations: Cosmological Interpretation}",
      journal = {ApJS},
         year = 2011,
        month = feb,
       volume = {192},
       number = {2},
          eid = {18},
        pages = {18},
          doi = {10.1088/0067-0049/192/2/18},
archivePrefix = {arXiv},
       eprint = {1001.4538},
 primaryClass = {astro-ph.CO},
       adsurl = {https://ui.adsabs.harvard.edu/abs/2011ApJS..192...18K}}

@ARTICLE{Lesgourgues_2012,
       author = {{Lesgourgues}, Julien and {Pastor}, Sergio},
        title = "{Neutrino mass from Cosmology}",
      journal = {Advances in High Energy Physics},
         year = {2012},
         month = {dec},
        volume = {2012},
       number = {1},
        pages = {608515},
archivePrefix = {arXiv},
       eprint = {1212.6154},
 primaryClass = {hep-ph},
          doi = {https://doi.org/10.1155/2012/608515},
       adsurl = {https://ui.adsabs.harvard.edu/abs/2012arXiv1212.6154L}}

@ARTICLE{Akita_2020,
       author = {{Akita}, Kensuke and {Yamaguchi}, Masahide},
        title = "{A precision calculation of relic neutrino decoupling}",
      journal = {JCAP},
         year = 2020,
        month = aug,
       volume = {2020},
       number = {8},
          eid = {012},
        pages = {012},
          doi = {10.1088/1475-7516/2020/08/012},
archivePrefix = {arXiv},
       eprint = {2005.07047},
 primaryClass = {hep-ph},
       adsurl = {https://ui.adsabs.harvard.edu/abs/2020JCAP...08..012A}}

@ARTICLE{Froustey_2020,
       author = {{Froustey}, Julien and {Pitrou}, Cyril and {Volpe}, Maria Cristina},
        title = "{Neutrino decoupling including flavour oscillations and primordial nucleosynthesis}",
      journal = {JCAP},
         year = 2020,
        month = dec,
       volume = {2020},
       number = {12},
          eid = {015},
        pages = {015},
          doi = {10.1088/1475-7516/2020/12/015},
archivePrefix = {arXiv},
       eprint = {2008.01074},
 primaryClass = {hep-ph},
       adsurl = {https://ui.adsabs.harvard.edu/abs/2020JCAP...12..015F}}

@ARTICLE{Bennett_2021,
       author = {{Bennett}, Jack J. and {Buldgen}, Gilles and {de Salas}, Pablo F. and {Drewes}, Marco and {Gariazzo}, Stefano and {Pastor}, Sergio and {Wong}, Yvonne Y.~Y.},
        title = "{Towards a precision calculation of the effective number of neutrinos N$_{eff}$ in the Standard Model. Part II. Neutrino decoupling in the presence of flavour oscillations and finite-temperature QED}",
      journal = {JCAP},
         year = 2021,
        month = apr,
       volume = {2021},
       number = {4},
          eid = {073},
        pages = {073},
          doi = {10.1088/1475-7516/2021/04/073},
archivePrefix = {arXiv},
       eprint = {2012.02726},
 primaryClass = {hep-ph},
       adsurl = {https://ui.adsabs.harvard.edu/abs/2021JCAP...04..073B}}

@ARTICLE{Planck_2020,
	author = {{Planck Collaboration VI}},
	title = "{Planck 2018 results. VI. Cosmological parameters}",
	journal = {A\&A},
	year = 2020,
	month = sep,
	volume = {641},
	eid = {A6},
	pages = {A6},
	doi = {10.1051/0004-6361/201833910},
	archivePrefix = {arXiv},
	eprint = {1807.06209},
	primaryClass = {astro-ph.CO},
	adsurl = {https://ui.adsabs.harvard.edu/abs/2020A\&A...641A...6P}}

@ARTICLE{Tristram_2024,
       author = {{Tristram}, M. and {Banday}, A.~J. and {Douspis}, M. and {Garrido}, X. and {G{\'o}rski}, K.~M. and {Henrot-Versill{\'e}}, S. and {Hergt}, L.~T. and {Ili{\'c}}, S. and {Keskitalo}, R. and {Lagache}, G. and {Lawrence}, C.~R. and {Partridge}, B. and {Scott}, D.},
        title = "{Cosmological parameters derived from the final Planck data release (PR4)}",
      journal = {A\&A},
         year = 2024,
        month = feb,
       volume = {682},
          eid = {A37},
        pages = {A37},
          doi = {10.1051/0004-6361/202348015},
archivePrefix = {arXiv},
       eprint = {2309.10034},
 primaryClass = {astro-ph.CO},
       adsurl = {https://ui.adsabs.harvard.edu/abs/2024A\&A...682A..37T}}

@ARTICLE{Louis_2025,
       author = {{The Atacama Cosmology Telescope collaboration}},
        title = "{The Atacama Cosmology Telescope: DR6 power spectra, likelihoods and {\ensuremath{\Lambda}}CDM parameters}",
      journal = {JCAP},
         year = 2025,
        month = nov,
       volume = {2025},
       number = {11},
          eid = {062},
        pages = {062},
          doi = {10.1088/1475-7516/2025/11/062},
archivePrefix = {arXiv},
       eprint = {2503.14452},
 primaryClass = {astro-ph.CO},
       adsurl = {https://ui.adsabs.harvard.edu/abs/2025JCAP...11..062L}}

@ARTICLE{Calabrese_2025,
       author = {{Calabrese}, Erminia and {Hill}, J. Colin and {Jense}, Hidde T. and {La Posta}, Adrien and {Abril-Cabezas}, Irene and {Addison}, Graeme E. and {Ade}, Peter A.~R. and {Aiola}, Simone and {Alford}, Tommy and {Alonso}, David and {Amiri}, Mandana and {An}, Rui and {Atkins}, Zachary and {Austermann}, Jason E. and {Barbavara}, Eleonora and {Barbieri}, Nicola and {Battaglia}, Nicholas and {Battistelli}, Elia Stefano and {Beall}, James A. and {Bean}, Rachel and {Beheshti}, Ali and {Beringue}, Benjamin and {Bhandarkar}, Tanay and {Biermann}, Emily and {Bolliet}, Boris and {Bond}, J Richard and {Capalbo}, Valentina and {Carrero}, Felipe and {Chen}, Shi-Fan and {Chesmore}, Grace and {Cho}, Hsiao-mei and {Choi}, Steve K. and {Clark}, Susan E. and {Cothard}, Nicholas F. and {Coughlin}, Kevin and {Coulton}, William and {Crichton}, Devin and {Crowley}, Kevin T. and {Darwish}, Omar and {Devlin}, Mark J. and {Dicker}, Simon and {Duell}, Cody J. and {Duff}, Shannon M. and {Duivenvoorden}, Adriaan J. and {Dunkley}, Jo and {Dunner}, Rolando and {Embil Villagra}, Carmen and {Fankhanel}, Max and {Farren}, Gerrit S. and {Ferraro}, Simone and {Foster}, Allen and {Freundt}, Rodrigo and {Fuzia}, Brittany and {Gallardo}, Patricio A. and {Garrido}, Xavier and {Gerbino}, Martina and {Giardiello}, Serena and {Gill}, Ajay and {Givans}, Jahmour and {Gluscevic}, Vera and {Goldstein}, Samuel and {Golec}, Joseph E. and {Gong}, Yulin and {Guan}, Yilun and {Halpern}, Mark and {Harrison}, Ian and {Hasselfield}, Matthew and {He}, Adam and {Healy}, Erin and {Henderson}, Shawn and {Hensley}, Brandon and {Herv{\'\i}as-Caimapo}, Carlos and {Hilton}, Gene C. and {Hilton}, Matt and {Hincks}, Adam D. and {Hlo{\v{z}}ek}, Ren{\'e}e and {Ho}, Shuay-Pwu Patty and {Hood}, John and {Hornecker}, Erika and {Huber}, Zachary B. and {Hubmayr}, Johannes and {Huffenberger}, Kevin M. and {Hughes}, John P. and {Ikape}, Margaret and {Irwin}, Kent and {Isopi}, Giovanni and {Joshi}, Neha and {Keller}, Ben and {Kim}, Joshua and {Knowles}, Kenda and {Koopman}, Brian J. and {Kosowsky}, Arthur and {Kramer}, Darby and {Kusiak}, Aleksandra and {Lague}, Alex and {Lakey}, Victoria and {Lattanzi}, Massimiliano and {Lee}, Eunseong and {Li}, Yaqiong and {Li}, Zack and {Limon}, Michele and {Lokken}, Martine and {Louis}, Thibaut and {Lungu}, Marius and {MacCrann}, Niall and {MacInnis}, Amanda and {Madhavacheril}, Mathew S. and {Maldonado}, Diego and {Maldonado}, Felipe and {Mallaby-Kay}, Maya and {Marques}, Gabriela A. and {van Marrewijk}, Joshiwa and {McCarthy}, Fiona and {McMahon}, Jeff and {Mehta}, Yogesh and {Menanteau}, Felipe and {Moodley}, Kavilan and {Morris}, Thomas W. and {Mroczkowski}, Tony and {Naess}, Sigurd and {Namikawa}, Toshiya and {Nati}, Federico and {Nerval}, Simran K. and {Newburgh}, Laura and {Nicola}, Andrina and {Niemack}, Michael D. and {Nolta}, Michael R. and {Orlowski-Scherer}, John and {Pagano}, Luca and {Page}, Lyman A. and {Pandey}, Shivam and {Partridge}, Bruce and {Perez Sarmiento}, Karen and {Prince}, Heather and {Puddu}, Roberto and {Qu}, Frank J. and {Ragavan}, Damien C. and {Ried Guachalla}, Bernardita and {Rogers}, Keir K. and {Rojas}, Felipe and {Sakuma}, Tai and {Schaan}, Emmanuel and {Schmitt}, Benjamin L. and {Sehgal}, Neelima and {Shaikh}, Shabbir and {Sherwin}, Blake D. and {Sierra}, Carlos and {Sievers}, Jon and {Sif{\'o}n}, Crist{\'o}bal and {Simon}, Sara and {Sonka}, Rita and {Spergel}, David N. and {Staggs}, Suzanne T. and {Storer}, Emilie and {Surrao}, Kristen and {Switzer}, Eric R. and {Tampier}, Niklas and {Thiele}, Leander and {Thornton}, Robert and {Trac}, Hy and {Tucker}, Carole and {Ullom}, Joel and {Vale}, Leila R. and {Van Engelen}, Alexander and {Van Lanen}, Jeff and {Vargas}, Cristian and {Vavagiakis}, Eve M. and {Wagoner}, Kasey and {Wang}, Yuhan and {Wenzl}, Lukas and {Wollack}, Edward J. and {Zheng}, Kaiwen},
        title = "{The Atacama Cosmology Telescope: DR6 Constraints on Extended Cosmological Models}",
      journal = {JCAP},
         year = 2025,
	   month = nov,
	  volume = {2025},
	  number = {11},
	     eid = {063},
	   pages = {063},
          doi = {10.1088/1475-7516/2025/11/063},
archivePrefix = {arXiv},
       eprint = {2503.14454},
 primaryClass = {astro-ph.CO},
       adsurl = {https://ui.adsabs.harvard.edu/abs/2025JCAP...11..063C}}

@ARTICLE{SPT_2026,
       author = {{SPT-3G Collaboration}},
        title = "{SPT-3G D1: CMB temperature and polarization power spectra and cosmology from 2019 and 2020 observations of the SPT-3G main field}",
      journal = {Physical Review D},
         year = 2026,
        month = apr,
       volume = {113},
       number = {8},
          eid = {083504},
        pages = {083504},
          doi = {10.1103/7wt3-9v2y},
archivePrefix = {arXiv},
       eprint = {2506.20707},
 primaryClass = {astro-ph.CO},
       adsurl = {https://ui.adsabs.harvard.edu/abs/2026PhRvD.113h3504C}}

@ARTICLE{Najera_2026,
       author = {{N{\'a}jera}, Jos{\'e} Antonio and {Banik}, Indranil and {Desmond}, Harry and {Kalaitzidis}, Vasileios},
        title = "{Background solutions to the Hubble tension in $f(Q)$ gravity and consistency with BAO measurements}",
      journal = {Galaxies},
         year = 2026,
        month = mar,
       volume = {14},
       number = {2},
          eid = {19},
        pages = {19},
archivePrefix = {arXiv},
       eprint = {2510.20964},
 primaryClass = {astro-ph.CO},
          doi = {10.3390/galaxies14020019},
       adsurl = {https://ui.adsabs.harvard.edu/abs/2026Galax..14...19N}}

@ARTICLE{Desmond_2026,
       author = {{Desmond}, Harry and {Stiskalek}, Richard and {N{\'a}jera}, Jos{\'e} Antonio and {Banik}, Indranil},
        title = "{The subtle statistics of the distance ladder: On the distance prior and selection effects}",
      journal = {MNRAS},
         year = 2026,
        month = aug,
       volume = {550},
       number = {2},
          eid = {stag1144},
        pages = {stag1144},
archivePrefix = {arXiv},
       eprint = {2511.03394},
 primaryClass = {astro-ph.CO},
          doi = {10.1093/mnras/stag1144},
       adsurl = {https://ui.adsabs.harvard.edu/abs/2026MNRAS.550f1144D}}

@ARTICLE{Stiskalek_2026_MW,
       author = {{Stiskalek}, Richard and {Riess}, Adam and {Desmond}, Harry and {Lavaux}, Guilhem and {Scolnic}, Dan},
        title = "{Forward-modelling Milky Way Cepheids: selection effects and physical priors in the Gaia-HST calibration}",
      journal = {MNRAS},
         year = 2026,
        month = sep,
          eid = {arXiv:2603.09880},
       volume = {in press},
          doi = {10.48550/arXiv.2603.09880},
archivePrefix = {arXiv},
       eprint = {2603.09880},
 primaryClass = {astro-ph.GA},
       adsurl = {https://ui.adsabs.harvard.edu/abs/2026arXiv260309880S}}

@ARTICLE{Percival_2002,
       author = {{Percival}, Will J. and {Sutherland}, Will and {Peacock}, John A. and {Baugh}, Carlton M. and {Bland-Hawthorn}, Joss and {Bridges}, Terry and {Cannon}, Russell and {Cole}, Shaun and {Colless}, Matthew and {Collins}, Chris and {Couch}, Warrick and {Dalton}, Gavin and {De Propris}, Roberto and {Driver}, Simon P. and {Efstathiou}, George and {Ellis}, Richard S. and {Frenk}, Carlos S. and {Glazebrook}, Karl and {Jackson}, Carole and {Lahav}, Ofer and {Lewis}, Ian and {Lumsden}, Stuart and {Maddox}, Steve and {Moody}, Stephen and {Norberg}, Peder and {Peterson}, Bruce A. and {Taylor}, Keith},
        title = "{Parameter constraints for flat cosmologies from cosmic microwave background and 2dFGRS power spectra}",
      journal = {MNRAS},
         year = 2002,
        month = dec,
       volume = {337},
       number = {3},
        pages = {1068-1080},
          doi = {10.1046/j.1365-8711.2002.06001.x},
archivePrefix = {arXiv},
       eprint = {astro-ph/0206256},
 primaryClass = {astro-ph},
       adsurl = {https://ui.adsabs.harvard.edu/abs/2002MNRAS.337.1068P}}

@ARTICLE{Kable_2019,
       author = {{Kable}, Joshua A. and {Addison}, Graeme E. and {Bennett}, Charles L.},
        title = "{Quantifying the CMB Degeneracy between the Matter Density and Hubble Constant in Current Experiments}",
      journal = {ApJ},
         year = 2019,
        month = jan,
       volume = {871},
       number = {1},
          eid = {77},
        pages = {77},
          doi = {10.3847/1538-4357/aaf56d},
archivePrefix = {arXiv},
       eprint = {1809.03983},
 primaryClass = {astro-ph.CO},
       adsurl = {https://ui.adsabs.harvard.edu/abs/2019ApJ...871...77K}}

@ARTICLE{Poulin_2023,
       author = {{Poulin}, Vivian and {Smith}, Tristan L. and {Karwal}, Tanvi},
        title = "{The Ups and Downs of Early Dark Energy solutions to the Hubble tension: A review of models, hints and constraints circa 2023}",
      journal = {Physics of the Dark Universe},
         year = 2023,
        month = dec,
       volume = {42},
          eid = {101348},
        pages = {101348},
          doi = {10.1016/j.dark.2023.101348},
archivePrefix = {arXiv},
       eprint = {2302.09032},
 primaryClass = {astro-ph.CO},
       adsurl = {https://ui.adsabs.harvard.edu/abs/2023PDU....4201348P}}

@ARTICLE{Keenan_2013,
	author = {{Keenan}, R.~C. and {Barger}, A.~J. and {Cowie}, L.~L.},
	title = "{Evidence for a \raisebox{-0.5ex}\textasciitilde300 Megaparsec Scale Under-density in the Local Galaxy Distribution}",
	journal = {ApJ},
	year = 2013,
	month = sep,
	volume = {775},
	number = {1},
	eid = {62},
	pages = {62},
	doi = {10.1088/0004-637X/775/1/62},
	archivePrefix = {arXiv},
	eprint = {1304.2884},
	primaryClass = {astro-ph.CO},
	adsurl = {https://ui.adsabs.harvard.edu/abs/2013ApJ...775...62K}}

@ARTICLE{Futamase_2026,
       author = {{Futamase}, Toshifumi and {Kojima}, Reiki and {Tomonaga}, Masanori},
        title = "{Reconstructing a large-scale matter-density contrast profile to reconcile Pantheon+ supernovae with DESI DR2 BAO in an inhomogeneous universe}",
      journal = {Physical Review D},
         year = 2026,
        month = may,
       volume = {113},
        issue = {10},
          eid = {103515},
        pages = {103515},
 archivePrefix = {arXiv},
       eprint = {2604.05466},
 primaryClass = {astro-ph.CO},
         doi = {10.1103/ldjj-rxc6},
       adsurl = {https://ui.adsabs.harvard.edu/abs/2026PhRvD.113j3515F}}

@ARTICLE{Secrest_2021,
       author = {{Secrest}, Nathan J. and {von Hausegger}, Sebastian and {Rameez}, Mohamed and {Mohayaee}, Roya and {Sarkar}, Subir and {Colin}, Jacques},
        title = "{A Test of the Cosmological Principle with Quasars}",
      journal = {ApJL},
         year = 2021,
        month = feb,
       volume = {908},
       number = {2},
          eid = {L51},
        pages = {L51},
          doi = {10.3847/2041-8213/abdd40},
archivePrefix = {arXiv},
       eprint = {2009.14826},
 primaryClass = {astro-ph.CO},
       adsurl = {https://ui.adsabs.harvard.edu/abs/2021ApJ...908L..51S}}

@ARTICLE{Secrest_2022,
       author = {{Secrest}, Nathan J. and {von Hausegger}, Sebastian and {Rameez}, Mohamed and {Mohayaee}, Roya and {Sarkar}, Subir},
        title = "{A Challenge to the Standard Cosmological Model}",
      journal = {ApJL},
         year = 2022,
        month = oct,
       volume = {937},
       number = {2},
          eid = {L31},
        pages = {L31},
          doi = {10.3847/2041-8213/ac88c0},
archivePrefix = {arXiv},
       eprint = {2206.05624},
 primaryClass = {astro-ph.CO},
       adsurl = {https://ui.adsabs.harvard.edu/abs/2022ApJ...937L..31S}}

@ARTICLE{Dam_2023,
       author = {{Dam}, Lawrence and {Lewis}, Geraint F. and {Brewer}, Brendon J.},
        title = "{Testing the cosmological principle with CatWISE quasars: a bayesian analysis of the number-count dipole}",
      journal = {MNRAS},
         year = 2023,
        month = oct,
       volume = {525},
       number = {1},
        pages = {231-245},
          doi = {10.1093/mnras/stad2322},
archivePrefix = {arXiv},
       eprint = {2212.07733},
 primaryClass = {astro-ph.CO},
       adsurl = {https://ui.adsabs.harvard.edu/abs/2023MNRAS.525..231D}}

@ARTICLE{Singal_2023,
       author = {{Singal}, Ashok K.},
        title = "{Discordance of dipole asymmetries seen in recent large radio surveys with the cosmological principle}",
      journal = {MNRAS},
         year = 2023,
        month = sep,
       volume = {524},
       number = {3},
        pages = {3636-3646},
          doi = {10.1093/mnras/stad2161},
archivePrefix = {arXiv},
       eprint = {2303.05141},
 primaryClass = {astro-ph.CO},
       adsurl = {https://ui.adsabs.harvard.edu/abs/2023MNRAS.524.3636S}}

@ARTICLE{Wagenveld_2024,
       author = {{Wagenveld}, J.~D. and {Kl{\"o}ckner}, H. -R. and {Gupta}, N. and {Sekhar}, S. and {Jagannathan}, P. and {Deka}, P.~P. and {Jose}, J. and {Balashev}, S.~A. and {Borgaonkar}, D. and {Chatterjee}, A. and {Combes}, F. and {Emig}, K.~L. and {Gaunekar}, A.~N. and {Hilton}, M. and {J{\'o}zsa}, G.~I.~G. and {Klutse}, D.~Y. and {Knowles}, K. and {Krogager}, J. -K. and {Momjian}, E. and {Muller}, S. and {Sikhosana}, S.~P.},
        title = "{The MeerKAT Absorption Line Survey Data Release 2: Wideband continuum catalogues and a measurement of the cosmic radio dipole}",
      journal = {A\&A},
         year = 2024,
        month = oct,
       volume = {690},
          eid = {A163},
        pages = {A163},
          doi = {10.1051/0004-6361/202450291},
archivePrefix = {arXiv},
       eprint = {2408.16619},
 primaryClass = {astro-ph.CO},
       adsurl = {https://ui.adsabs.harvard.edu/abs/2024A\&A...690A.163W}}

@ARTICLE{Wagenveld_2025,
       author = {{Wagenveld}, J.~D. and {von Hausegger}, S. and {Kl{\"o}ckner}, H.-R. and {Schwarz}, D.~J.},
        title = "{The kinematic contribution to the cosmic number count dipole}",
      journal = {A\&A},
         year = 2025,
        month = may,
       volume = {697},
          eid = {A112},
        pages = {A112},
          doi = {10.1051/0004-6361/202453397},
archivePrefix = {arXiv},
       eprint = {2503.02470},
 primaryClass = {astro-ph.CO},
       adsurl = {https://ui.adsabs.harvard.edu/abs/2025A\&A...697A.112W}}

@ARTICLE{Oayda_2026,
       author = {{Oayda}, Oliver T. and {Lewis}, Geraint F.},
        title = "{Wising up to CatWISE: using simulation-based inference to interpret the ecliptic bias and confirm the cosmic dipole excess}",
      journal = {MNRAS},
         year = 2026,
        month = mar,
       volume = {546},
       number = {4},
          eid = {stag248},
        pages = {stag248},
          doi = {10.1093/mnras/stag248},
archivePrefix = {arXiv},
       eprint = {2602.05070},
 primaryClass = {astro-ph.CO},
       adsurl = {https://ui.adsabs.harvard.edu/abs/2026MNRAS.546ag248O}}

@ARTICLE{Bashir_2026,
       author = {{Bashir}, Masroor and {Chingangbam}, Pravabati and {Appleby}, Stephen},
        title = "{The CatWISE2020 Quasar Dipole: A Reassessment of the Cosmic Dipole Anomaly}",
      journal = {ApJ},
         year = 2026,
        month = jun,
       volume = {1003},
       number = {2},
          eid = {162},
        pages = {162},
          doi = {10.3847/1538-4357/ae6588},
archivePrefix = {arXiv},
       eprint = {2511.00822},
 primaryClass = {astro-ph.CO},
       adsurl = {https://ui.adsabs.harvard.edu/abs/2026ApJ..1003..162B}}

@ARTICLE{Gandhi_2026,
       author = {{Gandhi}, Akash},
        title = "{Intrinsic dipole anisotropies and the Hubble tension: a unified theoretical framework for correlated large-scale modes}",
      journal = {MNRAS},
         year = 2026,
        month = may,
       volume = {548},
       number = {2},
          eid = {stag582},
        pages = {stag582},
          doi = {10.1093/mnras/stag582},
       adsurl = {https://ui.adsabs.harvard.edu/abs/2026MNRAS.548ag582G}}

@ARTICLE{Haslbauer_2020,
	author = {{Haslbauer}, Moritz and {Banik}, Indranil and {Kroupa}, Pavel},
	title = "{The KBC void and Hubble tension contradict {\ensuremath{\Lambda}}CDM on a Gpc scale - Milgromian dynamics as a possible solution}",
	journal = {MNRAS},
	year = 2020,
	month = oct,
	volume = {499},
	number = {2},
	pages = {2845-2883},
	doi = {10.1093/mnras/staa2348},
	archivePrefix = {arXiv},
	eprint = {2009.11292},
	primaryClass = {astro-ph.CO},
	adsurl = {https://ui.adsabs.harvard.edu/abs/2020MNRAS.499.2845H}}

@ARTICLE{Wu_2017,
       author = {{Wu}, Hao-Yi and {Huterer}, Dragan},
        title = "{Sample variance in the local measurements of the Hubble constant}",
      journal = {MNRAS},
         year = 2017,
        month = nov,
       volume = {471},
       number = {4},
        pages = {4946-4955},
          doi = {10.1093/mnras/stx1967},
archivePrefix = {arXiv},
       eprint = {1706.09723},
 primaryClass = {astro-ph.CO},
       adsurl = {https://ui.adsabs.harvard.edu/abs/2017MNRAS.471.4946W}}

@ARTICLE{Camarena_2018,
       author = {{Camarena}, David and {Marra}, Valerio},
        title = "{Impact of the cosmic variance on H$_{0}$ on cosmological analyses}",
      journal = {Physical Review D},
         year = 2018,
        month = jul,
       volume = {98},
       number = {2},
          eid = {023537},
        pages = {023537},
          doi = {10.1103/PhysRevD.98.023537},
archivePrefix = {arXiv},
       eprint = {1805.09900},
 primaryClass = {astro-ph.CO},
       adsurl = {https://ui.adsabs.harvard.edu/abs/2018PhRvD..98b3537C}}

@ARTICLE{Scolnic_2023,
       author = {{Scolnic}, Daniel and {Vincenzi}, Maria},
        title = "{The Role of Type Ia Supernovae in Constraining the Hubble Constant}",
      journal = {ArXiv e-prints},
         year = 2023,
        month = nov,
       volume = {Arxiv},
          doi = {10.48550/arXiv.2311.16830},
archivePrefix = {arXiv},
       eprint = {2311.16830},
 primaryClass = {astro-ph.CO},
       adsurl = {https://ui.adsabs.harvard.edu/abs/2023arXiv231116830S}}

@ARTICLE{Sunyaev_1980,
       author = {{Sunyaev}, R.~A. and {Zeldovich}, Ia. B.},
        title = "{Microwave background radiation as a probe of the contemporary structure and history of the universe}",
      journal = {ARA\&A},
         year = 1980,
        month = jan,
       volume = {18},
        pages = {537-560},
          doi = {10.1146/annurev.aa.18.090180.002541},
       adsurl = {https://ui.adsabs.harvard.edu/abs/1980ARA\&A..18..537S}}

@ARTICLE{Fixsen_2009,
       author = {{Fixsen}, D.~J.},
        title = "{The Temperature of the Cosmic Microwave Background}",
      journal = {ApJ},
         year = 2009,
        month = dec,
       volume = {707},
       number = {2},
        pages = {916-920},
          doi = {10.1088/0004-637X/707/2/916},
archivePrefix = {arXiv},
       eprint = {0911.1955},
 primaryClass = {astro-ph.CO},
       adsurl = {https://ui.adsabs.harvard.edu/abs/2009ApJ...707..916F}}

@ARTICLE{Yoo_2019,
       author = {{Yoo}, Jaiyul and {Mitsou}, Ermis and {Dirian}, Yves and {Durrer}, Ruth},
        title = "{Background photon temperature T {\textasciimacron}: A new cosmological Parameter?}",
      journal = {Physical Review D},
         year = 2019,
        month = sep,
       volume = {100},
       number = {6},
          eid = {063510},
        pages = {063510},
          doi = {10.1103/PhysRevD.100.063510},
archivePrefix = {arXiv},
       eprint = {1905.09288},
 primaryClass = {astro-ph.CO},
       adsurl = {https://ui.adsabs.harvard.edu/abs/2019PhRvD.100f3510Y}}

@ARTICLE{Ivanov_2020_TCMB,
       author = {{Ivanov}, Mikhail M. and {Ali-Ha{\"\i}moud}, Yacine and {Lesgourgues}, Julien},
        title = "{H$_{0}$ tension or T$_{0}$ tension?}",
      journal = {Physical Review D},
         year = 2020,
        month = sep,
       volume = {102},
       number = {6},
          eid = {063515},
        pages = {063515},
          doi = {10.1103/PhysRevD.102.063515},
archivePrefix = {arXiv},
       eprint = {2005.10656},
 primaryClass = {astro-ph.CO},
       adsurl = {https://ui.adsabs.harvard.edu/abs/2020PhRvD.102f3515I}}

@ARTICLE{Ding_2020,
       author = {{Ding}, QianHang and {Nakama}, Tomohiro and {Wang}, Yi},
        title = "{A gigaparsec-scale local void and the Hubble tension}",
      journal = {Science China Physics, Mechanics, and Astronomy},
         year = 2020,
        month = may,
       volume = {63},
       number = {9},
          eid = {290403},
        pages = {290403},
          doi = {10.1007/s11433-020-1531-0},
archivePrefix = {arXiv},
       eprint = {1912.12600},
 primaryClass = {astro-ph.CO},
       adsurl = {https://ui.adsabs.harvard.edu/abs/2020SCPMA..6390403D}}

@ARTICLE{Cai_2025,
       author = {{Cai}, Tingqi and {Ding}, Qianhang and {Wang}, Yi},
        title = "{Reconciling cosmic dipolar tensions with a gigaparsec void}",
      journal = {Physical Review D},
         year = 2025,
        month = may,
       volume = {111},
       number = {10},
          eid = {103502},
        pages = {103502},
          doi = {10.1103/PhysRevD.111.103502},
archivePrefix = {arXiv},
       eprint = {2211.06857},
 primaryClass = {astro-ph.CO},
       adsurl = {https://ui.adsabs.harvard.edu/abs/2025PhRvD.111j3502C}}

@ARTICLE{Krishnan_2020,
       author = {{Krishnan}, C. and {Colg{\'a}in}, E. {\'O}. and {Ruchika}, Sen, A.~A. and {Sheikh-Jabbari}, M.~M. and {Yang}, T.},
        title = "{Is there an early Universe solution to Hubble tension?}",
      journal = {Physical Review D},
         year = 2020,
        month = nov,
       volume = {102},
       number = {10},
          eid = {103525},
        pages = {103525},
          doi = {10.1103/PhysRevD.102.103525},
archivePrefix = {arXiv},
       eprint = {2002.06044},
 primaryClass = {astro-ph.CO},
       adsurl = {https://ui.adsabs.harvard.edu/abs/2020PhRvD.102j3525K}}

@ARTICLE{Krishnan_2021,
       author = {{Krishnan}, C. and {{\'O} Colg{\'a}in}, E. and {Sheikh-Jabbari}, M.~M. and {Yang}, Tao},
        title = "{Running Hubble tension and a H0 diagnostic}",
      journal = {Physical Review D},
         year = 2021,
        month = may,
       volume = {103},
       number = {10},
          eid = {103509},
        pages = {103509},
          doi = {10.1103/PhysRevD.103.103509},
archivePrefix = {arXiv},
       eprint = {2011.02858},
 primaryClass = {astro-ph.CO},
       adsurl = {https://ui.adsabs.harvard.edu/abs/2021PhRvD.103j3509K}}

@ARTICLE{Maddox_1990,
       author = {{Maddox}, S.~J. and {Efstathiou}, G. and {Sutherland}, W.~J. and {Loveday}, J.},
        title = "{Galaxy correlations on large scales}",
      journal = {MNRAS},
         year = 1990,
        month = jan,
       volume = {242},
        pages = {43-47},
          doi = {10.1093/mnras/242.1.43P},
       adsurl = {https://ui.adsabs.harvard.edu/abs/1990MNRAS.242P..43M}}

@INPROCEEDINGS{Shanks_1990,
       author = {{Shanks}, T.},
        title = "{Galaxy Count Models and the Extragalactic Background Light}",
    booktitle = {The Galactic and Extragalactic Background Radiation},
         year = 1990,
	series = {International Astronomical Union Symposium no. 139},
       editor = {{Bowyer}, Stuart and {Leinert}, Christoph},
    publisher = {Springer Dordrecht},
     location = {Dordrecht},
       volume = {139},
        month = aug,
        pages = {269-281},
       adsurl = {https://ui.adsabs.harvard.edu/abs/1990IAUS..139..269S}}

@ARTICLE{Huang_1997,
       author = {{Huang}, J. -S. and {Cowie}, L.~L. and {Gardner}, J.~P. and {Hu}, E.~M. and {Songaila}, A. and {Wainscoat} and {R.~J.}},
        title = "{The Hawaii K-Band Galaxy Survey. II. Bright K-Band Imaging}",
      journal = {ApJ},
         year = 1997,
        month = feb,
       volume = {476},
       number = {1},
        pages = {12-21},
          doi = {10.1086/303598},
archivePrefix = {arXiv},
       eprint = {astro-ph/9610084},
 primaryClass = {astro-ph},
       adsurl = {https://ui.adsabs.harvard.edu/abs/1997ApJ...476...12H}}

@ARTICLE{Busswell_2004,
       author = {{Busswell}, G.~S. and {Shanks}, T. and {Frith}, W.~J. and {Outram}, P.~J. and {Metcalfe}, N. and {Fong}, R.},
        title = "{The local hole in the galaxy distribution: new optical evidence}",
      journal = {MNRAS},
         year = 2004,
        month = nov,
       volume = {354},
       number = {4},
        pages = {991-1004},
          doi = {10.1111/j.1365-2966.2004.08217.x},
archivePrefix = {arXiv},
       eprint = {astro-ph/0302330},
 primaryClass = {astro-ph},
       adsurl = {https://ui.adsabs.harvard.edu/abs/2004MNRAS.354..991B}}

@ARTICLE{Frith_2003,
       author = {{Frith}, W.~J. and {Busswell}, G.~S. and {Fong}, R. and {Metcalfe}, N. and {Shanks}, T.},
        title = "{The local hole in the galaxy distribution: evidence from 2MASS}",
      journal = {MNRAS},
         year = 2003,
        month = nov,
       volume = {345},
       number = {3},
        pages = {1049-1056},
          doi = {10.1046/j.1365-8711.2003.07027.x},
archivePrefix = {arXiv},
       eprint = {astro-ph/0302331},
 primaryClass = {astro-ph},
       adsurl = {https://ui.adsabs.harvard.edu/abs/2003MNRAS.345.1049F}}

@ARTICLE{Frith_2005,
       author = {{Frith}, W.~J. and {Shanks}, T. and {Outram}, P.~J.},
        title = "{2MASS constraints on the local large-scale structure: a challenge to {\ensuremath{\Lambda}}CDM?}",
      journal = {MNRAS},
         year = 2005,
        month = aug,
       volume = {361},
       number = {2},
        pages = {701-709},
          doi = {10.1111/j.1365-2966.2005.09200.x},
archivePrefix = {arXiv},
       eprint = {astro-ph/0411204},
 primaryClass = {astro-ph},
       adsurl = {https://ui.adsabs.harvard.edu/abs/2005MNRAS.361..701F}}

@ARTICLE{Frith_2006,
       author = {{Frith}, W.~J. and {Metcalfe}, N. and {Shanks}, T.},
        title = "{New H-band galaxy number counts: a large local hole in the galaxy distribution}",
      journal = {MNRAS},
         year = 2006,
        month = oct,
       volume = {371},
       number = {4},
        pages = {1601-1609},
          doi = {10.1111/j.1365-2966.2006.10736.x},
archivePrefix = {arXiv},
       eprint = {astro-ph/0509875},
 primaryClass = {astro-ph},
       adsurl = {https://ui.adsabs.harvard.edu/abs/2006MNRAS.371.1601F}}

@ARTICLE{Whitbourn_2014,
       author = {{Whitbourn}, J.~R. and {Shanks}, T.},
        title = "{The local hole revealed by galaxy counts and redshifts}",
      journal = {MNRAS},
         year = 2014,
        month = jan,
       volume = {437},
       number = {3},
        pages = {2146-2162},
          doi = {10.1093/mnras/stt2024},
archivePrefix = {arXiv},
       eprint = {1307.4405},
 primaryClass = {astro-ph.CO},
       adsurl = {https://ui.adsabs.harvard.edu/abs/2014MNRAS.437.2146W}}

@ARTICLE{Whitbourn_2016,
       author = {{Whitbourn}, J.~R. and {Shanks}, T.},
        title = "{The galaxy luminosity function and the Local Hole}",
      journal = {MNRAS},
         year = 2016,
        month = jun,
       volume = {459},
       number = {1},
        pages = {496-507},
          doi = {10.1093/mnras/stw555},
archivePrefix = {arXiv},
       eprint = {1603.02322},
 primaryClass = {astro-ph.CO},
       adsurl = {https://ui.adsabs.harvard.edu/abs/2016MNRAS.459..496W}}

@ARTICLE{Wong_2022,
       author = {{Wong}, Jonathan H.~W. and {Shanks}, T. and {Metcalfe}, N. and {Whitbourn}, J.~R.},
        title = "{The local hole: a galaxy underdensity covering 90 per cent of sky to {\ensuremath{\approx}}200 Mpc}",
      journal = {MNRAS},
         year = 2022,
        month = apr,
       volume = {511},
       number = {4},
        pages = {5742-5755},
          doi = {10.1093/mnras/stac396},
archivePrefix = {arXiv},
       eprint = {2107.08505},
 primaryClass = {astro-ph.CO},
       adsurl = {https://ui.adsabs.harvard.edu/abs/2022MNRAS.511.5742W}}

@ARTICLE{Rubart_2013,
	author = {{Rubart}, M. and {Schwarz}, D.~J.},
	title = "{Cosmic radio dipole from NVSS and WENSS}",
	journal = {A\&A},
	year = 2013,
	month = jul,
	volume = {555},
	eid = {A117},
	pages = {A117},
	doi = {10.1051/0004-6361/201321215},
	archivePrefix = {arXiv},
	eprint = {1301.5559},
	primaryClass = {astro-ph.CO},
	adsurl = {https://ui.adsabs.harvard.edu/abs/2013A\&A...555A.117R}}

@ARTICLE{Rubart_2014,
       author = {{Rubart}, Matthias and {Bacon}, David and {Schwarz}, Dominik J.},
        title = "{Impact of local structure on the cosmic radio dipole}",
      journal = {A\&A},
         year = 2014,
        month = may,
       volume = {565},
          eid = {A111},
        pages = {A111},
          doi = {10.1051/0004-6361/201423583},
archivePrefix = {arXiv},
       eprint = {1402.0376},
 primaryClass = {astro-ph.CO},
       adsurl = {https://ui.adsabs.harvard.edu/abs/2014A\&A...565A.111R}}

@ARTICLE{Bohringer_2015,
	author = {{B{\"o}hringer}, Hans and {Chon}, Gayoung and {Bristow}, Martyn and {Collins}, Chris A.},
	title = "{The extended ROSAT-ESO Flux-Limited X-ray Galaxy Cluster Survey (REFLEX II). V. Exploring a local underdensity in the southern sky}",
	journal = {A\&A},
	year = 2015,
	month = feb,
	volume = {574},
	eid = {A26},
	pages = {A26},
	doi = {10.1051/0004-6361/201424817},
	archivePrefix = {arXiv},
	eprint = {1410.2172},
	primaryClass = {astro-ph.CO},
	adsurl = {https://ui.adsabs.harvard.edu/abs/2015A\&A...574A..26B}}

@ARTICLE{Bohringer_2020,
	author = {{B{\"o}hringer}, Hans and {Chon}, Gayoung and {Collins}, Chris A.},
	title = "{Observational evidence for a local underdensity in the Universe and its effect on the measurement of the Hubble constant}",
	journal = {A\&A},
	year = 2020,
	month = jan,
	volume = {633},
	eid = {A19},
	pages = {A19},
	doi = {10.1051/0004-6361/201936400},
	archivePrefix = {arXiv},
	eprint = {1907.12402},
	primaryClass = {astro-ph.CO},
	adsurl = {https://ui.adsabs.harvard.edu/abs/2020A\&A...633A..19B}}

@ARTICLE{Tully_2023_CF4,
       author = {{Tully}, R. Brent and {Kourkchi}, Ehsan and {Courtois}, H{\'e}l{\`e}ne M. and {Anand}, Gagandeep S. and {Blakeslee}, John P. and {Brout}, Dillon and {Jaeger}, Thomas de and {Dupuy}, Alexandra and {Guinet}, Daniel and {Howlett}, Cullan and {Jensen}, Joseph B. and {Pomar{\`e}de}, Daniel and {Rizzi}, Luca and {Rubin}, David and {Said}, Khaled and {Scolnic}, Daniel and {Stahl}, Benjamin E.},
        title = "{Cosmicflows-4}",
      journal = {ApJ},
         year = 2023,
        month = feb,
       volume = {944},
       number = {1},
          eid = {94},
        pages = {94},
          doi = {10.3847/1538-4357/ac94d8},
archivePrefix = {arXiv},
       eprint = {2209.11238},
 primaryClass = {astro-ph.CO},
       adsurl = {https://ui.adsabs.harvard.edu/abs/2023ApJ...944...94T}}

@ARTICLE{Mazurenko_2025,
       author = {{Mazurenko}, Sergij and {Banik}, Indranil and {Kroupa}, Pavel},
        title = "{The redshift dependence of the inferred H$_{0}$ in a local void solution to the Hubble tension}",
      journal = {MNRAS},
         year = 2025,
        month = feb,
       volume = {536},
       number = {4},
        pages = {3232-3241},
          doi = {10.1093/mnras/stae2758},
archivePrefix = {arXiv},
       eprint = {2311.17988},
 primaryClass = {astro-ph.CO},
       adsurl = {https://ui.adsabs.harvard.edu/abs/2025MNRAS.536.3232M}}

@ARTICLE{Banik_2025_BAO,
       author = {{Banik}, Indranil and {Kalaitzidis}, Vasileios},
        title = "{Testing the local void hypothesis using baryon acoustic oscillation measurements over the last 20 yr}",
      journal = {MNRAS},
         year = 2025,
        month = jun,
       volume = {540},
       number = {1},
        pages = {545-561},
          doi = {10.1093/mnras/staf781},
archivePrefix = {arXiv},
       eprint = {2501.17934},
 primaryClass = {astro-ph.CO},
       adsurl = {https://ui.adsabs.harvard.edu/abs/2025MNRAS.540..545B}}

@ARTICLE{Banik_2026_void,
       author = {{Banik}, Indranil and {Desmond}, Harry and {Kalaitzidis}, Vasileios and {Mazurenko}, Sergij},
        title = "{The local void model for the Hubble and BAO tensions}",
      journal = {ArXiv e-prints},
         year = 2026,
        month = feb,
          eid = {arXiv:2602.03928},
       volume = {Arxiv},
          doi = {10.48550/arXiv.2602.03928},
archivePrefix = {arXiv},
       eprint = {2602.03928},
 primaryClass = {astro-ph.CO},
       adsurl = {https://ui.adsabs.harvard.edu/abs/2026arXiv260203928B}}

@ARTICLE{Tao_2026,
       author = {{Tao}, Bojun and {Zhang}, Hong-Xin and {Wang}, Wenting and {Wang}, Enci and {Chen}, Guangwen and {Wang}, Huiyuan and {Chen}, Lijun and {Chen}, Qian-Hui and {Huang}, Song and {Kong}, Xu and {Rong}, Yu},
        title = "{The First Systematic Survey of Stellar Halos in High-inclination Galaxies Reveals Unusually Quiescent Merger Histories of Nearby Galaxies}",
      journal = {ApJS},
         year = 2026,
        month = may,
       volume = {284},
       number = {1},
          eid = {31},
        pages = {31},
          doi = {10.3847/1538-4365/ae48ec},
archivePrefix = {arXiv},
       eprint = {2602.20258},
 primaryClass = {astro-ph.GA},
       adsurl = {https://ui.adsabs.harvard.edu/abs/2026ApJS..284...31T}}

@ARTICLE{Stiskalek_2025_void,
       author = {{Stiskalek}, Richard and {Desmond}, Harry and {Banik}, Indranil},
        title = "{Testing the local supervoid solution to the Hubble tension with direct distance tracers}",
      journal = {MNRAS},
         year = 2025,
        month = oct,
       volume = {543},
       number = {2},
        pages = {1556-1573},
          doi = {10.1093/mnras/staf1571},
archivePrefix = {arXiv},
       eprint = {2506.10518},
 primaryClass = {astro-ph.CO},
       adsurl = {https://ui.adsabs.harvard.edu/abs/2025MNRAS.543.1556S}}

@ARTICLE{Stiskalek_2026_Cepheids,
       author = {{Stiskalek}, Richard and {Desmond}, Harry and {Tsaprazi}, Eleni and {Heavens}, Alan and {Lavaux}, Guilhem and {McAlpine}, Stuart and {Jasche}, Jens},
        title = "{1.8 per cent measurement of H$_{0}$ from Cepheids alone}",
      journal = {MNRAS},
         year = 2026,
        month = feb,
       volume = {546},
       number = {2},
          eid = {staf2260},
        pages = {staf2260},
          doi = {10.1093/mnras/staf2260},
archivePrefix = {arXiv},
       eprint = {2509.09665},
 primaryClass = {astro-ph.CO},
       adsurl = {https://ui.adsabs.harvard.edu/abs/2026MNRAS.546f2260S}}

@ARTICLE{Asencio_2021,
	author = {{Asencio}, Elena and {Banik}, Indranil and {Kroupa}, Pavel},
	title = "{A massive blow for {\ensuremath{\Lambda}}CDM - the high redshift, mass, and collision velocity of the interacting galaxy cluster El Gordo contradicts concordance cosmology}",
	journal = {MNRAS},
	year = 2021,
	month = feb,
    number = {4},
    volume = {500},
    pages = {5249-5267},
	doi = {10.1093/mnras/staa3441},
	adsurl = {https://ui.adsabs.harvard.edu/abs/2021MNRAS.500.5249A}}

@ARTICLE{Asencio_2023,
       author = {{Asencio}, E. and {Banik}, I. and {Kroupa}, P.},
        title = "{The El Gordo galaxy cluster challenges {$\Lambda$}CDM for any plausible collision velocity}",
      journal = {ApJ},
         year = 2023,
        month = sep,
       volume = {954},
       number = {2},
          eid = {162},
        pages = {162},
archivePrefix = {arXiv},
       eprint = {2308.00744},
 primaryClass = {astro-ph.CO},
          doi = {10.3847/1538-4357/ace62a},
       adsurl = {https://ui.adsabs.harvard.edu/abs/2023ApJ...954..162A}}

@ARTICLE{Russell_2026,
       author = {{Russell}, Alfie and {Banik}, Indranil and {Cray}, Oscar and {Zhao}, Hongsheng},
        title = "{How does a MOND cosmology fare on Gpc scales? {\ensuremath{-}} collisionless N-body simulations of {\ensuremath{\nu}}HDM}",
      journal = {MNRAS},
         year = 2026,
        month = apr,
       volume = {547},
       number = {3},
          eid = {stag399},
        pages = {stag399},
          doi = {10.1093/mnras/stag399},
       adsurl = {https://ui.adsabs.harvard.edu/abs/2026MNRAS.547ag399R}}

@ARTICLE{JUNO_2025,
       author = {{Juno Collaboration}},
        title = "{Potential to identify neutrino mass ordering with reactor antineutrinos at JUNO}",
      journal = {Chinese Physics C},
         year = 2025,
        month = mar,
       volume = {49},
       number = {3},
          eid = {033104},
        pages = {033104},
          doi = {10.1088/1674-1137/ad7f3e},
archivePrefix = {arXiv},
       eprint = {2405.18008},
 primaryClass = {hep-ex},
       adsurl = {https://ui.adsabs.harvard.edu/abs/2025ChPhC..49c3104A}}

@ARTICLE{Alpher_1948,
	author = {{Alpher}, R.~A. and {Bethe}, H. and {Gamow}, G.},
	title = "{The Origin of Chemical Elements}",
	journal = {Physical Review},
	year = 1948,
	month = apr,
	volume = {73},
	number = {7},
	pages = {803-804},
	doi = {10.1103/PhysRev.73.803},
	adsurl = {https://ui.adsabs.harvard.edu/abs/1948PhRv...73..803A}}

@ARTICLE{Tytler_2000,
       author = {{Tytler}, David and {O'Meara}, John M. and {Suzuki}, Nao and {Lubin}, Dan},
        title = "{Review of Big Bang Nucleosynthesis and Primordial Abundances}",
      journal = {Physica Scripta Volume T},
         year = 2000,
        month = jan,
       volume = {85},
        pages = {12},
          doi = {10.1238/Physica.Topical.085a00012},
archivePrefix = {arXiv},
       eprint = {astro-ph/0001318},
 primaryClass = {astro-ph},
       adsurl = {https://ui.adsabs.harvard.edu/abs/2000PhST...85...12T}}

@ARTICLE{Cyburt_2016,
	title = {Big bang nucleosynthesis: Present status},
	author = {Cyburt, Richard H. and Fields, Brian D. and Olive, Keith A. and Yeh, Tsung-Han},
	journal = {Reviews of Modern Physics},
	volume = {88},
	issue = {1},
	pages = {015004},
	numpages = {22},
	year = 2016,
	month = feb,
	publisher = {American Physical Society},
	doi = {10.1103/RevModPhys.88.015004},
	adsurl = {https://link.aps.org/doi/10.1103/RevModPhys.88.015004}}

@ARTICLE{Pitrou_2018,
       author = {{Pitrou}, Cyril and {Coc}, Alain and {Uzan}, Jean-Philippe and {Vangioni}, Elisabeth},
        title = "{Precision big bang nucleosynthesis with improved Helium-4 predictions}",
      journal = {Physics Reports},
         year = 2018,
        month = sep,
       volume = {754},
        pages = {1-66},
          doi = {10.1016/j.physrep.2018.04.005},
archivePrefix = {arXiv},
       eprint = {1801.08023},
 primaryClass = {astro-ph.CO},
       adsurl = {https://ui.adsabs.harvard.edu/abs/2018PhR...754....1P}}

@ARTICLE{Giovanetti_2025_BBN,
       author = {{Giovanetti}, Cara and {Lisanti}, Mariangela and {Liu}, Hongwan and {Mishra-Sharma}, Siddharth and {Ruderman}, Joshua T.},
        title = "{Cosmological parameter estimation with a joint-likelihood analysis of the cosmic microwave background and big bang nucleosynthesis}",
      journal = {Physical Review D},
         year = 2025,
        month = sep,
       volume = {112},
       number = {6},
          eid = {063530},
        pages = {063530},
          doi = {10.1103/wspy-s948},
archivePrefix = {arXiv},
       eprint = {2408.14531},
 primaryClass = {astro-ph.CO},
       adsurl = {https://ui.adsabs.harvard.edu/abs/2025PhRvD.112f3530G}}

@ARTICLE{Aver_2015,
       author = {{Aver}, Erik and {Olive}, Keith A. and {Skillman}, Evan D.},
        title = "{The effects of He I {\ensuremath{\lambda}}10830 on helium abundance determinations}",
      journal = {JCAP},
         year = 2015,
        month = jul,
       volume = {2015},
       number = {7},
        pages = {011-011},
          doi = {10.1088/1475-7516/2015/07/011},
archivePrefix = {arXiv},
       eprint = {1503.08146},
 primaryClass = {astro-ph.CO},
       adsurl = {https://ui.adsabs.harvard.edu/abs/2015JCAP...07..011A}}

@ARTICLE{Cooke_2018,
       author = {{Cooke}, Ryan J. and {Pettini}, Max and {Steidel}, Charles C.},
        title = "{One Percent Determination of the Primordial Deuterium Abundance}",
      journal = {ApJ},
         year = 2018,
        month = mar,
       volume = {855},
       number = {2},
          eid = {102},
        pages = {102},
          doi = {10.3847/1538-4357/aaab53},
archivePrefix = {arXiv},
       eprint = {1710.11129},
 primaryClass = {astro-ph.CO},
       adsurl = {https://ui.adsabs.harvard.edu/abs/2018ApJ...855..102C}}

@ARTICLE{Pettini_2026,
       author = {{Pettini}, Max and {Cooke}, Ryan},
        title = "{Precision cosmology with the lightest elements}",
      journal = {Ap\&SS},
         year = 2026,
        month = feb,
       volume = {371},
       number = {2},
          eid = {20},
        pages = {20},
          doi = {10.1007/s10509-026-04551-x},
archivePrefix = {arXiv},
       eprint = {2511.06275},
 primaryClass = {astro-ph.CO},
       adsurl = {https://ui.adsabs.harvard.edu/abs/2026Ap\&SS.371...20P}}

@ARTICLE{Giovanetti_2026,
       author = {{Giovanetti}, Cara},
        title = "{A generic $ω_b$ tension in early-time solutions to the Hubble tension}",
      journal = {ArXiv e-prints},
         year = 2026,
        month = apr,
       volume = {Arxiv},
          doi = {10.48550/arXiv.2604.05095},
archivePrefix = {arXiv},
       eprint = {2604.05095},
 primaryClass = {astro-ph.CO},
       adsurl = {https://ui.adsabs.harvard.edu/abs/2026arXiv260405095G}}

@ARTICLE{Launders_2026,
       author = {{Launders}, Timothy and {Giovanetti}, Cara and {Liu}, Hongwan},
        title = "{A data-driven prediction for the primordial deuterium abundance}",
      journal = {ArXiv e-prints},
         year = 2026,
        month = apr,
          eid = {arXiv:2604.16600},
       volume = {Arxiv},
archivePrefix = {arXiv},
       eprint = {2604.16600},
 primaryClass = {astro-ph.CO},
       adsurl = {https://ui.adsabs.harvard.edu/abs/2026arXiv260416600L}}

@ARTICLE{Efstathiou_1990,
	author = {{Efstathiou}, G. and {Sutherland}, W.~J. and {Maddox}, S.~J.},
	title = "{The cosmological constant and cold dark matter}",
	journal = {Nature},
	year = 1990,
	month = dec,
	volume = {348},
	number = {6303},
	pages = {705-707},
	doi = {10.1038/348705a0},
	adsurl = {https://ui.adsabs.harvard.edu/abs/1990Natur.348..705E}}

@ARTICLE{Ostriker_Steinhardt_1995,
	author = {{Ostriker}, J.~P. and {Steinhardt}, P.~J.},
	title = "{The observational case for a low-density Universe with a non-zero cosmological constant}",
	journal = {Nature},
	year = 1995,
	month = oct,
	volume = 377,
	pages = {600-602},
	doi = {10.1038/377600a0},
	adsurl = {http://adsabs.harvard.edu/abs/1995Natur.377..600O}}

@ARTICLE{Hu_1996,
       author = {{Hu}, Wayne and {Sugiyama}, Naoshi},
        title = "{Small-Scale Cosmological Perturbations: an Analytic Approach}",
      journal = {ApJ},
         year = 1996,
        month = nov,
       volume = {471},
        pages = {542},
          doi = {10.1086/177989},
archivePrefix = {arXiv},
       eprint = {astro-ph/9510117},
 primaryClass = {astro-ph},
       adsurl = {https://ui.adsabs.harvard.edu/abs/1996ApJ...471..542H}}

@ARTICLE{Lemos_2023,
       author = {{Lemos}, Pablo and {Lewis}, Antony},
        title = "{CMB constraints on the early Universe independent of late-time cosmology}",
      journal = {Physical Review D},
         year = 2023,
        month = "may",
       volume = {107},
       number = {10},
          eid = {103505},
        pages = {103505},
          doi = {10.1103/PhysRevD.107.103505},
archivePrefix = {arXiv},
       eprint = {2302.12911},
 primaryClass = {astro-ph.CO},
       adsurl = {https://ui.adsabs.harvard.edu/abs/2023PhRvD.107j3505L}}

@ARTICLE{Eisenstein_1998,
       author = {{Eisenstein}, Daniel J. and {Hu}, Wayne},
        title = "{Baryonic Features in the Matter Transfer Function}",
      journal = {ApJ},
         year = 1998,
        month = mar,
       volume = {496},
       number = {2},
        pages = {605-614},
          doi = {10.1086/305424},
archivePrefix = {arXiv},
       eprint = {astro-ph/9709112},
 primaryClass = {astro-ph},
       adsurl = {https://ui.adsabs.harvard.edu/abs/1998ApJ...496..605E}}

@ARTICLE{CLASS,
       author = {{Blas}, Diego and {Lesgourgues}, Julien and {Tram}, Thomas},
        title = "{The Cosmic Linear Anisotropy Solving System (CLASS). Part II: Approximation schemes}",
      journal = {JCAP},
         year = 2011,
        month = jul,
       volume = {2011},
       number = {7},
          eid = {034},
        pages = {034},
          doi = {10.1088/1475-7516/2011/07/034},
archivePrefix = {arXiv},
       eprint = {1104.2933},
 primaryClass = {astro-ph.CO},
       adsurl = {https://ui.adsabs.harvard.edu/abs/2011JCAP...07..034B}}

@ARTICLE{Poulin_2019,
       author = {{Poulin}, Vivian and {Smith}, Tristan L. and {Karwal}, Tanvi and {Kamionkowski}, Marc},
        title = "{Early Dark Energy can Resolve the Hubble Tension}",
      journal = {Physical Review Letters},
         year = 2019,
        month = jun,
       volume = {122},
       number = {22},
          eid = {221301},
        pages = {221301},
          doi = {10.1103/PhysRevLett.122.221301},
archivePrefix = {arXiv},
       eprint = {1811.04083},
 primaryClass = {astro-ph.CO},
       adsurl = {https://ui.adsabs.harvard.edu/abs/2019PhRvL.122v1301P}}

@ARTICLE{Jimenez_2002,
       author = {{Jimenez}, Raul and {Loeb}, Abraham},
        title = "{Constraining Cosmological Parameters Based on Relative Galaxy Ages}",
      journal = {ApJ},
         year = 2002,
        month = jul,
       volume = {573},
       number = {1},
        pages = {37-42},
          doi = {10.1086/340549},
archivePrefix = {arXiv},
       eprint = {astro-ph/0106145},
 primaryClass = {astro-ph},
       adsurl = {https://ui.adsabs.harvard.edu/abs/2002ApJ...573...37J}}

@ARTICLE{Jia_2023,
       author = {{Jia}, X.~D. and {Hu}, J.~P. and {Wang}, F.~Y.},
        title = "{Evidence of a decreasing trend for the Hubble constant}",
      journal = {A\&A},
         year = 2023,
        month = jun,
       volume = {674},
          eid = {A45},
        pages = {A45},
          doi = {10.1051/0004-6361/202346356},
archivePrefix = {arXiv},
       eprint = {2212.00238},
 primaryClass = {astro-ph.CO},
       adsurl = {https://ui.adsabs.harvard.edu/abs/2023A\&A...674A..45J}}

@ARTICLE{Jia_2025a,
       author = {{Jia}, X.~D. and {Hu}, J.~P. and {Yi}, S.~X. and {Wang}, F.~Y.},
        title = "{Uncorrelated Estimations of H$_{0}$ Redshift Evolution from DESI Baryon Acoustic Oscillation Observations}",
      journal = {ApJL},
         year = 2025,
        month = feb,
       volume = {979},
       number = {2},
          eid = {L34},
        pages = {L34},
          doi = {10.3847/2041-8213/ada94d},
archivePrefix = {arXiv},
       eprint = {2406.02019},
 primaryClass = {astro-ph.CO},
       adsurl = {https://ui.adsabs.harvard.edu/abs/2025ApJ...979L..34J}}

@ARTICLE{Jia_2025b,
       author = {{Jia}, X.~D. and {Hu}, J.~P. and {Gao}, D.~H. and {Yi}, S.~X. and {Wang}, F.~Y.},
        title = "{The Hubble Tension resolved by the DESI Baryon Acoustic Oscillations Measurements}",
      journal = {ApJL},
         year = 2025,
        month = nov,
       volume = {994},
       number = {1},
          eid = {L22},
        pages = {L22},
          doi = {10.3847/2041-8213/ae1965},
archivePrefix = {arXiv},
       eprint = {2509.17454},
 primaryClass = {astro-ph.CO},
       adsurl = {https://ui.adsabs.harvard.edu/abs/2025ApJ...994L..22J}}

@ARTICLE{Jia_2026,
       author = {{Jia}, X.~D. and {Dai}, X.~Y. and {Yang}, Y.~P. and {Wang}, F.~Y.},
        title = "{A Review on Resolving the Hubble Tension via Late-Universe Physics}",
      journal = {Galaxies},
         year = 2026,
        month = may,
       volume = {14},
       number = {3},
          eid = {55},
        pages = {55},
archivePrefix = {arXiv},
          doi = {https://doi.org/10.3390/galaxies14030055},
       adsurl = {https://ui.adsabs.harvard.edu/abs/2026Galax..14...55J}}

@ARTICLE{Lopez_2025,
       author = {{Lopez-Hernandez}, Mauricio and {De-Santiago}, Josue},
        title = "{Is there a dynamical tendency in H0 with late time measurements?}",
      journal = {JCAP},
         year = 2025,
        month = mar,
       volume = {2025},
       number = {3},
          eid = {026},
        pages = {026},
          doi = {10.1088/1475-7516/2025/03/026},
archivePrefix = {arXiv},
       eprint = {2411.00095},
 primaryClass = {astro-ph.CO},
       adsurl = {https://ui.adsabs.harvard.edu/abs/2025JCAP...03..026L}}

@ARTICLE{Moresco_2018,
       author = {{Moresco}, Michele and {Jimenez}, Raul and {Verde}, Licia and {Pozzetti}, Lucia and {Cimatti}, Andrea and {Citro}, Annalisa},
        title = "{Setting the Stage for Cosmic Chronometers. I. Assessing the Impact of Young Stellar Populations on Hubble Parameter Measurements}",
      journal = {ApJ},
         year = 2018,
        month = dec,
       volume = {868},
       number = {2},
          eid = {84},
        pages = {84},
          doi = {10.3847/1538-4357/aae829},
archivePrefix = {arXiv},
       eprint = {1804.05864},
 primaryClass = {astro-ph.CO},
       adsurl = {https://ui.adsabs.harvard.edu/abs/2018ApJ...868...84M}}

@ARTICLE{Moresco_2020,
       author = {{Moresco}, Michele and {Jimenez}, Raul and {Verde}, Licia and {Cimatti}, Andrea and {Pozzetti}, Lucia},
        title = "{Setting the Stage for Cosmic Chronometers. II. Impact of Stellar Population Synthesis Models Systematics and Full Covariance Matrix}",
      journal = {ApJ},
         year = 2020,
        month = jul,
       volume = {898},
       number = {1},
          eid = {82},
        pages = {82},
          doi = {10.3847/1538-4357/ab9eb0},
archivePrefix = {arXiv},
       eprint = {2003.07362},
 primaryClass = {astro-ph.GA},
       adsurl = {https://ui.adsabs.harvard.edu/abs/2020ApJ...898...82M}}

@ARTICLE{Moresco_2024,
       author = {{Moresco}, Michele},
        title = "{Measuring the expansion history of the Universe with cosmic chronometers}",
      journal = {ArXiv e-prints},
         year = 2024,
        month = dec,
          eid = {arXiv:2412.01994},
       volume = {Arxiv},
          doi = {10.48550/arXiv.2412.01994},
archivePrefix = {arXiv},
       eprint = {2412.01994},
 primaryClass = {astro-ph.CO},
       adsurl = {https://ui.adsabs.harvard.edu/abs/2024arXiv241201994M}}

@ARTICLE{Wang_2026,
       author = {{Wang}, Ze-fan and {Lei}, Lei and {Fan}, Yi-zhong},
        title = "{New H(z) Measurement at Redshift = 0.12 with DESI Data Release 1}",
      journal = {ApJ},
         year = 2026,
        month = may,
       volume = {1003},
       number = {1},
          eid = {81},
        pages = {81},
          doi = {10.3847/1538-4357/ae610a},
archivePrefix = {arXiv},
       eprint = {2601.07345},
 primaryClass = {astro-ph.CO},
       adsurl = {https://ui.adsabs.harvard.edu/abs/2026ApJ..1003...81W}}

@ARTICLE{Speagle_2020,
       author = {{Speagle}, Joshua S.},
        title = "{DYNESTY: a dynamic nested sampling package for estimating Bayesian posteriors and evidences}",
      journal = {MNRAS},
         year = 2020,
        month = "apr",
       volume = {493},
       number = {3},
        pages = {3132-3158},
          doi = {10.1093/mnras/staa278},
archivePrefix = {arXiv},
       eprint = {1904.02180},
 primaryClass = {astro-ph.IM},
       adsurl = {https://ui.adsabs.harvard.edu/abs/2020MNRAS.493.3132S}}

@ARTICLE{Koposov_2024,
        author = {Koposov, Sergey and Speagle, Josh and Barbary, Kyle and Ashton, Gregory and Bennett, Ed and Buchner, Johannes and Scheffler, Carl and Cook, Ben and Talbot, Colm and Guillochon, James and others},
  title        = {joshspeagle/dynesty: v2.1.4},
  year         = 2024,
  month        = jun,
  journal      = {Zenodo},
  volume       = {12537467},
  version      = {v2.1.4},
  doi          = {10.5281/zenodo.12537467},
  adsurl       = {https://doi.org/10.5281/zenodo.12537467}}

@INPROCEEDINGS{Skilling_2004,
       author = {{Skilling}, John},
        title = "{Nested Sampling}",
    booktitle = {Bayesian Inference and Maximum Entropy Methods in Science and Engineering: 24th International Workshop on Bayesian Inference and Maximum Entropy Methods in Science and Engineering},
         year = 2004,
       editor = {{Fischer}, Rainer and {Preuss}, Roland and {Toussaint}, Udo Von},
       series = {American Institute of Physics Conference Series},
       volume = {735},
        month = nov,
    publisher = {AIP},
        pages = {395-405},
          doi = {10.1063/1.1835238},
       adsurl = {https://ui.adsabs.harvard.edu/abs/2004AIPC..735..395S}}

@article{Skilling_2006,
   author = {{Skilling}, John},
    title = "{Nested sampling for general Bayesian computation}",
    journal = {Bayesian Analysis},
    year = 2006,
    volume = 1,
    number = 4,
    pages = {833-859},
    doi = {10.1214/06-BA127},
  adsurl = {https://doi.org/10.1214/06-BA127}}

@ARTICLE{Feroz_2009,
       author = {{Feroz}, F. and {Hobson}, M.~P. and {Bridges}, M.},
        title = "{MULTINEST: an efficient and robust Bayesian inference tool for cosmology and particle physics}",
      journal = {MNRAS},
         year = 2009,
        month = oct,
       volume = {398},
       number = {4},
        pages = {1601-1614},
          doi = {10.1111/j.1365-2966.2009.14548.x},
archivePrefix = {arXiv},
       eprint = {0809.3437},
 primaryClass = {astro-ph},
       adsurl = {https://ui.adsabs.harvard.edu/abs/2009MNRAS.398.1601F}}

@ARTICLE{Lewis_2025_GetDist,
       author = {{Lewis}, Antony},
        title = "{GetDist: a Python package for analysing Monte Carlo samples}",
      journal = {JCAP},
         year = 2025,
        month = aug,
       volume = {2025},
       number = {8},
          eid = {025},
        pages = {025},
          doi = {10.1088/1475-7516/2025/08/025},
archivePrefix = {arXiv},
       eprint = {1910.13970},
 primaryClass = {astro-ph.IM},
       adsurl = {https://ui.adsabs.harvard.edu/abs/2025JCAP...08..025L}}

@ARTICLE{Cimatti_2023,
       author = {Cimatti, A. and Moresco, M.},
        title = "{Revisiting oldest stars as cosmological probes}",
      journal = {ApJ},
         year = 2023,
        month = aug,
       volume = {953},
       number = {2},
          eid = {149},
        pages = {149},
archivePrefix = {arXiv},
       eprint = {2302.07899},
 primaryClass = {astro-ph.CO},
          doi = {10.3847/1538-4357/ace439},
       adsurl = {https://ui.adsabs.harvard.edu/abs/2023ApJ...953..149C}}

@ARTICLE{Cogato_2024,
       author = {{Cogato}, Fabrizio and {Moresco}, Michele and {Amati}, Lorenzo and {Cimatti}, Andrea},
        title = "{An analytical late-Universe approach to the weaving of modern cosmology}",
      journal = {MNRAS},
         year = 2024,
        month = jan,
       volume = {527},
       number = {3},
        pages = {4874-4888},
          doi = {10.1093/mnras/stad3546},
archivePrefix = {arXiv},
       eprint = {2309.01375},
 primaryClass = {astro-ph.CO},
       adsurl = {https://ui.adsabs.harvard.edu/abs/2024MNRAS.527.4874C}}

@ARTICLE{Guo_2025,
       author = {{Guo}, Wuzheng and {Wang}, Qiumin and {Cao}, Shuo and {Biesiada}, Marek and {Liu}, Tonghua and {Lian}, Yujie and {Jiang}, Xinyue and {Mu}, Chengsheng and {Cheng}, Dadian},
        title = "{Newest Measurements of Hubble Constant from DESI 2024 Baryon Acoustic Oscillation Observations}",
      journal = {ApJL},
         year = 2025,
        month = jan,
       volume = {978},
       number = {2},
          eid = {L33},
        pages = {L33},
          doi = {10.3847/2041-8213/ada37f},
archivePrefix = {arXiv},
       eprint = {2412.13045},
 primaryClass = {astro-ph.CO},
       adsurl = {https://ui.adsabs.harvard.edu/abs/2025ApJ...978L..33G}}

@ARTICLE{Bernal_2021,
       author = {{Bernal}, Jos{\'e} Luis and {Verde}, Licia and {Jimenez}, Raul and {Kamionkowski}, Marc and {Valcin}, David and {Wandelt}, Benjamin D.},
        title = "{Trouble beyond H$_{0}$ and the new cosmic triangles}",
      journal = {Physical Review D},
         year = 2021,
        month = may,
       volume = {103},
       number = {10},
          eid = {103533},
        pages = {103533},
          doi = {10.1103/PhysRevD.103.103533},
archivePrefix = {arXiv},
       eprint = {2102.05066},
 primaryClass = {astro-ph.CO},
       adsurl = {https://ui.adsabs.harvard.edu/abs/2021PhRvD.103j3533B}}

@ARTICLE{Valcin_2020,
       author = {{Valcin}, David and {Bernal}, Jos{\'e} Luis and {Jimenez}, Raul and {Verde}, Licia and {Wandelt}, Benjamin D.},
        title = "{Inferring the age of the universe with globular clusters}",
      journal = {JCAP},
         year = 2020,
        month = dec,
       volume = {2020},
       number = {12},
          eid = {002},
        pages = {002},
          doi = {10.1088/1475-7516/2020/12/002},
archivePrefix = {arXiv},
       eprint = {2007.06594},
 primaryClass = {astro-ph.CO},
       adsurl = {https://ui.adsabs.harvard.edu/abs/2020JCAP...12..002V}}

@ARTICLE{Valcin_2021,
       author = {{Valcin}, David and {Jimenez}, Raul and {Verde}, Licia and {Bernal}, Jos{\'e} Luis and {Wandelt}, Benjamin D.},
        title = "{The age of the Universe with globular clusters: reducing systematic uncertainties}",
      journal = {JCAP},
         year = 2021,
        month = aug,
       volume = {2021},
       number = {8},
          eid = {017},
        pages = {017},
          doi = {10.1088/1475-7516/2021/08/017},
archivePrefix = {arXiv},
       eprint = {2102.04486},
 primaryClass = {astro-ph.GA},
       adsurl = {https://ui.adsabs.harvard.edu/abs/2021JCAP...08..017V}}

@ARTICLE{Valcin_2025,
       author = {{Valcin}, David and {Jimenez}, Raul and {Seljak}, Uro{\v{s}} and {Verde}, Licia},
        title = "{The age of the universe with globular clusters. Part III. Gaia distances and hierarchical modeling}",
      journal = {JCAP},
         year = 2025,
        month = oct,
       volume = {2025},
       number = {10},
          eid = {030},
        pages = {030},
          doi = {10.1088/1475-7516/2025/10/030},
archivePrefix = {arXiv},
       eprint = {2503.19481},
 primaryClass = {astro-ph.CO},
       adsurl = {https://ui.adsabs.harvard.edu/abs/2025JCAP...10..030V}}

@ARTICLE{Valcin_2026,
       author = {{Valcin}, David and {Jimenez}, Raul and {Lardo}, Carmela and {Seljak}, Uro{\v{s}} and {Verde}, Licia},
        title = "{The Age of the Universe with Globular Clusters IV: Multiple Stellar Populations}",
      journal = {ArXiv e-prints},
         year = 2026,
        month = mar,
          eid = {arXiv:2603.04872},
       volume = {Arxiv},
          doi = {10.48550/arXiv.2603.04872},
archivePrefix = {arXiv},
       eprint = {2603.04872},
 primaryClass = {astro-ph.GA},
       adsurl = {https://ui.adsabs.harvard.edu/abs/2026arXiv260304872V}}

@ARTICLE{Xiang_2022,
       author = {{Xiang}, Maosheng and {Rix}, Hans-Walter},
        title = "{A time-resolved picture of our Milky Way's early formation history}",
      journal = {Nature},
         year = 2022,
        month = mar,
       volume = {603},
       number = {7902},
        pages = {599-603},
          doi = {10.1038/s41586-022-04496-5},
archivePrefix = {arXiv},
       eprint = {2203.12110},
 primaryClass = {astro-ph.GA},
       adsurl = {https://ui.adsabs.harvard.edu/abs/2022Natur.603..599X}}

@ARTICLE{Xiang_2025,
       author = {{Xiang}, Maosheng and {Rix}, Hans-Walter and {Yang}, Hang and {Liu}, Jifeng and {Huang}, Yang and {Frankel}, Neige},
        title = "{The formation and survival of the Milky Way's oldest stellar disk}",
      journal = {Nature Astronomy},
         year = 2025,
        month = jan,
       volume = {9},
        pages = {101-110},
          doi = {10.1038/s41550-024-02382-w},
archivePrefix = {arXiv},
       eprint = {2410.09705},
 primaryClass = {astro-ph.GA},
       adsurl = {https://ui.adsabs.harvard.edu/abs/2025NatAs...9..101X}}

@ARTICLE{Banik_2025_cosmology,
       author = {{Banik}, Indranil and {Samaras}, Nick},
        title = "{Constraints on the Hubble and matter density parameters with and without modelling the CMB anisotropies}",
      journal = {Astronomy},
         year = 2025,
        month = nov,
       volume = {4},
        pages = {24},
          doi = {10.3390/astronomy4040024},
archivePrefix = {arXiv},
       eprint = {2410.00804},
 primaryClass = {astro-ph.CO},
       adsurl = {https://ui.adsabs.harvard.edu/abs/2024arXiv241000804B}}

@ARTICLE{Banik_2026_age,
       author = {{Banik}, Indranil and {Kudakolawa Kaluarachchige}, Thenujaya and {Cookson}, Stephen and {Desmond}, Harry},
        title = "{The age of the Universe from a large sample of the oldest Galactic stars}",
      journal = {ArXiv e-prints},
         year = 2026,
        month = jul,
          eid = {arXiv:2607.00764},
       volume = {Arxiv},
          doi = {10.48550/arXiv.2607.00764},
archivePrefix = {arXiv},
       eprint = {2607.00764},
 primaryClass = {astro-ph.CO},
       adsurl = {https://ui.adsabs.harvard.edu/abs/2026arXiv260700764B}}

@ARTICLE{Montalban_2021,
       author = {{Montalb{\'a}n}, Josefina and {Mackereth}, J. Ted and {Miglio}, Andrea and {Vincenzo}, Fiorenzo and {Chiappini}, Cristina and {Buldgen}, Gael and {Mosser}, Beno{\^\i}t and {Noels}, Arlette and {Scuflaire}, Richard and {Vrard}, Mathieu and {Willett}, Emma and {Davies}, Guy R. and {Hall}, Oliver J. and {Nielsen}, Martin Bo and {Khan}, Saniya and {Rendle}, Ben M. and {van Rossem}, Walter E. and {Ferguson}, Jason W. and {Chaplin}, William J.},
        title = "{Chronologically dating the early assembly of the Milky Way}",
      journal = {Nature Astronomy},
         year = 2021,
        month = jan,
       volume = {5},
        pages = {640-647},
          doi = {10.1038/s41550-021-01347-7},
archivePrefix = {arXiv},
       eprint = {2006.01783},
 primaryClass = {astro-ph.GA},
       adsurl = {https://ui.adsabs.harvard.edu/abs/2021NatAs...5..640M}}

@ARTICLE{Limberg_2022,
       author = {{Limberg}, Guilherme and {Souza}, Stefano O. and {P{\'e}rez-Villegas}, Angeles and {Rossi}, Silvia and {Perottoni}, H{\'e}lio D. and {Santucci}, Rafael M.},
        title = "{Reconstructing the Disrupted Dwarf Galaxy Gaia-Sausage/Enceladus Using Its Stars and Globular Clusters}",
      journal = {ApJ},
         year = 2022,
        month = aug,
       volume = {935},
       number = {2},
          eid = {109},
        pages = {109},
          doi = {10.3847/1538-4357/ac8159},
archivePrefix = {arXiv},
       eprint = {2206.10505},
 primaryClass = {astro-ph.GA},
       adsurl = {https://ui.adsabs.harvard.edu/abs/2022ApJ...935..109L}}

@ARTICLE{Nepal_2024,
       author = {{Nepal}, S. and {Chiappini}, C. and {Queiroz}, A.~B. and {Guiglion}, G. and {Montalb{\'a}n}, J. and {Steinmetz}, M. and {Miglio}, A. and {Khalatyan}, A.},
        title = "{Discovery of the local counterpart of disc galaxies at z > 4: The oldest thin disc of the Milky Way using Gaia-RVS}",
      journal = {A\&A},
         year = 2024,
        month = aug,
       volume = {688},
          eid = {A167},
        pages = {A167},
          doi = {10.1051/0004-6361/202449445},
archivePrefix = {arXiv},
       eprint = {2402.00561},
 primaryClass = {astro-ph.GA},
       adsurl = {https://ui.adsabs.harvard.edu/abs/2024A\&A...688A.167N}}

@ARTICLE{Souza_2024,
       author = {{Souza}, S.~O. and {Libralato}, M. and {Nardiello}, D. and {Kerber}, L.~O. and {Ortolani}, S. and {P{\'e}rez-Villegas}, A. and {Oliveira}, R.~A.~P. and {Barbuy}, B. and {Bica}, E. and {Griggio}, M. and {Dias}, B.},
        title = "{Combined Gemini-South and HST photometric analysis of the globular cluster NGC 6558: The age of the metal-poor population of the Galactic bulge}",
      journal = {A\&A},
         year = 2024,
        month = oct,
       volume = {690},
          eid = {A37},
        pages = {A37},
          doi = {10.1051/0004-6361/202450795},
archivePrefix = {arXiv},
       eprint = {2407.15918},
 primaryClass = {astro-ph.GA},
       adsurl = {https://ui.adsabs.harvard.edu/abs/2024A\&A...690A..37S}}

@ARTICLE{Lundkvist_2025,
       author = {{Lundkvist}, M.~S. and {Larsen}, J.~R. and {Li}, Y. and {Winther}, M.~L. and {Bedding}, T.~R. and {Kjeldsen}, H. and {White}, T.~R. and {Nielsen}, M.~B. and {Buldgen}, G. and {Guillaume}, C. and {Stokholm}, A.~L. and {Huber}, D. and {R{\o}rsted}, J.~L. and {Mani}, P. and {Grundahl}, F.},
        title = "{Asteroseismic investigation of HD 140283: The Methuselah star}",
      journal = {A\&A},
         year = 2025,
        month = nov,
       volume = {703},
          eid = {A232},
        pages = {A232},
          doi = {10.1051/0004-6361/202556292},
archivePrefix = {arXiv},
       eprint = {2510.11532},
 primaryClass = {astro-ph.SR},
       adsurl = {https://ui.adsabs.harvard.edu/abs/2025A\&A...703A.232L}}

@ARTICLE{Shariat_2026,
       author = {{Shariat}, Cheyanne and {El-Badry}, Kareem and {Bhattacharjee}, Soumyadeep},
        title = "{How precisely can we measure the ages of subgiant and giant stars?}",
      journal = {The Open Journal of Astrophysics},
         year = 2026,
        month = may,
       volume = {9},
        pages = {62534},
          doi = {10.33232/001c.162534},
archivePrefix = {arXiv},
       eprint = {2510.08675},
 primaryClass = {astro-ph.SR},
       adsurl = {https://ui.adsabs.harvard.edu/abs/2026OJAp....962534S}}

@ARTICLE{Tomasetti_2026,
       author = {{Tomasetti}, Elena and {Chiappini}, Cristina and {Nepal}, Samir and {Moresco}, Michele and {Lardo}, Carmela and {Cimatti}, Andrea and {Anders}, Friedrich and {Queiroz}, Anna B.~A. and {Limberg}, Guilherme},
        title = "{The oldest Milky Way stars: New constraints on the age of the Universe and the Hubble constant}",
      journal = {A\&A},
         year = 2026,
        month = mar,
       volume = {707},
          eid = {A111},
        pages = {A111},
archivePrefix = {arXiv},
       eprint = {2509.02692},
 primaryClass = {astro-ph.CO},
          doi = {10.1051/0004-6361/202557038},
       adsurl = {https://ui.adsabs.harvard.edu/abs/2018A\&A...707A.111T}}

@ARTICLE{Sachs_1967,
       author = {{Sachs}, R.~K. and {Wolfe}, A.~M.},
        title = "{Perturbations of a Cosmological Model and Angular Variations of the Microwave Background}",
      journal = {ApJ},
         year = 1967,
        month = jan,
       volume = {147},
        pages = {73},
          doi = {10.1086/148982},
       adsurl = {https://ui.adsabs.harvard.edu/abs/1967ApJ...147...73S}}

@ARTICLE{Vagnozzi_2023,
       author = {{Vagnozzi}, Sunny},
        title = "{Seven Hints That Early-Time New Physics Alone Is Not Sufficient to Solve the Hubble Tension}",
      journal = {Universe},
         year = 2023,
        month = aug,
       volume = {9},
       number = {9},
        pages = {393},
          doi = {10.3390/universe9090393},
archivePrefix = {arXiv},
       eprint = {2308.16628},
 primaryClass = {astro-ph.CO},
       adsurl = {https://ui.adsabs.harvard.edu/abs/2023Univ....9..393V}}

@ARTICLE{Pesce_2020,
	author = {{Pesce}, D.~W. and {Braatz}, J.~A. and {Reid}, M.~J. and {Riess}, A.~G. and {Scolnic}, D. and {Condon}, J.~J. and {Gao}, F. and {Henkel}, C. and {Impellizzeri}, C.~M.~V. and {Kuo}, C.~Y. and {Lo}, K.~Y.},
	title = "{The Megamaser Cosmology Project. XIII. Combined Hubble Constant Constraints}",
	journal = {ApJ},
	year = 2020,
	month = mar,
	volume = {891},
	number = {1},
	eid = {L1},
	pages = {L1},
	doi = {10.3847/2041-8213/ab75f0},
	archivePrefix = {arXiv},
	eprint = {2001.09213},
	primaryClass = {astro-ph.CO},
	adsurl = {https://ui.adsabs.harvard.edu/abs/2020ApJ...891L...1P}}

@ARTICLE{Barua_2025,
       author = {{Barua}, Shubham and {Ramakrishnan}, Vyaas and {Desai}, Shantanu},
        title = "{Determination of Hubble constant from Megamaser Cosmology Project using profile likelihood}",
      journal = {Ap\&SS},
         year = 2025,
        month = jun,
       volume = {370},
       number = {6},
          eid = {62},
        pages = {62},
          doi = {10.1007/s10509-025-04454-3},
archivePrefix = {arXiv},
       eprint = {2502.11998},
 primaryClass = {astro-ph.IM},
       adsurl = {https://ui.adsabs.harvard.edu/abs/2025Ap\&SS.370...62B}}

@ARTICLE{Riess_2022_comprehensive,
       author = {{Riess}, Adam G. and {Yuan}, Wenlong and {Macri}, Lucas M. and {Scolnic}, Dan and {Brout}, Dillon and {Casertano}, Stefano and {Jones}, David O. and {Murakami}, Yukei and {Anand}, Gagandeep S. and {Breuval}, Louise and {Brink}, Thomas G. and {Filippenko}, Alexei V. and {Hoffmann}, Samantha and {Jha}, Saurabh W. and {D'arcy Kenworthy}, W. and {Mackenty}, John and {Stahl}, Benjamin E. and {Zheng}, WeiKang},
        title = "{A Comprehensive Measurement of the Local Value of the Hubble Constant with 1 km s$^{-1}$ Mpc$^{-1}$ Uncertainty from the Hubble Space Telescope and the SH0ES Team}",
      journal = {ApJL},
         year = 2022,
        month = jul,
       volume = {934},
       number = {1},
          eid = {L7},
        pages = {L7},
          doi = {10.3847/2041-8213/ac5c5b},
archivePrefix = {arXiv},
       eprint = {2112.04510},
 primaryClass = {astro-ph.CO},
       adsurl = {https://ui.adsabs.harvard.edu/abs/2022ApJ...934L...7R}}

@ARTICLE{Breuval_2024,
       author = {{Breuval}, Louise and {Riess}, Adam G. and {Casertano}, Stefano and {Yuan}, Wenlong and {Macri}, Lucas M. and {Romaniello}, Martino and {Murakami}, Yukei S. and {Scolnic}, Daniel and {Anand}, Gagandeep S. and {Soszy{\'n}ski}, Igor},
        title = "{Small Magellanic Cloud Cepheids Observed with the Hubble Space Telescope Provide a New Anchor for the SH0ES Distance Ladder}",
      journal = {ApJ},
         year = 2024,
        month = sep,
       volume = {973},
       number = {1},
          eid = {30},
        pages = {30},
          doi = {10.3847/1538-4357/ad630e},
archivePrefix = {arXiv},
       eprint = {2404.08038},
 primaryClass = {astro-ph.CO},
       adsurl = {https://ui.adsabs.harvard.edu/abs/2024ApJ...973...30B}}

@ARTICLE{Vogl_2025,
       author = {{Vogl}, C. and {Taubenberger}, S. and {Cs{\"o}rnyei}, G. and {Leibundgut}, B. and {Kerzendorf}, W.~E. and {Sim}, S.~A. and {Peterson}, E.~R. and {Courtois}, H.~M. and {Blondin}, S. and {Fl{\"o}rs}, A. and {Holas}, A. and {Shields}, J.~V. and {Spyromilio}, J. and {Suyu}, S.~H. and {Hillebrandt}, W.},
        title = "{No rungs attached: A distance-ladder-free determination of the Hubble constant through type II supernova spectral modelling}",
      journal = {A\&A},
         year = 2025,
        month = oct,
       volume = {702},
          eid = {A41},
        pages = {A41},
          doi = {10.1051/0004-6361/202452910},
archivePrefix = {arXiv},
       eprint = {2411.04968},
 primaryClass = {astro-ph.CO},
       adsurl = {https://ui.adsabs.harvard.edu/abs/2025A\&A...702A..41V}}

@ARTICLE{Valentino_2025,
       author = {{Di Valentino}, Eleonora and {Said}, Jackson Levi and {Riess}, Adam and {Pollo}, Agnieszka and {Poulin}, Vivian and {G{\'o}mez-Valent}, Adri{\`a} and {Weltman}, Amanda and {Palmese}, Antonella and {Huang}, Caroline D. and {van de Bruck}, Carsten and {Saraf}, Chandra Shekhar and {Kuo}, Cheng-Yu and {Uhlemann}, Cora and {Grand{\'o}n}, Daniela and {Paz}, Dante and {Eckert}, Dominique and {Teixeira}, Elsa M. and {Saridakis}, Emmanuel N. and {Colg{\'a}in}, Eoin {\'O}. and {Beutler}, Florian and {Niedermann}, Florian and {Bajardi}, Francesco and {Barenboim}, Gabriela and {Gubitosi}, Giulia and {Musella}, Ilaria and {Banik}, Indranil and {Szapudi}, Istvan and {Singal}, Jack and {Cases}, Jaume Haro and {Chluba}, Jens and {Torrado}, Jes{\'u}s and {Mifsud}, Jurgen and {Jedamzik}, Karsten and {Said}, Khaled and {Dialektopoulos}, Konstantinos and {Herold}, Laura and {Perivolaropoulos}, Leandros and {Zu}, Lei and {Galbany}, Llu{\'\i}s and {Breuval}, Louise and {Visinelli}, Luca and {Escamilla}, Luis A. and {Anchordoqui}, Luis A. and {Sheikh-Jabbari}, M.~M. and {Lembo}, Margherita and {Dainotti}, Maria Giovanna and {Vincenzi}, Maria and {Asgari}, Marika and {Gerbino}, Martina and {Forconi}, Matteo and {Cantiello}, Michele and {Moresco}, Michele and {Benetti}, Micol and {Sch{\"o}neberg}, Nils and {Akarsu}, {\"O}zg{\"u}r and {Nunes}, Rafael C. and {Bernardo}, Reginald Christian and {Ch{\'a}vez}, Ricardo and {Anderson}, Richard I. and {Watkins}, Richard and {Capozziello}, Salvatore and {Li}, Siyang and {Vagnozzi}, Sunny and {Pan}, Supriya and {Treu}, Tommaso and {Irsic}, Vid and {Handley}, Will and {Giar{\`e}}, William and {Murakami}, Yukei and {Banihashemi}, Abdolali and {Poudou}, Ad{\`e}le and {Heavens}, Alan and {Kogut}, Alan and {Domi}, Alba and {Lenart}, Aleksander {\L}ukasz and {Melchiorri}, Alessandro and {Vadal{\`a}}, Alessandro and {Amon}, Alexandra and {Rivera}, Alexander Bonilla and {Reeves}, Alexander and {Zhuk}, Alexander and {Bonanno}, Alfio and {{\"O}vg{\"u}n}, Ali and {Pisani}, Alice and {Talebian}, Alireza and {Abebe}, Amare and {Aboubrahim}, Amin and {Gonz{\'a}lez Mor{\'a}n}, Ana Luisa and {Kov{\'a}cs}, Andr{\'a}s and {Lymperis}, Andreas and {Papatriantafyllou}, Andreas and {Liddle}, Andrew R. and {Paliathanasis}, Andronikos and {Borowiec}, Andrzej and {Yadav}, Anil Kumar and {Yadav}, Anita and {Sen}, Anjan Ananda and {William}, Anjitha John and {Davis}, Anne Christine and {Shajib}, Anowar J. and {Walters}, Anthony and {Lonappan}, Anto Idicherian and {Chudaykin}, Anton and {Capodagli}, Antonio and {da Silva}, Antonio and {De Felice}, Antonio and {Racioppi}, Antonio and {Oficial}, Araceli Soler and {Montiel}, Ariadna and {Favale}, Arianna and {Bernui}, Armando and {Velasco}, Arrianne Crystal and {Heinesen}, Asta and {Bakopoulos}, Athanasios and {Chatzistavrakidis}, Athanasios and {Khanpour}, Bahman and {Sathyaprakash}, Bangalore S. and {Zgirski}, Bartek and {L'Huillier}, Benjamin and {Famaey}, Benoit and {Jain}, Bhuvnesh and {Zhang}, Bing and {Karmakar}, Biswajit and {Dragovich}, Branko and {Thomas}, Brooks and {Correa}, Carlos and {Boiza}, Carlos G. and {Marques}, Catarina and {Escamilla-Rivera}, Celia and {Tzerefos}, Charalampos and {Zhang}, Chi and {De Leo}, Chiara and {Pfeifer}, Christian and {Lee}, Christine and {Venter}, Christo and {Gomes}, Cl{\'a}udio and {Roque De bom}, Clecio and {Moreno-Pulido}, Cristian and {Iosifidis}, Damianos and {Grin}, Dan and {Blixt}, Daniel and {Scolnic}, Dan and {Oriti}, Daniele and {Dobrycheva}, Daria and {Bettoni}, Dario and {Benisty}, David and {Fern{\'a}ndez-Arenas}, David and {Wiltshire}, David L. and {Sanchez Cid}, David and {Tamayo}, David and {Valls-Gabaud}, David and {Pedrotti}, Davide and {Wang}, Deng and {Staicova}, Denitsa and {Totolou}, Despoina and {Rubiera-Garcia}, Diego and {Milakovi{\'c}}, Dinko and {Pesce}, Dominic W. and {Sluse}, Dominique and {Borka}, Du{\v{s}}ko and {Yusofi}, Ebrahim and {Giusarma}, Elena and {Terlevich}, Elena and {Tomasetti}, Elena and {Vagenas}, Elias C. and {Fazzari}, Elisa and {Ferreira}, Elisa G.~M. and {Barakovic}, Elvis and {Dimastrogiovanni}, Emanuela and {Holm}, Emil Brinch and {Mottola}, Emil and {{\"O}z{\"u}lker}, Emre and {Specogna}, Enrico and {Brocato}, Enzo and {Jensko}, Erik and {Enriquez}, Erika Antonette and {Bhatia}, Esha and {Bresolin}, Fabio and {Avila}, Felipe and {Bouch{\`e}}, Filippo and {Bombacigno}, Flavio and {Anagnostopoulos}, Fotios K. and {Pace}, Francesco and {Sorrenti}, Francesco and {Lobo}, Francisco S.~N. and {Courbin}, Fr{\'e}d{\'e}ric and {Hansen}, Frode K. and {Sloan}, Greg and {Farrugia}, Gabriel and {Lynch}, Gabriel and {Garcia-Arroyo}, Gabriela and {Raimondo}, Gabriella and {Lambiase}, Gaetano and {Anand}, Gagandeep S. and {Poulot}, Gaspard and {Leon}, Genly and {Kouniatalis}, Gerasimos and {Nardini}, Germano and {Cs{\"o}rnyei}, G{\'e}za and {Galloni}, Giacomo},
        title = "{The CosmoVerse White Paper: Addressing observational tensions in cosmology with systematics and fundamental physics}",
      journal = {Physics of the Dark Universe},
         year = 2025,
        month = sep,
       volume = {49},
          eid = {101965},
        pages = {101965},
          doi = {10.1016/j.dark.2025.101965},
archivePrefix = {arXiv},
       eprint = {2504.01669},
 primaryClass = {astro-ph.CO},
       adsurl = {https://ui.adsabs.harvard.edu/abs/2025PDU....4901965D}}

@ARTICLE{H0DN_2026,
       author = {{H0DN Collaboration}},
        title = "{The Local Distance Network: a community consensus report on the measurement of the Hubble constant at 1\% precision}",
      journal = {A\&A},
         year = 2026,
        month = apr,
       volume = {708},
          eid = {A166},
        pages = {A166},
archivePrefix = {arXiv},
       eprint = {2510.23823},
 primaryClass = {astro-ph.CO},
          doi = {10.1051/0004-6361/202557993},
       adsurl = {https://ui.adsabs.harvard.edu/abs/2026A\&A...708A.166H}}

@ARTICLE{Leavitt_1912,
       author = {{Leavitt}, Henrietta S. and {Pickering}, Edward C.},
        title = "{Periods of 25 Variable Stars in the Small Magellanic Cloud.}",
      journal = {Harvard College Observatory Circular},
         year = 1912,
        month = mar,
       volume = {173},
        pages = {1-3},
       adsurl = {https://ui.adsabs.harvard.edu/abs/1912HarCi.173....1L}}

@ARTICLE{DESI_2025,
       author = {{DESI Collaboration}},
        title = "{DESI DR2 results. II. Measurements of baryon acoustic oscillations and cosmological constraints}",
      journal = {Physical Review D},
         year = 2025,
        month = oct,
       volume = {112},
       number = {8},
          eid = {083515},
        pages = {083515},
          doi = {10.1103/tr6y-kpc6},
archivePrefix = {arXiv},
       eprint = {2503.14738},
 primaryClass = {astro-ph.CO},
       adsurl = {https://ui.adsabs.harvard.edu/abs/2025PhRvD.112h3515A}}

@ARTICLE{Farshad_2026,
       author = {{Kamalinejad}, Farshad and {Slepian}, Zachary and {Krolewski}, Alex and {Greco}, Alessandro and {Ortol{\'a} Leonard}, William and {Chellino}, Jessica and {Reinhard}, Matthew and {Fern{\'a}ndez-Garc{\'\i}a}, Elena and {Prada}, Francisco and {Aguilar}, J. and {Ahlen}, S. and {Anand}, A. and {Bebek}, C. and {Bianchi}, D. and {Brooks}, D. and {Claybaugh}, T. and {Cuceu}, A. and {Dawson}, K.~S. and {de la Macorra}, A. and {Demina}, R. and {Doel}, P. and {Edelstein}, J. and {Forero-Romero}, J.~E. and {Gazta{\~n}aga}, E. and {Gontcho}, S. Gontcho A and {Gutierrez}, G. and {Herrera-Alcantar}, H.~K. and {Honscheid}, K. and {Howlett}, C. and {Huterer}, D. and {Ishak}, M. and {Joyce}, R. and {Juneau}, S. and {Kirkby}, D. and {Kisner}, T. and {Kremin}, A. and {Lahav}, O. and {Lamman}, C. and {Landriau}, M. and {Le Guillou}, L. and {Manera}, M. and {Meisner}, A. and {Miquel}, R. and {Newman}, J.~A. and {Percival}, W.~J. and {Poppett}, C. and {P{\'e}rez-R{\`a}fols}, I. and {Samushia}, L. and {Sanchez}, E. and {Schlegel}, D. and {Schubnell}, M. and {Seo}, H. and {Silber}, J. and {Sprayberry}, D. and {Tarl{\'e}}, G. and {Weaver}, B.~A. and {Zhao}, C. and {Zou}, H.},
        title = "{First Detection of the Baryon Acoustic Oscillation (BAO) Feature in the 3-Point Correlation Function of DESI DR1 Luminous Red Galaxies}",
      journal = {ArXiv e-prints},
         year = 2026,
        month = feb,
          eid = {arXiv:2602.16134},
       volume = {Arxiv},
          doi = {10.48550/arXiv.2602.16134},
archivePrefix = {arXiv},
       eprint = {2602.16134},
 primaryClass = {astro-ph.CO},
       adsurl = {https://ui.adsabs.harvard.edu/abs/2026arXiv260216134K}}

@ARTICLE{Chen_2024_BAO,
       author = {{Chen}, S. -F. and {Howlett}, C. and {White}, M. and {McDonald}, P. and {Ross}, A.~J. and {Seo}, H. -J. and {Padmanabhan}, N. and {Aguilar}, J. and {Ahlen}, S. and {Alam}, S. and {Alves}, O. and {Andrade}, U. and {Blum}, R. and {Brooks}, D. and {Chen}, X. and {Cole}, S. and {Dawson}, K. and {de la Macorra}, A. and {Dey}, A. and {Ding}, Z. and {Doel}, P. and {Ferraro}, S. and {Font-Ribera}, A. and {Forero-S{\'a}nchez}, D. and {Forero-Romero}, J.~E. and {Garcia-Quintero}, C. and {Gazta{\~n}aga}, E. and {Gontcho}, S.~G.~A. and {Hanif}, M.~M.~S. and {Honscheid}, K. and {Kisner}, T. and {Kremin}, A. and {Lambert}, A. and {Landriau}, M. and {Levi}, M.~E. and {Manera}, M. and {Meisner}, A. and {Mena-Fern{\'a}ndez}, J. and {Miquel}, R. and {Munoz-Gutierrez}, A. and {Paillas}, E. and {Palanque-Delabrouille}, N. and {Percival}, W.~J. and {P{\'e}rez-Fern{\'a}ndez}, A. and {Prada}, F. and {Rashkovetskyi}, M. and {Rezaie}, M. and {Rosado-Marin}, A. and {Rossi}, G. and {Ruggeri}, R. and {Sanchez}, E. and {Schlegel}, D. and {Silber}, J. and {Tarl{\'e}}, G. and {Vargas-Maga{\~n}a}, M. and {Weaver}, B.~A. and {Yu}, J. and {Yuan}, S. and {Zhou}, R. and {Zhou}, Z.},
        title = "{Baryon acoustic oscillation theory and modelling systematics for the DESI 2024 results}",
      journal = {MNRAS},
         year = 2024,
        month = oct,
       volume = {534},
       number = {1},
        pages = {544-574},
          doi = {10.1093/mnras/stae2090},
archivePrefix = {arXiv},
       eprint = {2402.14070},
 primaryClass = {astro-ph.CO},
       adsurl = {https://ui.adsabs.harvard.edu/abs/2024MNRAS.534..544C}}

@ARTICLE{Lewis_2025,
       author = {{Lewis}, Antony and {Chamberlain}, Ewan},
        title = "{Understanding acoustic scale observations: the one-sided fight against {\ensuremath{\Lambda}}}",
      journal = {JCAP},
         year = 2025,
        month = may,
       volume = {2025},
       number = {5},
          eid = {065},
        pages = {065},
          doi = {10.1088/1475-7516/2025/05/065},
archivePrefix = {arXiv},
       eprint = {2412.13894},
 primaryClass = {astro-ph.CO},
       adsurl = {https://ui.adsabs.harvard.edu/abs/2025JCAP...05..065L}}

@ARTICLE{Mirpoorian_2025,
       author = {{Mirpoorian}, Seyed Hamidreza and {Jedamzik}, Karsten and {Pogosian}, Levon},
        title = "{Is dynamical dark energy necessary? DESI BAO and modified recombination}",
      journal = {JCAP},
         year = 2025,
        month = dec,
       volume = {2025},
       number = {12},
          eid = {050},
        pages = {050},
          doi = {10.1088/1475-7516/2025/12/050},
archivePrefix = {arXiv},
       eprint = {2504.15274},
 primaryClass = {astro-ph.CO},
       adsurl = {https://ui.adsabs.harvard.edu/abs/2025JCAP...12..050M}}

@ARTICLE{DES_2024_SNe,
       author = {{DES Collaboration}},
        title = "{The Dark Energy Survey: Cosmology Results with {\ensuremath{\sim}}1500 New High-redshift Type Ia Supernovae Using the Full 5 yr Data Set}",
      journal = {ApJL},
         year = 2024,
        month = sep,
       volume = {973},
       number = {1},
          eid = {L14},
        pages = {L14},
          doi = {10.3847/2041-8213/ad6f9f},
archivePrefix = {arXiv},
       eprint = {2401.02929},
 primaryClass = {astro-ph.CO},
       adsurl = {https://ui.adsabs.harvard.edu/abs/2024ApJ...973L..14D}}

@ARTICLE{Popovic_2026,
       author = {{Popovic}, B. and {Shah}, P. and {Kenworthy}, W.~D. and {Kessler}, R. and {Davis}, T.~M. and {Goobar}, A. and {Scolnic}, D. and {Vincenzi}, M. and {Wiseman}, P. and {Chen}, R. and {Charleton}, E. and {Acevedo}, M. and {Armstrong}, P. and {Boyd}, B.~M. and {Brout}, D. and {Camilleri}, R. and {Frieman}, J. and {Galbany}, L. and {Grayling}, M. and {Kelsey}, L. and {Rose}, B. and {S{\'a}nchez}, B. and {Lee}, J. and {M{\"o}ller}, A. and {Smith}, M. and {Sullivan}, M. and {Shiamtanis}, N. and {Alarcon}, A. and {Allam}, S.~S. and {Andrade-Oliveira}, F. and {Avila}, S. and {Bacon}, D. and {Blazek}, J. and {Bocquet}, S. and {Brooks}, D. and {Burke}, D.~L. and {Rosell}, A. Carnero and {Carretero}, J. and {Cawthon}, R. and {da Costa}, L.~N. and {Pereira}, M.~E. da Silva and {Diehl}, H.~T. and {Dodelson}, S. and {Doel}, P. and {Everett}, S. and {Frohmaier}, C. and {Garc{\'\i}a-Bellido}, J. and {Gruen}, D. and {Gutierrez}, G. and {Herner}, K. and {Hinton}, S.~R. and {Hollowood}, D.~L. and {Honscheid}, K. and {Huterer}, D. and {James}, D.~J. and {Jeffrey}, N. and {Kuehn}, K. and {Lahav}, O. and {Lee}, S. and {Lidman}, C. and {Marshall}, J.~L. and {Mena-Fern{\'a}ndez}, J. and {Menanteau}, F. and {Miquel}, R. and {Muir}, J. and {Myles}, J. and {Ogando}, R.~L.~C. and {Paterno}, M. and {Malag{\'o}n}, A.~A. Plazas and {Porredon}, A. and {Prat}, J. and {Nichol}, R.~C. and {Romer}, A.~K. and {Roodman}, A. and {Sanchez}, E. and {Cid}, D. Sanchez and {Sevilla-Noarbe}, I. and {Suchyta}, E. and {Swanson}, M.~E.~C. and {To}, C. and {Tucker}, D.~L. and {Walker}, A.~R. and {Weaverdyck}, N. and {Aguena}, M.},
        title = "{The Dark Energy Survey supernova program: a reanalysis of cosmology results and evidence for evolving dark energy with an updated Type Ia supernova calibration}",
      journal = {MNRAS},
         year = 2026,
        month = jun,
       volume = {548},
       number = {4},
          eid = {stag632},
        pages = {stag632},
          doi = {10.1093/mnras/stag632},
archivePrefix = {arXiv},
       eprint = {2511.07517},
 primaryClass = {astro-ph.CO},
       adsurl = {https://ui.adsabs.harvard.edu/abs/2026MNRAS.548ag632P}}

@ARTICLE{Milgrom_1983,
	author = {{Milgrom}, M.},
	title = "{A modification of the Newtonian dynamics as a possible alternative to the hidden mass hypothesis}",
	journal = {ApJ},
	year = 1983,
	month = jul,
	volume = 270,
	pages = {365-370},
	doi = {10.1086/161130},
	adsurl = {http://adsabs.harvard.edu/abs/1983ApJ...270..365M}}
\bsp 
\label{lastpage}
\end{document}